\documentclass[longauth]{aa}  

\usepackage{graphicx}
\usepackage{txfonts}
\usepackage{natbib}
\usepackage{xcolor}
\usepackage[colorlinks=true,linkcolor=blue,citecolor=blue,filecolor=blue,urlcolor=blue]{hyperref}

\begin{document}

   \title{JWST Observations of Young protoStars (JOYS)}

   \subtitle{Overview of program and early results}

   \author{E.F. van Dishoeck
          \inst{1,2}
          \and
          {\L}. Tychoniec \inst{1}
          \and
          W.~R.~M. Rocha\inst{1,3}
          \and
          K. Slavicinska\inst{1,3}
          \and
          L. Francis\inst{1}
          \and
          M.~L. van Gelder\inst{1}
          \and
          T.~P. Ray\inst{4}
          \and
          H. Beuther\inst{5}
          \and
          A. Caratti o Garatti\inst{6}
          \and
          N.~G.~C. Brunken\inst{1}
          \and
          Y. Chen\inst{1}
          \and
          R. Devaraj\inst{4}
          \and
          V.~C. Geers\inst{7}
          \and
          C. Gieser\inst{5,2}
          \and
          T.~P. Greene\inst{8}
          \and
          K. Justtanont\inst{9}
          \and
          V.~J.~M. Le Gouellec\inst{8,10,11}
          \and
          P.~J. Kavanagh\inst{12}
          \and
          P.~D. Klaassen\inst{7}
          \and
          A.~G.~M. Janssen\inst{1}
          \and
          M.~G. Navarro\inst{13}
          \and
          P. Nazari\inst{14}
          \and
          S. Notsu\inst{15,16}
          \and
          G. Perotti\inst{5,17}
          \and
          M.~E. Ressler\inst{18}
          \and
          S.~D. Reyes\inst{5}
          \and
          A.~D. Sellek\inst{1}
          \and
          B. Tabone\inst{19}
          \and
          C.\ Tap\inst{1}
          \and
          N.C.M.A.\ Theijssen\inst{1}
          \and
          L. Colina\inst{20}
          \and
          M. G\"udel\inst{21,22}
          \and
          Th. Henning\inst{5}
          \and
          P.-O. Lagage\inst{23}
          \and
          G. \"Ostlin\inst{24}
          \and
          B. Vandenbussche\inst{25}
          \and
          G.~S.Wright\inst{7}
        }

          \institute{Leiden Observatory, Leiden University, P.O. Box 9513,
            2300 RA Leiden, The Netherlands \\
              \email{ewine@strw.leidenuniv.nl}
              \and
              Max Planck Institut f\"ur Extraterrestrische Physik (MPE), Giessenbachstrasse 1, 85748 Garching, Germany
              \and
              Laboratory for Astrophysics, Leiden Observatory, Leiden University , P.O. Box 9513, 2300 RA Leiden, The Netherlands
              \and
              School of Cosmic Physics, Dublin Institute for Advanced Studies, 31 Fitzwilliam Place, D02 XF86, Dublin, Ireland
              \and
              Max Planck Institute for Astronomy, K\"onigstuhl 17, 69117 Heidelberg, Germany
              \and
              INAF-Osservatorio Astronomico di Capodimonte, Salita Moiariello 16, 80131 Napoli, Italy
              \and
              UK Astronomy Technology Centre, Royal Observatory Edinburgh, Blackford Hill, Edinburgh EH9 3HJ, UK
              \and
              Space Science and Astrobiology Division, NASA’s Ames Research Center, Moffett Field, CA 94035, USA
              \and
              Department of Space, Earth and Environment, Chalmers University of Technology, Onsala Space Observatory, 439 92 Onsala, Sweden
              \and
              Institut de Ciencies de l’Espai (ICE-CSIC), Campus UAB, Carrer de Can Magrans S/N, E-08193 Cerdanyola del Valles, Catalonia
              \and
              Institut d’Estudis Espacials de Catalunya (IEEC), c/ Gran
              Capit\'a, 2-4, 08034 Barcelona, Spain
             \and
              Department of Physics, Maynooth University, Maynooth, Co. Kildare, Ireland
              \and
INAF—Osservatorio Astronomico di Roma, Via di Frascati 33, 00078 Monte Porzio Catone, Italy
              \and
              European Southern Observatory (ESO), Karl-Schwarzschild-Strasse 2, 1780 85748 Garching, Germany
              \and
              Department of Earth and Planetary Science, Graduate School of Science, University of Tokyo, 7-3-1 Hongo, Bunkyo-ku, Tokyo 113-0033, Japan
              \and
Star and Planet Formation Laboratory, RIKEN Cluster for Pioneering Research,
2-1 Hirosawa, Wako, Saitama 351-0198, Japan
              \and
Niels Bohr Institute, University of Copenhagen, NBB BA2, Jagtvej
155A, 2200 Copenhagen, Denmark
              \and
              Jet Propulsion Laboratory, California Institute of Technology, 4800 Oak Grove Drive, Pasadena, CA 91109, USA
              \and
Universit\'e Paris-Saclay, CNRS, Institut d’Astrophysique Spatiale, 91405 Orsay, France
              \and
               Centro de Astrobiolog{\i}a (CAB) CSIC-INTA, Ctra. de Ajalvir km 4, Torrej{\o}n de Ard{\o}z, 28850, Madrid, Spain
               \and
              Department of Astrophysics, University of Vienna, T\"urkenschanzstrasse 17, A-1180 Vienna, Austria
              \and
              ETH Z\"urich, Institute for Particle Physics and Astrophysics, Wolfgang-Pauli-Strasse 27, 8093 Zürich, Switzerland
              \and
Universit\'e Paris-Saclay, Universit\'e Paris Cit\'e, CEA, CNRS, AIM,
91191, Gif-sur-Yvette, France
               \and
Department of Astronomy, Oskar Klein Centre, Stockholm University, AlbaNova University Center, 10691 Stockholm, Sweden
               \and
Institute of Astronomy, KU Leuven, Celestijnenlaan 200D, 3001
Leuven, Belgium
}

   \date{Received March 10 2025; accepted May 5}

 
  \abstract
  {The embedded phase of star formation is a crucial period in the
    development of a young star when the system still accretes matter,
    emerges from its natal cloud with assistance from powerful jets and
    outflows, and forms a disk, thus setting the stage for the birth of a
    planetary system. The mid-infrared spectral line observations now
    possible with unprecedented sensitivity, spectral resolution, and
    sharpness from the {\it James Webb Space Telescope} (JWST) are
    key for probing many of the physical and chemical processes on
    sub-arcsecond scales that occur in highly extincted regions. They 
    provide unique diagnostics and complement millimeter
    observations.}
  {The aim of the JWST Observations of Young protoStars (JOYS) program is to
    address a wide variety of topics ranging from protostellar
    accretion and the nature of primeval jets, winds, and outflows to
    the chemistry of gas and ice in hot cores and cold dense
    protostellar environments to the characteristics of the embedded
    disks. We introduce the JOYS program and show representative
    results. }
  {The JWST Mid-InfraRed Instrument (MIRI) Medium Resolution
    Spectrometer (MRS) Integral Field Unit (IFU) 5--28 $\mu$m maps of
    17 low-mass targets (23 if binary components are counted
    individually) and six high-mass protostellar sources were taken
    with resolving powers $R=\lambda/\Delta \lambda=1500-4000$. We
    used small mosaics ranging from $1\times 1$ to $3\times 3$ MRS
    tiles to cover $\sim 4''$ to $20''$ fields of view, providing
    spectral imaging on spatial scales down to $\sim$30 au (low mass)
    and $\sim$600 au (high mass). For HH 211, the complete $\sim 1'$
    blue outflow lobe was mapped with the MRS. Atomic lines were
    interpreted with published shock models, whereas molecular lines
    were analyzed with simple rotation diagrams and local
    thermodynamic equilibrium slab models. We stress the importance of
    taking infrared pumping into account. Inferred abundance ratios
    were compared with detailed hot core chemical models including
    X-rays, whereas ice spectra were fit through comparison with
    laboratory spectra.}
  {The JWST MIRI-MRS spectra show a wide variety of features, with
    their spatial distribution providing key insight into their
    physical origin. The atomic line maps differ among refractory
    (e.g., Fe), semi-refractory (e.g., S), and volatile elements
    (e.g., Ne) and are linked to their different levels of depletion
    and local (shock) conditions. Jets are prominently seen in lines
    of [Fe II] and other refractory elements, whereas the pure
    rotational H$_2$ lines probe hot ($\sim 1000$ K) and warm
    (few$\times 10^2$ K) gas inside the cavity, as well as gas
    associated with jets, entrained outflows, and cavity walls for
    both low- and high-mass sources.  Wide-angle winds are found in
    low-$J$ H$_2$ lines. Nested stratified jet structures containing
    an inner ionized core with an outer molecular layer are commonly
    seen in the youngest sources. While [S I] follows the jet as seen
    in [Fe II] in the youngest protostars, it is different in more
    evolved sources, where it is concentrated on source. Noble gas
    lines such as [Ne II] 12.8 $\mu$m reveal a mix of jet shock and
    photoionized emission.  The H I recombination lines serve as a
    measure of protostellar accretion rates but are also associated
    with more extended jets. Gaseous molecular emission (CO$_2$,
    C$_2$H$_2$, HCN, H$_2$O, CH$_4$, SO$_2$, SiO) is seen toward
    several sources, but it is cool compared with what is found in
    more evolved disks, with excitation temperatures of only 100--250
    K, and likely associated with the warm inner envelopes (``hot
    cores'') . Along the outflow, CO$_2$ is often extended, thus
    contrasting with C$_2$H$_2$, which is usually centered on
    source. Water emission is commonly detected on source, even if
    relatively weak. Off source, it is seen only in the highest
    density shocks, such as those associated with NGC 1333
    IRAS4B. Some sources show gaseous molecular lines in absorption,
    including NH$_3$ in one case. Deep ice features are seen toward
    the protostars, revealing not just the major ice components but
    also ions (as part of salts) and complex organic molecules, with
    comparable abundances from low- to high-mass sources. The relative
    abundances of some gas and ice species are similar, which is
    consistent with ice sublimation in hot cores. We present a second
    detection of HDO ice in a solar-mass source, with an HDO/H$_2$O
    ice ratio of $\sim$0.4\%, thus providing a link with HDO/H$_2$O in
    disks and comets. A deep search for solid O$_2$ suggests that it
    is not a significant oxygen reservoir. Only a few embedded Class I
    disks show the same forest of water lines as Class II disks. This
    may be due to significant dust extinction of the upper layers in
    young disks caused by less settling of small dust as well as radial drift
    bringing in fresh dust. }
  {This paper illustrates the wide range of science questions that a
    single MIRI-MRS IFU data set can address. Our data suggest many
    similarities between low- and high-mass sources. Large source
    samples across evolutionary stages and luminosities are needed to
    further develop these diagnostics of the physics and chemistry of
    protostellar systems.
  }

   \keywords{star formation -- outflows and jets --
                circumstellar disks -- protostars: accretion --
                atomic and molecular spectra -- astrochemistry
               }

   \maketitle
%

\nolinenumbers
   
\section{Introduction}
\label{sec:intro}

The embedded phase of star formation, which is typically up to a few
$\times 10^5$ yr after cloud collapse, is a critical and highly active
period in the evolution of a protostellar system. It represents the
moment when the source gathers most of its mass and is emerging from
its dense environment \citep{Andre00,Evans09,Dunham14,Fischer17}. In
addition to accretion onto the protostar itself, many other physical
processes occur simultaneously in its immediate surroundings: infall
from the natal cloud and collapsing envelope onto the young disk; jets
and winds being ejected from the star-disk system; outflows sweeping
up and shocking the material; and ultraviolet (UV) photons and X-rays
heating, ionizing, and dissociating the gas
\citep[e.g.,][]{Arce07,Bally16,Tobin24}. Young disks are massive and
still spreading radially \citep[e.g.,][]{Hueso05,Ohashi23}. Notably,
planet formation has already started at this early stage
\citep{Manara18,Tychoniec20}. Chemical processes range from freeze-out
and ice chemistry in the cold outer parts to ice sublimation and high
temperature chemistry in ``hot cores'' and shocks
\citep[e.g.,][]{vanDishoeck98,Jorgensen20,Ceccarelli23}. Ultimately,
the early pre- and protostellar stages provide the chemical building
blocks of disks, comets, and planets
\citep[e.g.,][]{Caselli12,Oberg21}.

However, because of tens to hundreds of magnitudes of visual
extinction in the inner regions, these processes and materials can
often only be probed at infrared and millimeter-centimeter
wavelengths. The Atacama Large Millimeter/submillimeter Array (ALMA)
has allowed protostellar systems to be imaged in the warm and cold(er)
molecular gas and dust continuum at subarcsec scales, thus enabling
their physical and chemical structure to be probed
\citep[e.g.,][]{Tobin24}.  However, many key processes, as traced by
diagnostic lines of warm-to-hot (shocked) gas and of solid state
material, can only be observed at mid- and far-infrared wavelengths.

Pre-JWST observations with the {\it Infrared Space Observatory} (ISO),
the {\it Spitzer Space Telescope}, the {\it Herschel Space
  Observatory}, and ground-based facilities (e.g., VLT, Keck, IRTF)
have made great progress in finding and characterizing protostars
\citep[e.g.,][]{Furlan08,Evans09,Dunham14}; in using atomic, H$_2$,
and CO lines to probe shock physics
\citep[e.g.,][]{Rosenthal00,Dionatos09,Watson16,Karska18}; in tracing
H$_2$O from clouds to disks \citep[e.g.,][]{vanDishoeck21}; and in
making an inventory of interstellar ices \citep[e.g.,][]{Boogert15}.
However, all of these observations have so far been hampered by either
poor spatial resolution, poor spectral resolution, and/or low
sensitivity. Optical and near-infrared imaging and spectroscopic data
from the ground and with HST have provided high spatial and spectral
resolution information on jets and outflows but only in a limited
number of tracers at positions with low extinction further away from
the source \citep[e.g.,][]{McCaughrean94,Giannini15}. While radio
observations can probe ionized jets and masers in the inner regions of
deeply embedded sources \citep[e.g.,][]{Anglada18,Ray21,Moscadelli22},
they lack many other diagnostics. As a result, our understanding of
the physical and chemical structure of this early deeply embedded
protostellar phase is still incomplete.

The {\it James Webb Space Telescope} (JWST) \citep{Rigby23,Gardner23},
in particular its Mid-InfraRed Instrument (MIRI)
\citep{Rieke15,Wright23}, allows for the next big leap in star
formation and protostellar research by bridging the millimeter and
near-infrared wavelength regime and providing unique information. The
spatial resolution of MIRI of 0.19, 0.58, and 0.96$''$ at 5, 15, and
25 $\mu$m, respectively, corresponds to 29, 87, and 144 au at typical
distances of low-mass protostars ($\sim$150 pc, $<10^2$ L$_{\odot}$)
and 570, 1740, and 2800 au at that of high-mass protostars ($\sim$ 3
kpc, $>10^4$ L$_\odot$). These scales are well matched to the sizes of
their envelopes and disks.  The Medium Resolution Spectrometer (MRS)
enables mapping of the inner parts of jets and outflows through its
Integral Field Unit (IFU) \citep{Wells15,Argyriou23}. The MRS spectral
resolving power of $R=\lambda/\Delta \lambda \sim 1500-4000$ is much
higher than that of {\it Spitzer} (which had
$R$=50--100 at $<$10 $\mu$m; $R=600$ at $>$10 $\mu$m), thus boosting
line-to-continuum ratios of gas-phase lines and allowing solid-state
bands to be fully resolved.  Combined with mJy sensitivity for IFU
spectroscopy, this means that the inner parts of collapsing envelopes
and the outflows of protostars can be dissected on subarcsec scales in
the mid-infrared for the first time. Nearly all previous mid-infrared
spectral line studies were based on single pointings with 5--40$''$
aperture sizes.

The near-infrared spectrograph NIRSpec on JWST \citep{Boker23} has
similar capabilities as MIRI-MRS, but it covers the 1--5 $\mu$m range at
$R$ up to 2700. Its IFU spectrometer is also well suited for studying
the inner regions of protostars at greatly enhanced sensitivity than
was possible before \citep[e.g.,][]{Federman24}. Several protostellar
studies with JWST, including our program, therefore include both
MIRI-MRS 5--28 $\mu$m and NIRSpec-IFU 3--5 $\mu$m data (see references
below). This paper, however, focuses almost exclusively on the MIRI data.

\begin{figure}[t]
\begin{centering}
\includegraphics[width=9cm]{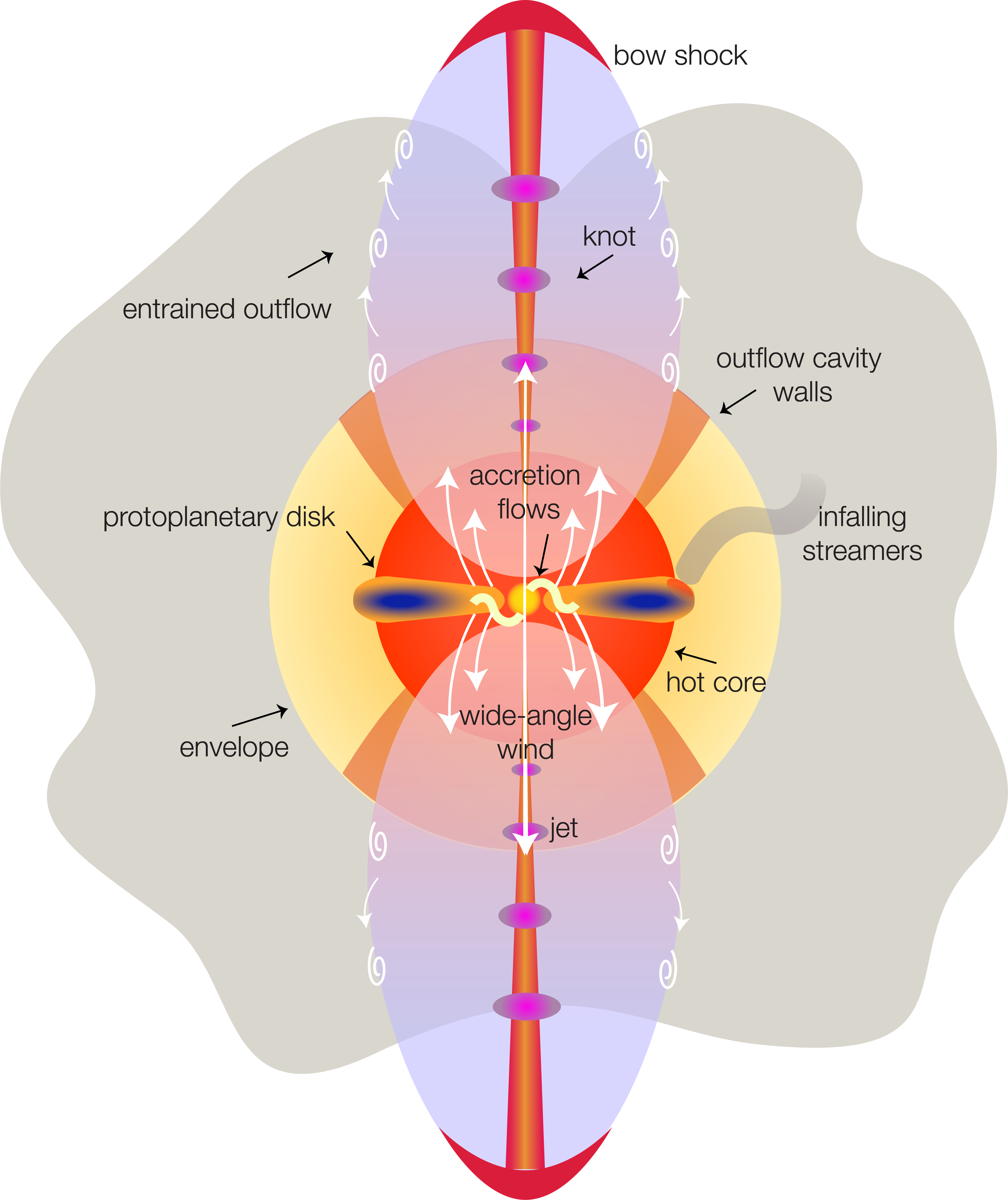}
\caption{Schematic of a protostellar source with the various physical
  components studied in this work indicated.}
\end{centering}
\label{fig:protostar-cartoon}
\end{figure}

\begin{figure*}[t]
\begin{centering}
\includegraphics[width=15cm]{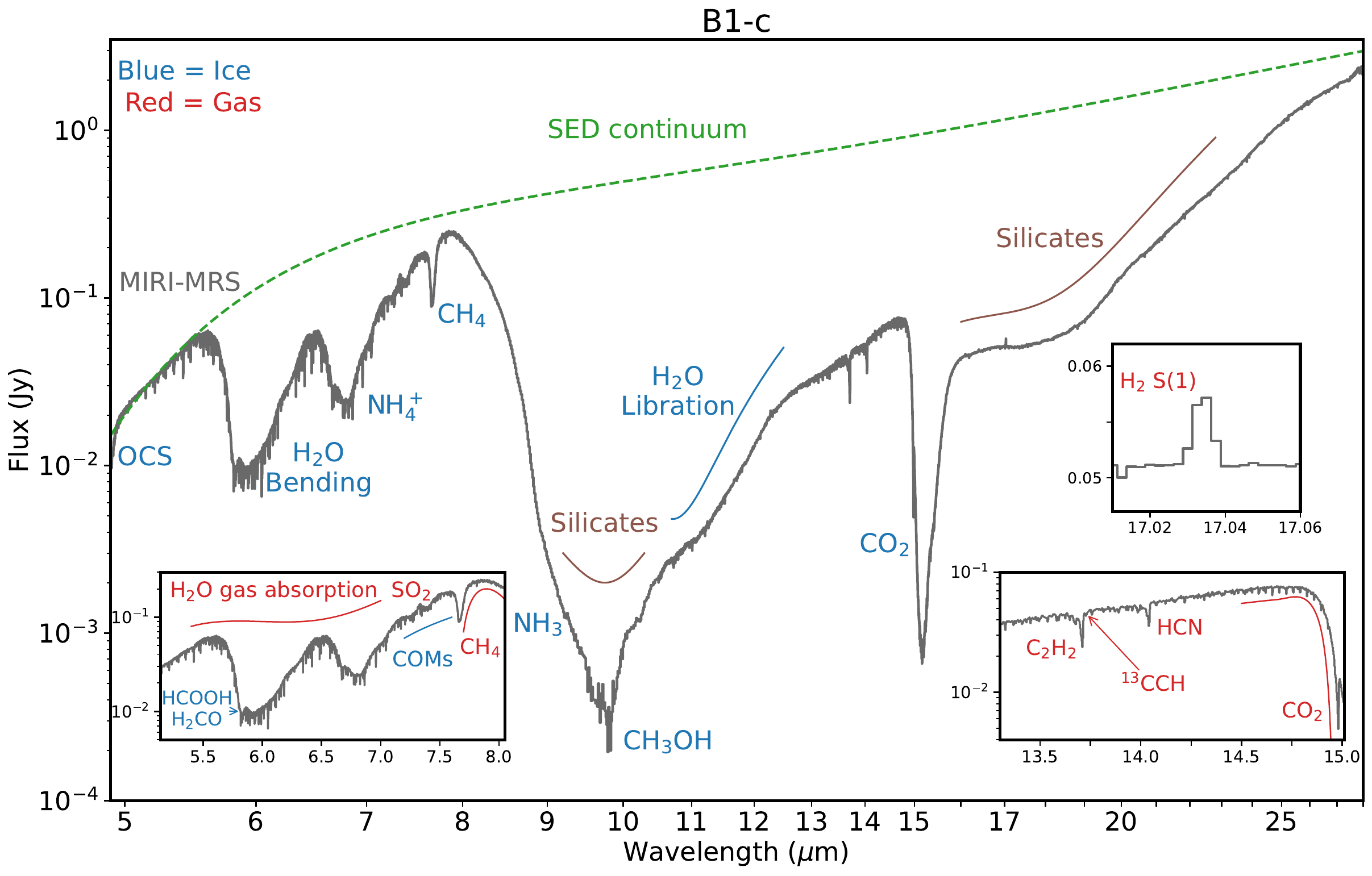}
\includegraphics[width=15cm]{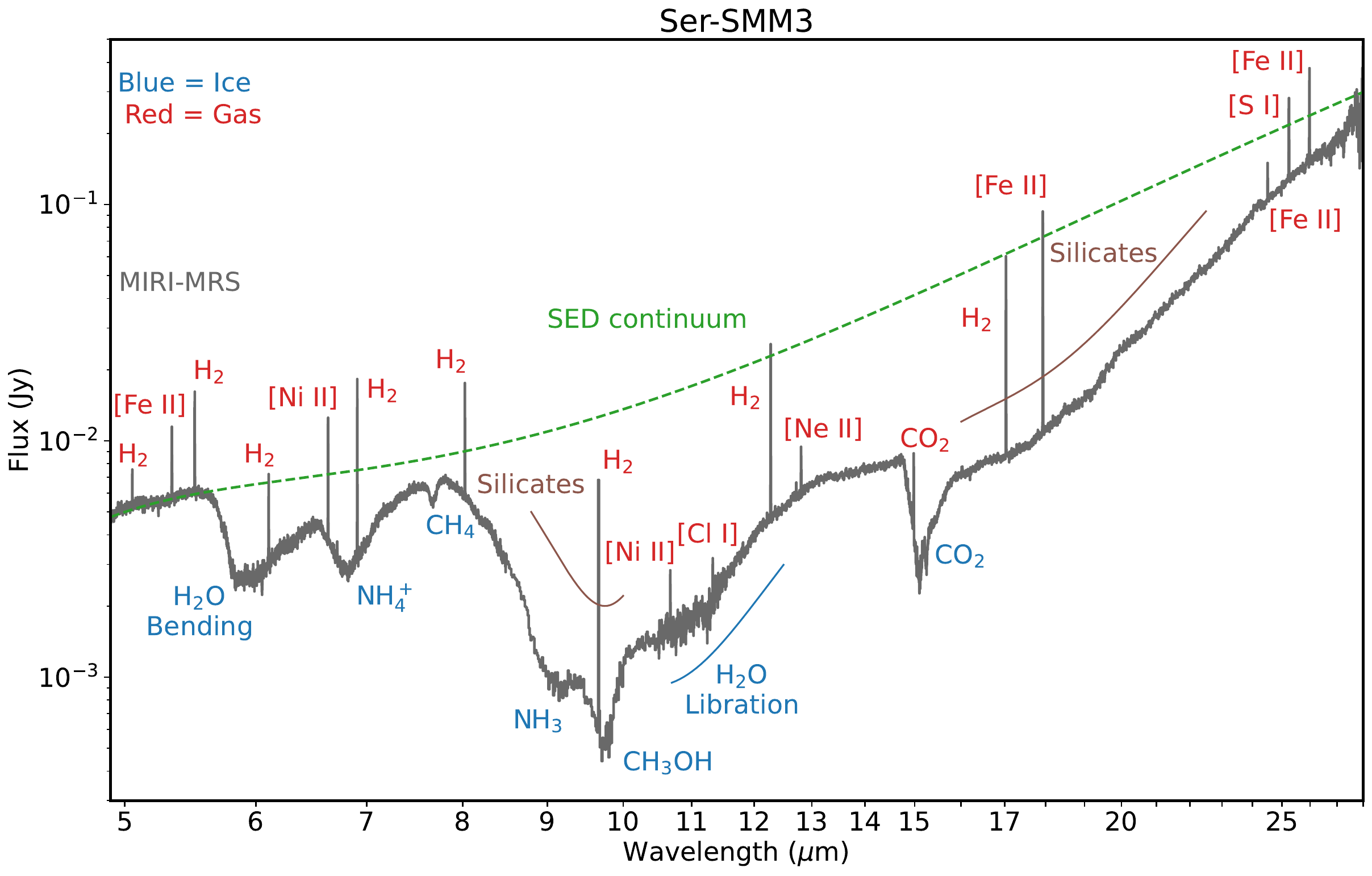}
\caption{Spectra obtained with JWST MIRI-MRS of the low-mass Class 0 protostars
  B1-c (top) and Serpens SMM3 (bottom) illustrating the different
  molecular gas (red) and ice (blue) features that can be observed and
  analyzed. Different components of the protostellar
  system are enhanced in the panels. The B1-c spectrum (top) highlights the
  gas-phase molecular lines in this source. Atomic emission lines are
  present but not very strong. In
  contrast, both H$_2$ and atomic emission lines are very prominent
  for Serpens SMM3 (bottom). At the bottom of the silicate feature at
  10 $\mu$m, close to the noise limit, the spectra have been binned by
  a factor of four to enhance the $S/N$. (See Fig.\ B.3 and B.11 of
  \citet{vanGelder24overview} for individual MIRI-MRS sub-bands.)}
\label{fig:B1c-overview}
\end{centering}
\end{figure*}

In this paper, we outline the MIRI Guaranteed Time Observations (GTO)
``JWST Observations of Young protoStars'' (JOYS) program centered on
MIRI-MRS 5--28 $\mu$m observations of protostellar sources from low
($<$1 L$_{\odot}$) to high luminosity ($>10^4$ L$_{\odot}$) and from
the earliest deeply embedded stages where the envelope mass is much
larger than the stellar mass -- Class 0 for low-mass and infrared dark
clouds (IRDCs) for high-mass protostars -- to the transitional stage
where the star and disk are fully assembled and only a tenuous
envelope is left (Class I/II for low-mass protostars)
\citep{vanDishoeck23,Beuther23}.  Our program 1290 (PI: E.F. van
Dishoeck) uses a single observational approach for a set of 23 (17
low-mass plus 6 high-mass) protostellar targets (the total is 32 if
resolved low- and high-mass binary components are counted
individually) in order to address the scientific topics listed in
Sect.~\ref{sec:background}. For each source, single IFU images or small
mosaics (up to $3\times 3$) were taken covering the central protostar,
its immediate envelope structure, and the inner region of the jets,
winds, and outflows (Fig.~\ref{fig:protostar-cartoon}). The field of
view (FoV) of the IFU varies from the shortest to the longest
wavelengths between $3.2''\times 3.7''$ and $6.6''\times 7.7''$, so
the $3\times 3$ mosaic covers a $10-20''$ region depending on
wavelength, thus corresponding to scales of $\sim 1000$ (low-mass) to
$\sim 10000$ (high-mass) au.  One source, Herbig-Haro (HH) 211, is
covered over the full extent of its blue outflow lobe and part of the
red lobe, $\sim 1'$ in length, by the MRS in program 1257 (PI: T.\
Ray) \citep{Ray23,Caratti24}.  The resulting spectral images provide a
spatially resolved census of the molecular, atomic, and ionized
species covered in the 5--28 $\mu$m range
(Fig.~\ref{fig:B1c-overview}). Nearly all IFU pixels have a rich
mid-infrared spectrum, making this program ideally suited for the IFU.

The 56-hr JOYS MIRI European Consortium GTO program (PIDs 1290 +
1257)\footnote{{\tt miri.strw.leidenuniv.nl}} is being carried out in
close collaboration with three other programs: the MIRI GTO program
1236 (PI: M.\ Ressler) covering ten protostellar Class 0 and I
binaries in Perseus (12.7 hr), the GTO program 1186 (PI: T.\ Greene)
obtaining NIRSpec IFU 1--5 $\mu$m observations of two Class 0
protostars in Serpens (11.9 hr) that are part of JOYS, and the General
Observer (GO) program 1960 (PI: E.F. van Dishoeck) targeting most of
the low-mass protostars with NIRSpec IFU spectroscopy (21.5
hr). Together, these five programs are denoted as JOYS+. In this
paper, we only focus on the program overview and representative
results from MIRI observations taken within the JOYS program (PIDs
1290 + 1257), but we note opportunities for future studies with the
full JOYS+ data set.

Several other JWST MIRI and NIRSpec GO programs focused on protostars
have been carried out in JWST Cycle 1, most notably the Investigating
Protostellar Accretion (IPA) program (PI: T. Megeath, PID 1802, 65.1
hr) \citep{Federman24,Rubinstein24,Narang24,Brunken24,Tyagi25}; the
CORINOS program (PI: Y. Yang, PID 2151, 24.6 hr)
\citep{Yang22,Salyk24,Okoda25}; PROJECT-J (PI: B. Nisini, PID 1706,
23.1 hr) \citep{Nisini24}; a deep MIRI+NIRSpec spectrum of L1527, a
source that is also contained in JOYS (PI: J. Tobin, PID 1798, 7.6
hr); the IceAge Early Release Science (ERS; PI: M. McClure, PID 1309,
33.9 hr) \citep{McClure23,Rocha24Ced}; and It'sCOMplicated GO programs
(PI: M. McClure, PID 1854, 17.7 hr) \citep{McClure25}.  We fold the
initial publications from these programs into the discussion of the
JOYS program results. We also make a comparison of young disks studied
within JOYS with the more mature Class II disks studied in the MIRI
mid-INfrared Disk Survey (MINDS) GTO program (PI: Th. Henning, PID
1282) \citep{Kamp23,vanDishoeck23,Henning24}. In addition, we make a link
with the rich molecular spectroscopy now seen in extragalactic sources
such as part of the Mid-Infrared Characterisation of Nearby Iconic
galaxy Centers (MICONIC) MIRI GTO program \citep{Buiten25}.

The outline of this paper is as follows.  Sect.\ 2 summarizes the
science cases and terminology used in this paper as well as the
different diagnostic features at mid-infrared wavelengths.  Sect.\ 3
presents the observational strategy and data reduction details.  The
subsequent sections present early and new results concerning the
different science topics highlighted in Sect.~\ref{sec:components} and
Fig. ~\ref{fig:protostar-cartoon}. Sect.\ 4 focuses on protostellar
accretion rates and variability, while Sect.\ 5 centers on images of
jets, winds, and outflows as traced by different volatile,
semi-refractory, and refractory species, and it presents a comparison
of different derivations of mass-loss rates. Sect.\ 6 summarizes the
gas-phase molecules seen in warm envelopes (hot cores), dense
molecular shocks, and winds in the context of different chemical
models. Sect.\ 7 highlights the detection of ices in the cold outer
envelopes, including the detection of icy complex molecules and HDO as
well as a deep search for O$_2$ ice. Sect.\ 8 focuses on the lack of
clear signatures of emission from young embedded disks in most
sources. The low-mass Class 0 sources Serpens SMM3 and B1-c and the
high-mass source IRAS 18089-1732 are used as representative examples
to illustrate the various results.  Figure~\ref{fig:B1c-overview}
provides example protostellar spectra with different gas and ice
features identified in the two low-mass sources; the spectrum of the
high-mass source IRAS 18089-1732 (hereafter IRAS 18089) is presented
in Sect.~\ref{sec:ices}. More detailed background information on each of
these science cases is presented in Sect.~\ref{sec:background} and
Appendix~\ref{sec:app_background} in order to avoid interrupting the
flow of the results. Appendix~\ref{sec:app_IRAS23385} provides an
example of science that can be done with the parallel imaging obtained
in this program.

The overall aim of this paper is to highlight the rich and diverse
science that a single MIRI-MRS data set can address. Future papers
will go into more depth regarding each individual topic across the full
JOYS(+) sample, as is done for the gas-phase molecular lines in
\citet{vanGelder24overview}.

\section{Infrared spectroscopy of protostellar systems}
\label{sec:components}

\subsection{Terminology}
\label{sec:terminology}

Figure~\ref{fig:protostar-cartoon} summarizes the different components
of a protostellar system. In this paper, we use the term ``jet'' to
refer to the high-velocity ($\gtrsim$50 km s$^{-1}$) and highly
collimated (an opening angle of less than a few degrees) material that
is directly ejected from the immediate vicinity of the protostellar
embryo by MHD winds and recollimated into a jet. The term ``outflow''
is mainly used to describe the cold entrained outflow material that is
imaged in low-$J$ CO millimeter lines. However, the term ``outflow''
is sometimes also loosely adopted to refer to the entire
jet+wind+(entrained-)outflow system, both in this paper and in the
literature. The term ``wind'' refers to lower velocity ($\sim$10 km
s$^{-1}$), wide-angle material that, similarly to the jet, is directly
launched from the star-disk system. There is considerable discussion
in the literature on the different types of winds: stellar winds
versus disk winds, with the latter category containing both
photoevaporative, X- and MHD disk winds depending on the launching
mechanism. The jet and wind can be part of a single
physical phenomenon that produces a velocity, temperature and
ionization gradient that decreases away from the jet axis and is
therefore revealed in different tracers. Some wind material can also
be mixed in from the surrounding slower-moving outflow.

Internally in the jet, varying ejection velocities may lead to shocks
that manifest themselves as ``knots,'' also called ``internal working
surfaces.'' At the tip of the outflow when the jet impacts the
interstellar medium at rest, they are usually called ``bow shocks,''
although such curved structures can also be found along the jet.  The
term ``shell'' indicates a curved paraboloidal structure produced when
a wind propagates into the surrounding medium, but we do not use it
here because of the possible confusion with internal working surfaces
and bow shocks.

The term ``protostellar embryo'' is used to indicate the young forming
star itself, whereas the term ``core'' is reserved for the
interstellar cloud core out of which the system forms. Strictly
speaking, the term ``protostar'' refers to the protostellar embryo
only, but is often used loosely in the literature (and in this paper)
to refer to the entire ``protostellar system'' consisting of embryo +
jet + wind + outflow + envelope system. For more evolved Class II (T
Tauri) sources, the young star is indicated with the term ``pre-main
sequence star'' (see papers and reviews by
\citealt{Ferreira06,Arce07,Frank14,Bally16,Ray21,Pascucci23,Bacciotti25}
for further descriptions of these terms).

The protostellar embryo is surrounded by a centrally concentrated
collapsing envelope extending to thousands of au consisting of dense
gas and dust that is heated by the protostellar luminosity and thus
has a decreasing temperature and density with radius
\citep{Jorgensen02,Tobin24}. ``Streamers'' are velocity-coherent
elongated narrow gas structures that may provide fresh material to the
protostellar system \citep[e.g.,][]{Valdivia24}. The inner warm part
of the envelope where ices sublimate ($T\gtrsim 100$~K) is called the
``hot core,'' sometimes also named ``hot corino'' for low-mass
sources. It is distinguished from dense shock-heated or UV-heated
gas. The term ``disk'' is reserved for disk-like elongated structures
for which the gas has close to Keplerian rotation. Elongated
structures seen in dust emission on larger scales are called ``tori.''
If infalling material reaches the disk, for example through streamers,
this may create a slow, dense ``accretion shock.''

\subsection{Mid-infrared spectroscopy}
\label{sec:irdiagnostics}

The mid-infrared regime is very rich spectroscopically, containing a
wealth of atomic, ionic and molecular gas lines, as well as Polycyclic
Aromatic Hydrocarbon (PAH), ice and silicate bands that cannot be
observed at any other wavelength (Fig.~\ref{fig:B1c-overview})
\citep[see review by][]{vanDishoeck04}. Early JWST summaries
are by \citet[][their Table~1]{Yang22,Nisini24} and
\citet{vanGelder24overview}.  The MIRI spatial and spectral resolution
allows one to distinguish the physical processes producing them:
spatially extended outflows and jets can readily be separated from
compact hot cores and disks, and gas-phase lines can be distinguished
from ices.  Key diagnostics in the MIRI 5-28 $\mu$m range are
described in the following paragraphs (see more detailed background
information and references in individual subsections of
Sect.~\ref{sec:background} and Appendix B).

\paragraph{H$_2$:} The mid-infrared pure rotational lines within $v$=0 and 1
  are excited in warm gas up to 2000 K and down to $\sim$100~K,
  probing and imaging its temperature structure and mass. In contrast,
  near-infrared ro-vibrational H$_2$ lines at 2 $\mu$m only probe the
  hottest gas and are often too extincted to be observed in these
  regions. Combined with NIRSpec observations at 2.5--5 $\mu$m, the
  H$_2$ excitation can also be used to distinguish shocks versus UV
  photon-dominated regions (PDRs).

\paragraph{Atomic lines:} Several [Fe II], [Fe I], [S I], [Ne II], and
  [Ne III] lines are commonly seen and cover refractory,
  semi-refractory, and volatile elements. Together with H$_2$, they are
  important diagnostics of dissociative $J-$ versus non-dissociative
  $C-$ type shocks (including shock speed and density) and whether or
  not dust destruction has taken place. The [Ne III]/[Ne II] ratio is
  a diagnostic of ionization state and thus the ionizing
  source. Lines from lower abundance elements such as [Ar II], [Ar
  III], [Ni II], [Co II], and [Cl I] provide additional information on
  the physical structure and elemental depletion of refractory and
  more volatile elements.

  \paragraph{H I recombination lines:} These lines
  provide one of the few possibilities to measure the accretion rate
  of hot (5000--15000 K) gas onto the growing protostellar embryo in
  deeply embedded sources (e.g., H I 7-6 (Humphries $\alpha$), 6-5 and
  8-6). They can also trace extended jet emission, especially the
  high-energy transitions such as 4-3 (Paschen $\beta$) and 7-4
  (Brackett $\gamma$).

\paragraph{Symmetric molecules:} Molecules without a permanent dipole
moment such as CH$_4$, C$_2$H$_2$, CH$_3^{(+)}$, and CO$_2$ cannot be
studied by traditional millimeter techniques but are uniquely
accessible to infrared studies. These are also among the most abundant
C- and O-containing species and are thus important to assess the C and
O chemistry and budgets.

\paragraph{Other molecular lines:} In addition to the symmetric
  molecules, gaseous CO, H$_2$O, HCN, SO$_2$, CS, NH$_3$, and SiO
  emission or absorption can be detected. Their lines arise from the
  inner regions of embedded disks, from the inner warm envelopes
  (``hot cores'') or from non-dissociative shocks. In particular,
  H$_2$O, a molecule that is difficult to observe from the ground due
  to the Earth's atmosphere, can be probed with MIRI through both its
  ro-vibrational band at 6 $\mu$m as well as the high-lying pure
  rotational lines at 12--28 $\mu$m. The excitation of these molecules
  provides a good indicator of gas temperature, whereas their
  abundances probe high temperature gas-phase chemistry and ice
  sublimation.  Radiative pumping by infrared continuum radiation from warm
  dust also plays a role in producing these lines.

  \paragraph{OH:} The OH molecule represents a special case since its
  mid-infrared lines at 9--11 $\mu$m originate from very high energy
  levels that are being populated following water
  photodissociation. This so-called prompt emission provides a measure
  of the UV field combined with the number of photodissociating water
  molecules. At longer mid-infrared wavelengths, OH emission is
  produced by chemical pumping through the O + H$_2$ reaction.

\paragraph{HD:} Several HD lines are covered by MIRI, which together
  with H$_2$ allow the [D]/[H] ratio in dense protostellar
  environments to be determined and compared to the more diffuse ISM
  studied with ultraviolet absorption lines, thereby providing
  information on the amount of D locked up in grains. By targeting
  sources at different galactocentric radii, constraints on the
  [D]/[H] gradient across the Galaxy can be obtained.

\paragraph{Ices:} All dominant ice components in the cold outer
  envelope are probed by JWST (MIRI + NIRSpec) at 2.3--20 $\mu$m.
  Narrow ice bands such as that of CH$_4$ can now be fully resolved,
  in contrast with {\it Spitzer}.  MIRI also enables the
  identification of weak absorption bands in the critical 5--9 $\mu$m
  region due to complex organic molecules and ammonium salts. The ice
  profiles of molecules such as $^{(13)}$CO$_2$ provide information on
  the ice environment and temperature history, whereas the gas/ice
  ratio of species such as CH$_4$, H$_2$O and CO$_2$ can be used to
  test hot core scenarios of ice sublimation at their snowlines. With
  the sensitivity of JWST, ice mapping against the extended
  mid-infrared continuum is possible.

\paragraph{Solids:} The vibrational bands of silicates, oxides and
  carbides occur uniquely at mid-infrared wavelengths. Their band
  profiles are sensitive to grain composition and whether or not they
  have been crystallized due to heating. The silicate optical depth
  provides an independent measure of extinction to the source. See
  Appendix~\ref{sec:app_extinction} for a more in-depth summary of
  extinction determinations.

\paragraph{PAHs:} The PAH features are commonly seen in PDRs associated
  with high-mass star-forming regions, where they measure the
  strength of the local UV field. In contrast, PAH bands are weak or
  absent from embedded protostellar sources, both low- and high mass,
  with only a very low level of PAH emission seen in the surrounding
  cloud. Likely explanations include freeze-out into ices and
  coagulation to larger PAH systems that do not emit, rather than a
  lack of UV radiation to excite them \citep{Geers09}. PAHs
  are not further discussed in this paper except briefly in
  Appendix~\ref{sec:app_IRAS23385}.

\paragraph{Mid-infrared SED:} The infrared part of the spectral energy
  distribution (SED) of a protostellar system is most sensitive to
  geometry and the presence of disks.  JWST allows for detection of more
  deeply embedded sources that were too weak for {\it Spitzer}. The
  wavelength regions on both sides of the 10\,$\mu$m silicate feature
  are particular transparent, allowing one to peer deep into the
  protostellar source. Scattering rather than thermal emission
  dominates at shorter wavelengths: its contribution as function of
  wavelength can now be constrained thanks to the combined spatial and
  spectral information of MIRI's and NIRSpec's IFUs.\\

The NIRSpec 1--5 $\mu$m data are not analyzed here but they add higher
lying atomic (e.g., [Fe II]) and H I recombination lines as well as a
forest of H$_2$ and CO ro-vibrational lines that together probe hotter
gas than studied in the MIRI range \citep[see][for IPA
  program]{Federman24,Rubinstein24}. As noted above, several key ice
features also occur in the NIRSpec range, most notably H$_2$O, CO and
CO$_2$ \citep[e.g.,][]{Brunken24iso,Tyagi25}.

\subsection{Science drivers and background}
\label{sec:background}

This section describes the main science drivers and unique
contributions that JWST MIRI can bring to studies of each of the
protostellar components labeled in
Figure~\ref{fig:protostar-cartoon}. Detailed results from related JWST
programs mentioned in Sect.~\ref{sec:intro} are cited throughout Sect.\
4--8. The terminology used is that of Sect.~\ref{sec:terminology},
whereas more detailed background information and references can be
found in Appendix~\ref{sec:app_background}.

\subsubsection{Protostellar accretion}

The young protostar itself is still growing in the embedded phase but
the rate at which accretion from disk to protostellar embryo occurs is
uncertain and difficult to measure, yet it is a crucial missing
parameter in many protostellar evolution studies.  The optical lines
such as H$\alpha$ and Br$\gamma$, traditionally used to measure
accretion rates for Class II sources \citep{Hartmann16,Manara23}, are
often too extincted, although a few Class 0 sources have been detected
at 2 $\mu$m \citep{LeGouellec24}, sometimes assisted by looking down
outflow cavities. The mid-infrared H I recombination lines have been
proposed as one of the few direct measurements of accretion rates of
deeply embedded sources, in particular the H I 7--6 line at 12.37
$\mu$m \citep{Rigliaco15}. These H I lines can now be surveyed by JWST
with unprecedented sensitivity \citep{Beuther23,Tofflemire25}.

The accretion process is known to be episodic with rates varying by
orders of magnitude over relatively short timescales
\citep[e.g.,][]{Fischer23}. This variability can manifest itself
through ``bullets'' or ``knots'' in the jets that are caused by varying
ejection velocities linked to discrete peaks in accretion
activity. JWST provides opportunities to observe these knots and their
movements much closer to the protostar than before \citep{Ray23,Federman24}.

\subsubsection{Protostellar jets, winds, and outflows}

Jets and outflows are detected in all embedded protostellar systems,
from low- to high-mass protostars
\citep[e.g.,][]{Beuther02,Ray07,Frank14,Bally16,Pascucci23}. Outflow-
and accretion activity are linked, so outflows are expected to be most
powerful in the earliest highly extincted stages, i.e., the Class 0
phase for low-mass protostars and the IRDCs for
high-mass sources. JWST is unique in its ability to probe the physics
of these earliest protostellar outflows at mid-infrared wavelengths,
especially the highly collimated underlying jet driving the outflow,
as illustrated by the many atomic, H$_2$ and other diagnostics
highlighted in Sect.~\ref{sec:irdiagnostics} and seen in
Figure~\ref{fig:B1c-overview}.

The ISO-SWS \citep[e.g.,][]{Cernicharo00}, {\it Spitzer-}IRS
\citep[e.g.,][]{Maret09,Dionatos09,Watson16} and {\it Herschel-}HIFI
and PACS \citep[e.g.,][]{Karska18,vanDishoeck21} data have illustrated the
potential of mid- and far-infrared studies to probe the warm (few hundred K)
and hot (1000--2000 K) gas associated with jets, winds, and outflows but
were limited to spatial resolutions of $5-10''$ or larger. Two
temperature components have typically been found in the H$_2$ and CO
excitation. For low-mass protostars, a transition from molecular to
atomic to ionized jets appears as the sources evolve from the Class 0
to the Class II stage \citep{Nisini15}. JWST MIRI improves the spatial
resolution by factors of $>$10--100 and can now study the base of the
wind on $<$100 au scales, most notably with H$_2$, showing whether it
is collimated or not.

A related area of interest concerns identifying the physical processes
that actually heat the gas as seen in H$_2$, as they have not been
identified. At bow-shock and jet knot positions it is clear that
mechanical heating through high-velocity shocks is responsible
\citep[e.g.,][]{Hollenbach89}. However, winds launched from the disk
close to the protostar can be heated by other mechanisms. For MHD
winds, the ambipolar ion-neutral friction dominates
\citep[e.g.,][]{Panoglou12}, whereas for photoevaporative winds, (E)UV
radiation and X-rays heat the gas \citep{Pascucci23}. Ultraviolet
heating can also be important for gas within or at outflow cavity
walls at larger distance from the protostar \citep[e.g.,][]{Spaans95},
whereas winds can impact those walls and create local shocks. Thus,
the dominant heating mechanisms of warm and hot H$_2$ can change from
position to position in outflows, and a combination of mechanisms can
contribute at any position.

Both JWST MIRI and NIRSpec have the ability to image the stratification of
the different velocity and temperature components of jets, winds, and
outflows at different evolutionary stages and thereby probe the different
physical processes
\citep[e.g.,][]{Caratti24,Federman24,Tychoniec24,Delabrosse24,Pascucci25}.
MIRI can also address the question whether dust is launched in
jets and winds in the youngest protostars, and whether those dust
grains are subsequently destroyed by shocks, enriching the gas in
refractory elements
\citep{Podio06,Anderson13,Giannini15,Delabrosse24}. This, in turn,
provides information on the launching mechanisms and disk radius where
the jet or wind is launched.

\subsubsection{Hot cores and dense molecular shocks: Chemistry}

In the inner regions of dense protostellar envelopes, ices sublimate
into the gas at temperatures dictated by their binding energies. Above
$\sim$100 K, the temperature of the H$_2$O snowline, most molecules
are in the gas where they may be processed further by high temperature
gas-phase reactions. In contrast with ALMA, JWST can probe molecules
like CO$_2$, C$_2$H$_2$ and CH$_4$ that have no permanent dipole
moment but that are key molecules in gas-ice chemistry. Together with
other simple molecules (see Sect.~\ref{sec:irdiagnostics}), they can test
hot core chemistry models consisting of ice sublimation and high temperature
gas-phase chemistry \citep[e.g.,][]{Doty02}, including irradiation by
UV and X-rays \citep{Bruderer09b,Notsu21}. Early pioneering high
spectral resolution ground-based data
\citep[e.g.,][]{Lacy89,Evans91,Barr20} and space-based observations
\citep[e.g.,][]{Lahuis00,Sonnentrucker07,Indriolo15exes} were limited
mostly to high-mass protostars. JWST MIRI-MRS now opens the
possibility to study gas-phase lines in nearby low-mass sources with
high enough spectral resolution to detect weak emission lines on top
of a strong continuum, and with high enough spatial resolution to
locate the emission either with the hot core or with the more extended
outflow. The detection of isotopologs such as $^{13}$CO$_2$ and
$^{13}$CCH$_2$ allow for more accurate determinations of the optical depth
of the lines and thus column densities (see
  \citet{vanGelder24overview} for protostars and
  \citet{Grant23,Tabone23,Colmenares24} for disks).  A particularly
interesting question is the chemistry of C$_2$H$_2$ in hot gas and its
sensitivity to both temperature, carbon abundance and X-rays
\citep{Walsh15}, also in connection with high abundances of C$_2$H$_2$
and other hydrocarbon molecules found with JWST in warm gas in Class
II protoplanetary disks, especially those around very low-mass stars
\citep{Tabone23}.

At positions offset from the protostellar source, bright emission from
simple molecules can also be found, especially at bow-shock positions
\citep[e.g.,][]{Tappe12} and at dense jet-knot positions
\citep[e.g.,][]{Neufeld24}. Indeed, JWST has demonstrated that the
so-called green fuzzies detected with {\it Spitzer}-IRAC Band 2 at
4.6 $\mu$m in both low- and high-mass protostars
\citep[e.g.,][]{Noriega04,Cyganowski08} may well be dominated by CO
ro-vibrational emission rather than H$_2$ \citep{Ray23}. Thus, JWST
can now test models of molecular infrared emission from dense
molecular shocks \citep[e.g.,][]{Hollenbach89,Kaufman96} and
investigate differences between high temperature hot core and shock
chemistry. JWST also opens up the possibility to investigate the
chemistry of wide-angle winds close to the protostar.

\subsubsection{Cold outer envelopes: Ices}
\label{sec:backgroundices}

The bulk of protostellar envelopes is cold enough that most molecules
are frozen out as ices onto the dust grains. These ices are therefore
a major reservoir of the heavy elements \citep[see review
by][]{Boogert15}. ISO-SWS, {\it Spitzer} and ground-based telescopes
have provided ice inventories toward the brightest, usually more
evolved, protostellar sources, both low- and high-mass. JWST MIRI and
NIRSpec can now survey the coldest and ice-richest protostars
\citep[e.g.,][]{Brunken24} as well as dense pre-stellar clouds
\citep{McClure23}.

The higher spectral resolution of JWST allows ice band profiles to be fully
resolved, especially of molecules like CH$_4$ in the critical 5--10
$\mu$m range where {\it Spitzer}-IRS lacked resolution; these bands in
turn provide diagnostic information on ice environment and heating
\citep[e.g.,][]{Pontoppidan08,Oberg11}. JWST's sensitivity results in
high $S/N$ spectra on weak sources in which not only simple but also
more complex organic molecules can be identified \citep{Rocha24}, as
hinted at in earlier data \citep[e.g.,][]{Schutte99}. This in turn
allows for tests of their formation, especially whether complex molecules
are formed in ices or whether they are mostly the product of
high-temperature gas-phase chemistry. Other species in ices that are
well suited for renewed studies with JWST include ammonium salts,
O$_2$ and HDO \citep{Slavicinska24HDO,Slavicinska25sulfur}. The
HDO/H$_2$O ice ratio is particularly interesting as a tracer of
the inheritance of water from clouds to comets
\citep{Altwegg19,Aikawa24}, but has so far been tested only
  indirectly with gas-phase water in hot cores that is thought to
represent sublimated ices \citep{Persson14,Jensen19}.

\subsubsection{Embedded disks}
\label{sec:backgrounddisks}

The physical and chemical properties of disks in the protostellar
phase are still poorly constrained \citep{Tobin24}. Due to their
higher accretion rates, embedded disks are warmer and therefore have
more molecules in the gas \citep{vantHoff20}. Also, infall of envelope
material onto the disk through streamers may still take place causing
weak shocks \citep[e.g.,][]{Pineda23,Podio24}. JWST MIRI can search
for signatures of these accretion shocks, especially through
sulfur-bearing species such as SO$_2$ and [S I] lines.

ALMA is making great strides in studying embedded disks
\citep[e.g.,][]{Harsono18,Ohashi23} but cannot probe the inner few au
of disks. Mid-infrared observations are particularly well suited for
studying the gas in the planet-forming zones of disks, as is being
amply demonstrated by the rich and diverse spectra obtained in early
JWST observations of Class II disks
\citep[e.g.,][]{Gasman23,Banzatti23,Xie23,Arabhavi24,Temmink24H2O}.
Pioneering ground-based CO 4.7 $\mu$m high spectral resolution data
have revealed Keplerian rotation from the inner parts of Class II
disks \citep[e.g.,][]{Najita03,Brown13,Banzatti22} as well as from
several Class I protostars \citep{Herczeg11} and from embedded
high-mass sources \citep[e.g.,][at 2.3 $\mu$m]{Ilee14}. These data
demonstrate that gas in the upper layers of disks in the inner few au
can be detected. A main question that MIRI-MRS can address is whether
these young disks show a similarly rich chemistry as their more mature
counterparts. MIRI and NIRSpec can also detect other interesting lines such as
H$_2$ and [Ne II] to study the role of winds in the mass loss from
young disks \citep[e.g.,][]{Tychoniec24,Delabrosse24}.

\begin{table*}[t]
    \centering
\caption{JOYS source sample and physical properties.}
\label{tab:sources}
{\small
\begin{tabular}{lrrrrrrrrrrr}
\hline
  \hline
Source & RA$^a$ & Dec$^a$ & $d$ & $L_{\rm bol}$ & $T_{\rm bol}$ & Class & $M_{\rm env}$ & Binary & Other names & Ref. \\
       & [J2000] & [J2000] & (pc) & (L$_\odot$) & (K) & & (M$_\odot$) & $''$  \\
\hline
IRAS4B & 03:29:12.02 & +31:13:08.0 & 293 & 6.8 & 28 & 0 & 4.7 &  10.7 & Per-emb 13 & 1-5  \\
IRAS4A1$^b$ & 03:29:10.54 & +31:13:30.9 & 293 & 14.1  & 34 & 0 & 8.7 &  1.8 & Per-emb 12 & 1-5  \\
IRAS4A2$^b$ & 03:29:10.43 & +31:13:32.1  & 293& 14.1 & 34 & 0 & 8.7 &  1.8 &  Per-emb 12 & 1-5  \\
B1-c & 03:33:17.88 & +31:09:31.8 & 293 & 5.0 & 48 & 0 & 5.3  &  -  &  Per-emb 29 & 1-5  \\
B1-b & 03:33:20.34  & +31:07:21.4  & 293 & 0.23 & 157 & I & 2.1 &  14 &  Per-emb 41 & 1,2,5  \\
B1-a-1 & 03:33:16.67  & +31:07:54.9 & 293 & 2.3  & 113 & I & 1.5 &  0.4   &  Per-emb 40 & 1-3,5  \\
B1-a-2 & 03:33:16.68  & +31:07:55.3  & 293 & 2.3 & 113 & I & 1.5 &  0.4 &  Per-emb 40 & 1-3,5  \\
L1448-mm & 03:25:38.88  & +30:44:05.3 & 293 & 8.5 & 49 & 0  & 3.9 &  8.1  &  Per-emb 26 & 1-5  \\
Per-emb 8 & 03:44:43.98  & +32:01:35.2 & 321 & 4.5 & 45 & 0  & 1.0 &  9.6 & -  &  1-5  \\
TMC1-W      &  04:41:12.69 & +25:46:34.7 & 142 & 0.7 & 161 & I  & 0.2 & 0.6 & IRAS 04381+2540 & 3,4,6,7\\
TMC1-E      &  04:41:12.73 & +25:46:34.8 & 142 & 0.7 & 161 & I  & 0.2  & 0.6 & IRAS 04381+2540 & 3,4,6,7\\
TMC1A      &  04:39:35.20 & +25:41:44.2 & 142 & 2.7 &  189 & I  & 0.2  & - & IRAS 04365+2535 & 3,4,6,7\\
L1527 IRS     &  04:39:53.88 & +26:03:09.5 & 142 & 3.1 &  79 & I  & 0.9  & - & IRAS 04368+2557 & 3,4,6,7\\
BHR71 IRS1     &  12:01:36.50  & -65:08:49.4 & 200 & 14.7 &  68 & 0  & 19  & 15.7 & IRAS 11590$-$6452 & 8-11\\
BHR71 IRS2     &  12:01:34.01   & -65:08:48.0 & 200 & 1.7 &  38 & 0  & 19  &  15.7 & - & 8-11\\
Ser-SMM1-a  &  18:29:49.81  &  +1:15:20.4 & 436 & 109 & 39 & 0 & 58  &  2.0 & Ser-emb 6 & 2,3,4,12,13 \\
Ser-SMM1-b1     &  18:29:49.68  & +1:15:21.1 & 436 & 109 & 39 & 0  & 58  &  0.3 &- & 2,3,4,12,13 \\
Ser-SMM1-b2    &  18:29:49.66  & +1:15:21.2 & 436 & 109 & 39 & 0  & 58  &  0.3 &-  & 2,3,4,12,13 \\
Ser-SMM3      &  18:29:59.31 & +1:14:00.3  & 436  & 27.5 & 37 & 0  & 11.5 & - & - & 2, 3, 4, 14 \\
Ser-S68N-N     &  18:29:48.13 & +1:16:44.6  & 436 & $>$6.0 & 58 & 0  & 10.4 & 1.4  &  - & 2, 5, 15, 16  \\
Ser-S68N-S     &  18:29:48.09 & +1:16:43.3  & 436 & $>$6.0 & 58 & 0  & 10.4 & 1.4  & Ser-emb 8 & 2, 5, 15, 16  \\
  Ser-emb-8(N)$^b$     &  18:29:48.73 & +1:16:55.6  & 436 & 1.8 & - & 0 & -  & 6.6 & SerpM-S68Nb & 2,15,17,18\\
  HH 211 & 03:43:56.81 & +32:00:50.2 & 321 & 4.1 & 27 &  0 & 2.4 & -           & Per-emb 1 & 1, 2, 25\\
\hline
  G28IRS2 & 18:42:51.99 & -3:59:54.0 & 4510 & 10120 & -   & IRDC/ & 3288 & - & G28P2 & 22, 23\\
        &                &           &      &       & - & HMPO \\
G28P1 & 18:42:50.59 & -4:03:16.3 & 4510 & 682 & -       & IRDC & 4276 &  - & - & 22, 23 \\
G28S & 18:42:46.45 & -4:04:15.2 & 4510 & 364 & -        & IRDC & 2296 &  - & - &22, 23 \\
IRAS23385 & 23:40:54.49 & +61:10:27.4 & 4900 & 3170 & -  & HMPO & 220 &  -& Mol160 & 20, 21\\
IRAS18089 & 18:11:51.24 & -17:31:30.4 & 2340 & 15723$^*$ & -& HMC & 1100$^*$ &  - & G12.89+0.49 & 23, 24\\
G31 & 18:47:34.33 & -1:12:45.5 & 5160 & 69984 & -       & HMC & 7889 & - & - & 22, 23 \\
  \hline
  \\
\end{tabular}
}
1- \citet{Tobin16}, 2- \citet{OrtizLeon18} , 3- \citet{Karska18}, 4- \citet{Kristensen12}, 5- \citet{Enoch09},  6- \citet{Krolikowski21}, 7- \citet{vantHoff20}, 8- \citet{Yang20}, 9- \citet{Seidensticker89}, 10- \citet{Tobin19}, 11- \citet{Yang17}, 12- \citet{Hull17}, 13- ALMA 2015.1.00354.S, 14- 2017.1.01350.S, 15- \citet{LeGouellec19}, 16 - 2015.1.00768.S, 18- \citet{Podio21}, 19- 2019.1.00931.S, 20- \citet{Beuther23}, 21- \citet{Molinari08}, 22- \citet{Wang08}, 23- \citet{Urquhart18}, 24- \citet{Xu11}, 25- \citet{Sadavoy14}

* - distance corrected. Properties scalable with distance are
corrected according to updated distance measurement.  $^a$ Coordinates
of millimeter interferometry source position; MIRI IFU pointing is
usually offset from source toward blue outflow lobe, see
Table~\ref{tab:sources2} for IFU pointings.  $^b$ Source continuum
center not covered in MIRI 1290 observations.

\end{table*}

\section{Observations and methods}
\label{sec:observations}

\subsection{Source sample}
\label{sec:sample}

Table~\ref{tab:sources} summarizes the main properties of the sources
observed with MIRI-MRS in the JOYS program. All low- and high-mass
sources have been well studied and characterized in great detail prior
to JWST, including with {\it Spitzer}, {\it Herschel} and
submillimeter (ALMA, NOEMA) ground-based data. For example, many of
them were part of the {\it Herschel} WISH program studying water from
pre-stellar cores to disks \citep{vanDishoeck21}. They were chosen for
their ability to address the different science cases outlined in Sect.\ 2
and 4--8.

\begin{figure}
\begin{centering}
\includegraphics[width=9cm]{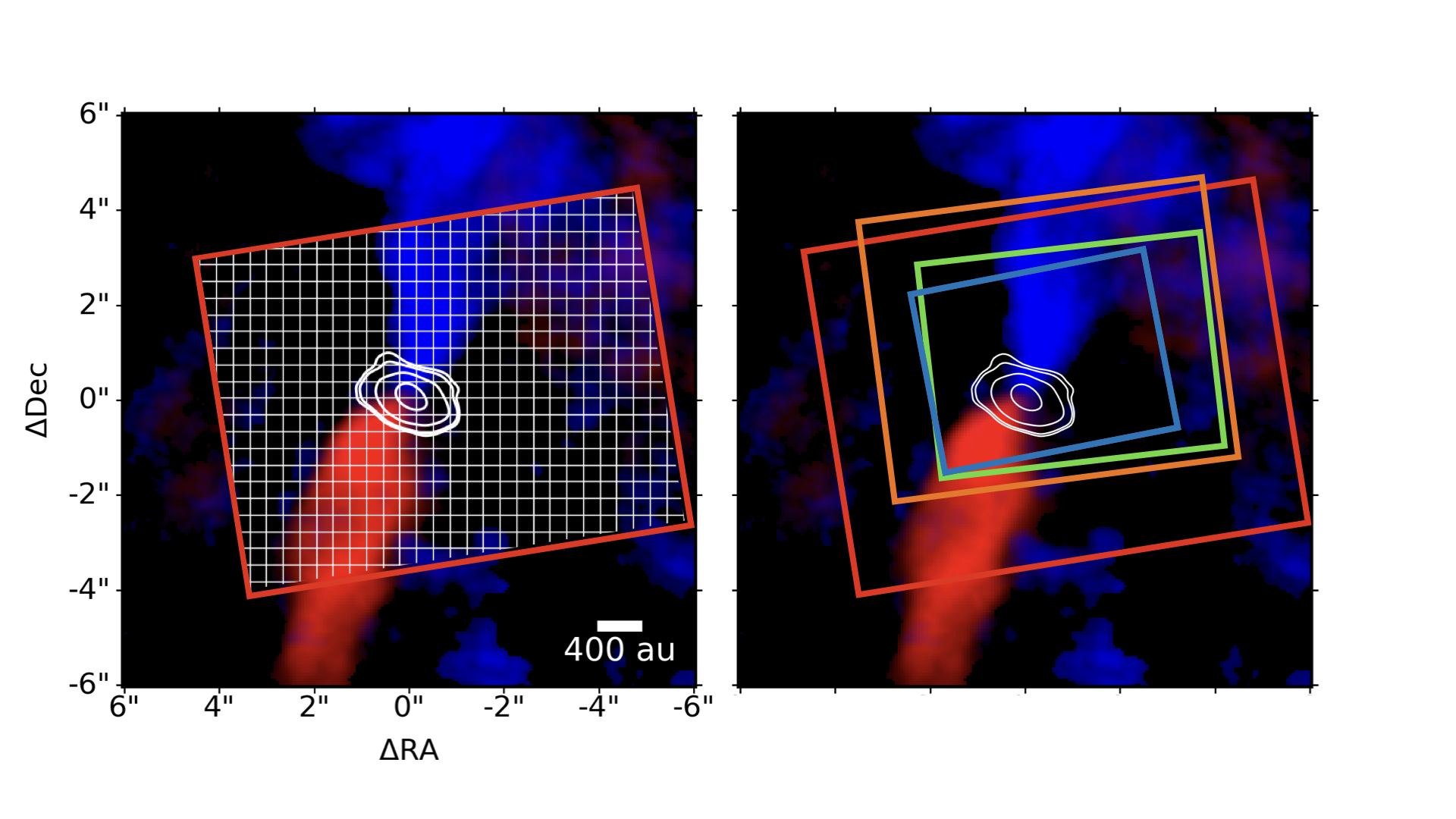}
\caption{Right: MIRI-MRS IFU footprint for channels 1 (blue), 2
  (green), 3 (orange), and 4 (red) overlaid on the ALMA Serpens SMM3 CO
  2--1 outflow map using data from \citet{Tychoniec21}. Two dithers
  were used. The positions of the IFU have been chosen on purpose to
  be slightly offset toward the blue outflow lobe for the shortest
  wavelength channels, but still covering some part of the red outflow
  lobe. The white contours indicate the 1.3 millimeter continuum
  showing the dust disk. Left: Illustration of the grid of individual
  spaxels for channel 4.  }
\label{fig:MIRI-footprint}
\end{centering}
\end{figure}

The low-mass sample consists of nearby sources at distances $<$500 pc
that cover both the deeply embedded Class 0 and more evolved Class I
phases with bolometric luminosities ranging from 0.2 to $>$100
L$_\odot$ and bolometric temperatures from $<30$ K to 200 K
\citep{vanGelder24overview}. They were furthermore selected to belong
to only three star-forming regions -- Perseus, Taurus and Serpens --
to minimize slew overheads, with two pointings in the isolated BHR 71
cloud added. Some sources were chosen because they have prominent
outflows and jets (e.g., L1448-mm, Ser-SMM1, BHR71). To further study
outflow physics, two dedicated pointings at ``knot'' positions in the
blue-shifted outflow lobes well offset from NGC 1333 IRAS4A and
Ser-emb8(N) were added. The protostar position of NGC 1333 IRAS4A is
covered in PID 1236 as part of JOYS+; hence this source is not further
included in the analysis here. Moreover, the entire blue outflow lobe
and a small portion of the red lobe of HH 211 were imaged with
MIRI-MRS in program 1257. These three Class 0 sources (HH 211,
  NGC 1333 IRAS4A and Ser-emb8(N)) are so deeply embedded that they
are not detected even with JWST.  Other sources were selected
because of their very deep ice features facilitating the search for
new minor ice species (e.g., B1-c). SVS4-5 is a special case because
it is a ``background'' Class I/II source behind or inside the envelope
of the Class 0 source Ser-SMM4 probing ices at very high densities
\citep{Pontoppidan04,Perotti20} and is not included in
Table~\ref{tab:sources}.  A number of sources with confirmed young
disks based on Keplerian rotation were included as well (e.g., L1527,
TMC1, TMC1A).

Several sources are close binaries that can be resolved by MIRI at its
shortest wavelengths, allowing for comparison of their outflow and ice
properties on scales of $<$1000 au. In total 17 low-mass sources (23
sources counting binaries individually) were targeted. Of these 17
(23) sources, four (six) are Class I and one source is borderline Class 0/I
(L1527). We note that program 1236 (PI: M. Ressler) is focused on binary
sources in Perseus so the comparison of binary properties will be part of
future JOYS+ publications.

For the high-mass sources, three Infrared Dark Clouds and three High-Mass
Protostellar Objects (HMPOs) and hot molecular cores (HMCs) were
chosen.  Their properties and distances are included in
Table~\ref{tab:sources}. The MIRI-MRS data reveal several of these
sources to be binaries or multiples: there are at least nine mid-IR
sources detected in the six high-mass MRS target fields: IRAS
23385+6053 (hereafter IRAS 23385) displays a binary \citep{Beuther23},
G28 P1 two distant sources, IRAS 18089-1732 (hereafter IRAS 18089) at
least three sources, and G28S shows no sources at 5 $\mu$m. G28 IRS2
and G31 show one source each.  As for the low-mass sources, the
high-mass set has been selected to cover a broad range of masses and
evolutionary stages without saturating the MIRI-MRS. We note that many
nearby well-known high-mass protostars studied with the ISO-SWS are
too bright for JWST.

\subsection{MIRI-IFU observations}

The coordinates in Table~\ref{tab:sources} represent those of the
protostellar source positions as found from millimeter
interferometry. For the low-mass protostars, the actual MIRI-MRS IFU
pointing positions are on purpose slightly offset from the millimeter
continuum source position to cover a larger fraction of the
less-extincted blue-shifted part of the outflow in the IFU, in
addition to still catching some part of the red-shifted outflow.  For
TMC1A, the IFU has a larger offset from the source position to capture
a larger fraction of its blue wind as imaged by ALMA
\citep{Bjerkeli16}.  All MRS IFU pointing coordinates are summarized
in Table~\ref{tab:sources2}. The mid-infrared peak positions of the
low-mass sources derived from the MRS data are listed in Table B.1 of
\citet{vanGelder24overview}.  A single NIRSpec IFU spectrum of the
source B1-c with the G395M mode is taken as well within program 1290
to obtain a full inventory of this ice-rich source including HDO ice.

The sources within one cloud were observed in a single uninterrupted
sequence with one background (dark) position observed in each cloud to
characterize detector artifacts and subtract the telescope
background. For Perseus, this background was taken to be the
position of the millimeter source B1-bS which is so deeply embedded
that it is undetected with JWST. For Taurus, the background was taken
with a single dither position, whereas for the other three cases a
2-point dither was used.

All sources have been observed with the full MIRI-MRS spectral
coverage using the three grating settings (A, B, C) to observe the
5--28 $\mu$m range spread over four channels (1--4) of the MIRI IFU
\citep{Wells15,Wright23}. A 2-point dither pattern for extended
sources was adopted for most sources, except for B1-c and Ser-SMM1A
for which a 4-point dither pattern was used. The typical integration
time is 200 s per grating setting, except for L1527 which used 1000 s
per grating and B1-c which had 2000 s in gratings A and C and 4000 s
in grating B. This latter setting was chosen because grating B covers
the deep silicate absorption around 10 $\mu$m where the source flux is
much lower.
The IFU FoV varies between $3.2''\times 3.7''$ (channel 1)
and $6.6''\times 7.7''$ (channel 4). Figure~\ref{fig:MIRI-footprint}
illustrates the IFU footprint of the different channels for Serpens
SMM3, covering the central object, its immediate envelope structure
and the inner region of the blue outflow and jet.

For some sources, we made small mosaics of the blue outflow lobe
ranging from 1$\times$2 to $3\times 3$, thus mapping regions up to $12-20''$
in size. These observational details are provided in
Table~\ref{tab:sources2} in Appendix A; we note that they are
sometimes listed as separate pointings in the archive.  The total
extent of the HH~211 blue lobe and a small portion of its red-shifted
one ($0.95'\times 0.22'$) has been observed with $12\times 2$ spatial
settings \citep{Caratti24}.  More details of the observations and
mid-infrared continuum images of all low-mass sources in each of the
four IFU channels are presented in Figures B.1-B.18 of
\citet{vanGelder24overview} and in \citet{Caratti24} for HH 211.

Most of the JOYS sources were observed in Fall 2023. Only the
high-mass source IRAS 23385+6053 (3000 L$_\odot$, 9 M$_\odot$) was
observed early in the JWST mission, in August 2022. Hence, most of the
early JOYS analysis has focused on this source
\citep{Beuther23,Gieser23,Francis24}.  Located at a distance of 4.9
kpc in the direction of the outer Galaxy, this puts the source at a
galactocentric radius of 11 kpc, making IRAS 23385 a unique laboratory
for studying star formation in the outer Galaxy.  This paper provides
one of the first opportunities to look at similarities and differences
among the larger JOYS sample from low to high mass.

\subsection{MIRI-MRS data reduction}

The MIRI-MRS data were reduced and calibrated using the JWST
calibration pipeline version 1.13.4 \citep{Bushouse24} using reference
context {\tt jwst$\_$1188.pmap}. The steps adopted are the same as
those in \citet{vanGelder24overview} of which only a short summary
is provided here. The raw data were processed through all three
steps of the JWST calibration pipeline. This included the subtraction
of the dedicated background on the detector level, where astronomical
emission in the background was masked so that it was not subtracted in
the science data. Furthermore, the fringe flat for extended sources
\citep{Crouzet25} and the 2D residual fringe correction (Kavanagh et
al. in prep.) were also applied on the detector level. Prior to
building the cubes, an additional bad pixel map was created using the
Vortex Imaging Package (VIP) version 4.1
\citep[][]{Christiaens23}. Last, the final data cubes were constructed
for each channel and sub-band separately with the outlier rejection and
master background steps switched off. Spectra were extracted manually
from selected positions in the datacubes using either an aperture with
a constant diameter as function of wavelength (``circle'') or an
aperture with its diameter increasing with wavelength (``cone'')
following the size of the point spread function \citep[PSF;][]{Law23}.

Table C.1 of \citet{vanGelder24overview} provides 1$\sigma$ noise
level estimates in each of the MRS sub-bands for the JOYS low-mass
targets at the continuum source position(s). They typically range from
0.1 mJy in channels 1--3 to several mJy in channel 4, with some
sources (e.g., B1-a-NS and TMC1A) having higher noise levels due to
their strong continuum or being located at the edge of the FoV. Noise
levels are similar for the high-mass sources because of their similar
integration times.

\subsection{Simultaneous imaging}

During the MIRI-MRS observations, the imager was turned on to
simultaneously observe a field of $74''\times 113''$ offset by
$\sim 1'$ from the MRS position at a position angle determined by the
time of observations. This simultaneous imaging provides not only
serendipitous science such as finding new deeply embedded protostars
in the cloud, but it also serves as the absolute astrometric
calibration of the MRS data by comparing with the positions of field
stars that are in the Gaia catalog. Indeed, for the initial JOYS
source IRAS 23385+6053, for which the source position was well known,
an adjustment of 1.61$''$ in RA and 0.35$''$ in Dec in the telescope
coordinates was needed based on comparison with the Gaia stars.

The imager has a plate scale of 0.11$''$ \citep{Bouchet15} and
provides diffraction-limited imaging in a number of broadband
filters. For the JOYS program, simultaneous imaging was obtained in
the F1500W broadband filter centered at 15.0 $\mu$m (FWHM 2.92 $\mu$m)
\citep{Beuther23}. At that wavelength, the MIRI PSF is 0.49$''$ FWHM,
providing nearly an order of magnitude sharper images than was
possible with previous space instruments. The 15 $\mu$m filter was
chosen to be away from PAH features, and to be intermediate between
the {\it Spitzer} IRAS 8 $\mu$m and MIPS 24 $\mu$m bands to fill in
that part of the SED.  Being at a wavelength where the telescope
background does not yet contribute, it also provides a good compromise
between high angular resolution and good sensitivity at the longer
MIRI wavelengths that can probe most deeply into the clouds. We note, however, that
this filter is centered at the CO$_2$ ice band, which can
absorb a significant fraction of the protostellar continuum along the
line of sight through the envelope.

\begin{figure}[h!]
\begin{centering}
\includegraphics[width=9cm]{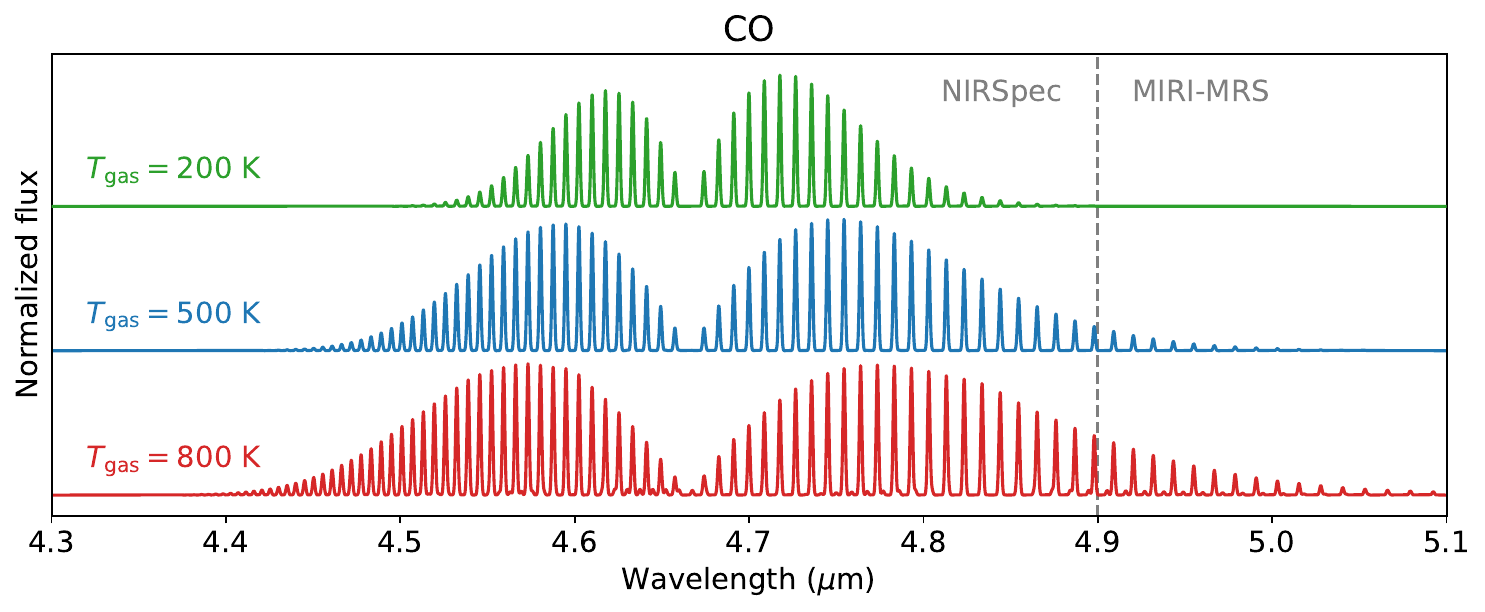}
\includegraphics[width=9cm]{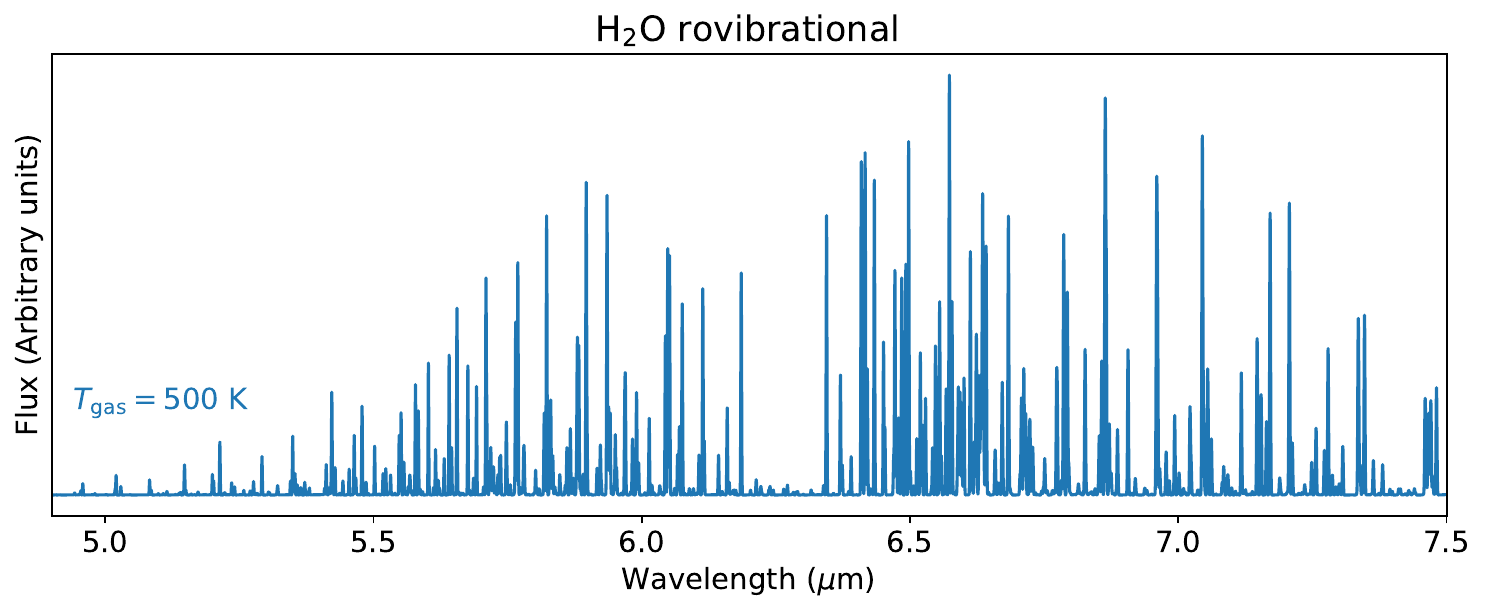}
\includegraphics[width=9cm]{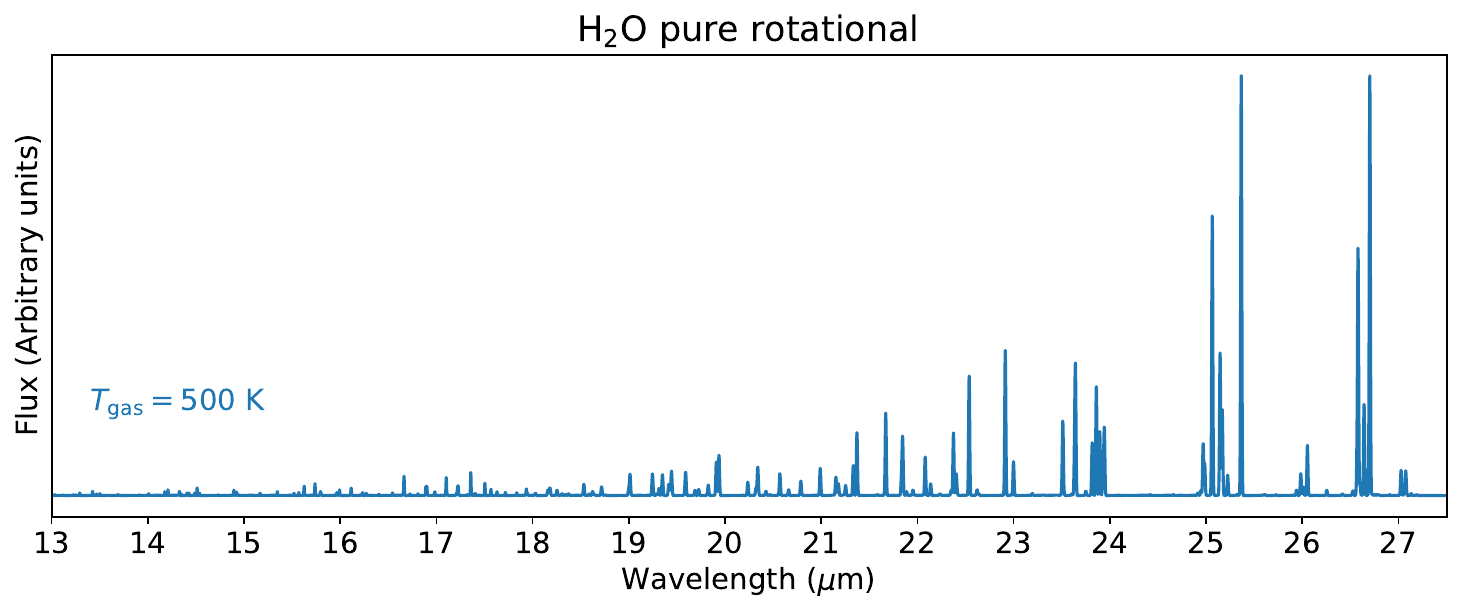}
\caption{Examples of normalized mid-infrared slab model molecular
  emission spectra in the optically thin regime. From top to bottom,
  CO at 500 K compared with 200 and 800 K, H$_2$O ro-vibrational lines
  at 5--7 $\mu$m, and H$_2$O pure rotational lines at 13--28 $\mu$m,
  both at 500 K.}
\label{fig:molecules1}
\end{centering}
\end{figure}

The MIRI images were processed through all three steps of the JWST
calibration pipeline using reference context {\tt jwst$\_$1235.pmap}
adopting the default parameters in each step. In the final step, the
{\tt tweakreg} step was switched off since very few Gaia stars are
available in the FoV. The background level was estimated and
subtracted by setting the {\tt skymethod} option to local in the {\tt
  skymatch} step. An example of the science that can be done with
these simultaneous images to find new protostars in the outer Galaxy
is provided in Appendix~\ref{sec:app_IRAS23385}.

\begin{figure}[tbh]
\begin{centering}
\includegraphics[width=8.5cm]{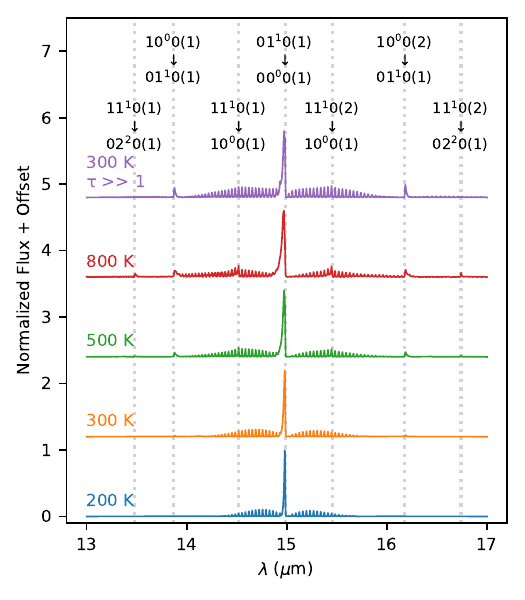}
\caption{Simulated normalized CO$_2$ mid-infrared spectra at different
  temperatures for optically thin emission (blue to red). Notable are the
  broadening of the $Q$ branch and the appearance of hot bands (i.e.,
  transitions between vibrationally excited states) with increasing
  temperature. These hot bands also become visible at lower
  temperatures when the CO$_2$ $Q-$branch emission becomes optically
  thick (purple).}
\label{fig:molecules2}
\end{centering}
\end{figure}

\subsection{Analysis of gas-phase spectra}
\label{sec:gasanalysis}

The analysis of atomic lines and ices seen in the MIRI spectra is
described in the various scientific sections. Here we summarize the
analysis of gas-phase molecular lines to infer column densities and
temperatures, following \citet{Francis24,vanGelder24overview,Salyk24}
and papers studying molecular emission lines from disks \citep[e.g.,][and refs
cited]{Salyk11,Banzatti25,Temmink24H2O}.

In the simplest case of spectrally resolved pure absorption and
assuming an isothermal slab model, the measured optical depths are
directly related to column densities. For unresolved lines such as
with JWST, an intrinsic line width needs to be assumed and a
curve-of-growth type of analysis to be performed
\citep{Boonman03,Barr20}; typical values are a FWHM of a few km
s$^{-1}$ up to 10--20 km s$^{-1}$
\citep{vanGelder24overview,McClure25}. It is usually assumed that the
covering fraction of the absorbers against the continuum is unity and
that the gas is at a single temperature, but this does not need to be
the case \citep{Knez09,Li24}. Proper radiative transfer may therefore
be needed \citep{Gonzalez02,Lacy13}.

Mid-infrared emission lines are also usually interpreted with
isothermal slab models fitted to continuum-subtracted spectra.  The
emission is assumed to originate in a plane-parallel cylindrical slab
of radius $R$ and column density $N$. The radius $R$ should not be
viewed as a disk or envelope radius but rather represents the emitting
area $\pi R^2$ having any shape at any location in the beam.  The gas
is taken to be at a single excitation temperature $T_{\rm ex}$, and
Figure~\ref{fig:molecules1} shows typical ro-vibrational spectra of CO
and H$_2$O at 200--800 K. We note that MIRI only probes the tail of
the CO $v$=1-0 $P$ branch at high temperatures, and that the 5--28
$\mu$m range covers both the H$_2$O ro-vibrational bending mode lines
as well as a forest of pure rotational water
lines. Figure~\ref{fig:molecules2} illustrates how the CO$_2$ spectrum
changes with temperature and optical depth \citep[see
also][]{Cami00,Bosman17}. Emission lines can be strongly affected by
extinction from ices and dust (see Sect.~\ref{sec:hotcore} and
Appendix~\ref{sec:app_extinction} for details).

If collisions dominate the excitation, $T_{\rm ex}$ will approach the
kinetic gas temperature $T_{\rm kin}$, but the critical densities for
this assumption to be valid are high, typically greater than $10^{12}$
cm$^{-3}$ for vibrational lines \citep[e.g.,][]{Bruderer15}. Such high
densities may only be achieved in the inner few au of disks
\citep{Salyk09,Carr11}. Non-LTE effects will be important for regions
with lower densities such as in shocks or in hot cores, where
densities are typically $10^6$ to $10^8$ cm$^{-3}$ on scales of 50 au
\citep{Kristensen12,Notsu21}.  However, in this case strong infrared
radiative pumping through ro-vibrational transitions, by a radiation
field whose intensity at the pumping wavelengths can be characterized
by a temperature $T_{\rm rad}$, can boost the vibrational emission
\citep[see e.g.,][for the cases of HCN and
SO$_2$]{Bruderer15,vanGelder24}.

In contrast, the rotational distribution within the $v=0$ vibrational
ground state can be thermalized by collisions at lower densities, and
this rotational distribution, characterized by $T_{\rm rot}$, is
largely conserved in the vibrationally excited states through infrared
pumping and can thus reflect the kinetic temperature $T_{\rm kin}$.
For molecules with a dipole moment, the density to achieve
$T_{\rm rot}$ close to $T_{\rm kin}$ is around $10^8-10^9$ cm$^{-3}$,
whereas the rotational level populations of molecules without a dipole
moment such as CO$_2$ are thermalized at much lower densities. In
other words, $T_{\rm vib}$ is usually larger than $T_{\rm rot}$, with
the former being controlled by radiative pumping at $T_{\rm rad}$ and
the latter being closer to the gas temperature $T_{\rm kin}$.

\section{Onset of star formation: Protostars and accretion}
\label{sec:onset}

\subsection{Probing the fossil accretion record from imaging}

JWST's ability to probe the deeply embedded Class 0 stage is
illustrated by the HH 211 outflow ($d$=321 pc,
\citealt{Ortiz18}). Despite the extent and power of this outflow, as
highighted by prominent shocked atomic, H${\rm _2}$ and CO emission at
infrared and millimeter wavelengths
\citep[e.g.,][]{McCaughrean94,Gueth99,Lee09,Dionatos18}, the current
stellar mass as determined from kinematic studies is still comparable
to that of a brown dwarf, only 0.08 M$_\odot$ \citep{Lee15}.

\begin{figure}
\begin{centering}
\includegraphics[width=8cm]{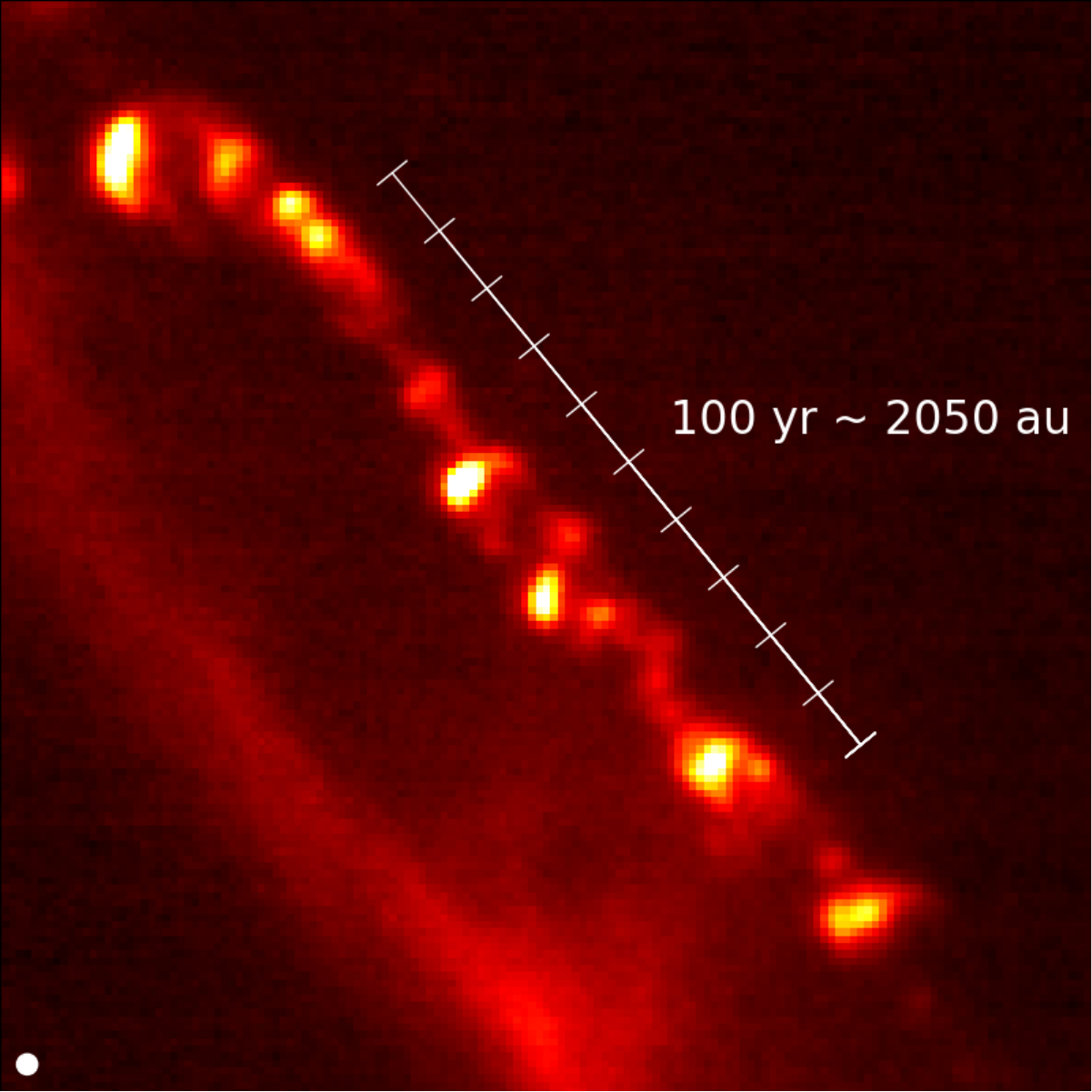}
\caption{Blow-up of the H$_2$ knots in part of the HH 211 jet in the
  blue outflow lobe imaged with the NIRCam F212N filter
  \citep{Ray23}. This figure illustrates the typical separation
  between knots and thus implicitly the timescales between accretion
  bursts. The ruler shows the distance of jet gas covered in 100 yr
  when traveling at 100 km s$^{-1}$, in 10 sections of 10 yr each.}
\label{fig:HH211knots}
\end{centering}
\end{figure}

JWST-NIRCam infrared imaging of HH 211, taken as part of JOYS, has
revealed its jet and outflow structure with unprecedented detail in
both molecular (H$_2$, CO) and atomic ([Fe II]) lines \citep{Ray23}.
Because of its youth, the outflow is still compact and can readily be
imaged in its entirety with JWST.  The NIRCam images reveal an
abundance of H$_2$ knots in the innermost highly extincted collimated
jet which were previously undetected and/or unresolved (see
Figure~\ref{fig:HH211knots}). The central source of HH 211,
i.e., the protostar itself, remains undetected even with JWST's
sensitivity, hidden behind $A_V>$100 mag of extinction.

These H$_2$ knots provide a fossil record of the recent accretion
history of the protostar.  Comparison with archival VLT-ISAAC K-band
images taken 20 years earlier demonstrates JWST's capabilities to
measure the proper motions and thus the 3D kinematics of the outflow
of this very young protostar \citep{Ray23}.  The tangential flow
velocity has been measured to be $\sim$100 km s$^{-1}$ so the
timescale to cross the entire length of the blue outflow lobe
(1$' \sim$ 18300 au at 321 pc) is only $\sim$1000 years.  Here we use
the data from \citet{Ray23} to show in Figure~\ref{fig:HH211knots}
that the spacing and structure of the jet knots is as short as $\sim$
5 yr. Taken together, this seems to imply that this protostar has
built up its current mass of 0.08 M$_\odot$ over a period of
$\sim$1000 yr in multiple bursts every 5--10 yr, and is still growing
in mass.

\subsection{H I recombination lines as a tracer of accretion}
\label{sec:accretion}

\noindent
\paragraph{High mass protostars.}
MIRI's potential to measure accretion rates for high-mass protostars
has been provided by the analysis of IRAS 23385+6053 ($d$=4.9 kpc,
Table~\ref{tab:sources}) \citep{Beuther23,Gieser23}, a region forming
a 9 M$_{\odot}$ high-mass protostar \citep{Cesaroni19}. The high
spatial resolution of MIRI reveals that this source is actually a
binary system (labeled A and B) with a projected linear separation of
$\sim$3300 au that can be resolved in the MRS short wavelength
channels. Only the A source coincides with the extension of the
millimeter continuum peak.  The H I 7-6 Humphreys $\alpha$ line is
detected at a 3-4$\sigma$ level toward sources A+B.  Follow up
analyses reveal that most of the ionization comes from source A (Reyes
et al., in prep.). Using the \citet{Rigliaco15} relation for this line
calibrated for low-mass T Tauri stars, an accretion luminosity of 140
L$_\odot$ can be inferred, which in turn gives an accretion rate of
$2.6 \times 10^{-6}$ M$_{\odot}$ yr$^{-1}$. Correcting this number for
an estimated visual extinction of 30--40 mag based on the depth of the
silicate feature, and using the \citet{McClure09} extinction law,
increases the rate to $\sim 10^{-4}$ M$_{\odot}$ yr$^{-1}$. For such a
high extinction, the bulk of the A+B sources' luminosity of
$3\times 10^3$ L$_\odot$ is indeed due to accretion. An analysis of H
I in the full JOYS high-mass sample will be presented in Reyes et al.\
(in prep.).

\begin{table*}[h!]
\centering
\caption{Hydrogen line measurements and accretion rates.}
\label{tab:accretion}
\begin{tabular}{lccccccccc}
\hline \hline
  Source & $A_V$ & $L_{\rm bol}$ & $M_{\star}$ & $R_{\star}$ & $L_{\rm acc}/L_{\rm bol}$ & Flux H I 7-6 & $L_{\rm acc}$  & $\dot{M}_{\rm acc}$ &
$\dot{M}_{\rm acc}$   \\ [5pt]
 & & & & & & & from H I 7-6 & from H I 7-6  & from $L_{\rm bol}$   \\
         & (mag) & (L$_{\odot}$) & (M$_{\odot}$) & (R$_{\odot}$) & &
(erg cm$^{-2}$ s$^{-1}$) & (L$_{\odot}$) & (M$_{\odot}$ yr$^{-1}$)
& (M$_{\odot}$ yr$^{-1}$) \\
  \hline 
  TMC1-W & 18 & 0.35 & 0.20 & 2.5 & 0.5 & $ 4.0 \pm 1.1\times10^{-15} $ & $1.2
 \times 10^{-2}$ & $4.9 \times 10^{-9}$ & $7.2 \times 10^{-8}$ \\
  TMC1A & 33 & 2.70 & 0.56 & 2.5 & 0.5 & $ 6.8 \pm 2.7\times10^{-14} $ & $4.5
 \times 10^{0}$ & $6.6 \times 10^{-7}$ & $2.0 \times 10^{-7}$ \\
  SerpSMM3 & 50 & 27.5 & 0.50 & 4.0 & 1.0 & $< 5.4 \times 10^{-16}$ &
 $< 2.0 \times 10^{-2}$ & $< 5.2 \times 10^{-9}$ & $7.2 \times 10^{-6}$\\
\hline
\end{tabular}
\tablefoot{$A_V$ is measured from the silicate feature using
  $\tau_{9.7}\ \times$ {18.5.}  $L_{\rm bol}$ is divided by the number
  of components in multiple systems. For TMC1A and TMC1-W the measured
  values of stellar properties are used, for SMM3 fiducial Class 0
  values are adopted. $L_{\rm acc}/L_{\rm bol}$ is not a measured
  value but an assumption that goes into estimating
  ${\dot M}_{\rm acc}$ from $L_{\rm bol}$. All values have been corrected
  for extinction. }
\end{table*}

\noindent
\paragraph{Low-mass protostars.}
An early JOYS example for low-mass protostars is provided by TMC1, a
Class I binary located in Taurus at 142 pc distance
\citep{Tychoniec24}. Several H I recombination lines are detected
toward both sources, with the strongest lines seen toward TMC1-W. The
fact that these lines are only seen on top of the infrared continuum,
i.e., at the source position, suggests that there is no or little jet
contribution.  Table~\ref{tab:accretion} lists the inferred accretion
rate, using the relation by \citet{Rigliaco15} and correcting for
extinction and contamination by water lines (see below and update by
Tychoniec, in prep.).  The value of $5\times 10^{-9}$ M$_{\odot}$
yr$^{-1}$ is an order of magnitude lower than that derived from the
bolometric luminosity $L_{\rm bol}$ also listed in
Table~\ref{tab:accretion} obtained assuming the relation
$ {\dot M}_{\rm acc} = f R_* L_{\rm bol}/ G M_* $ holds. Here
$f=L_{\rm acc}/L_{\rm bol}$ is the fraction of the luminosity that is
ascribed to accretion.  The low value found for TMC1-W suggests that
it is currently in a low accretion state, perhaps similar to the
low-mass Class 0 source IRAS 16253-2429 studied with JWST in the IPA
program \citep{Narang24}.

\begin{figure}
\begin{centering}
\includegraphics[width=8cm]{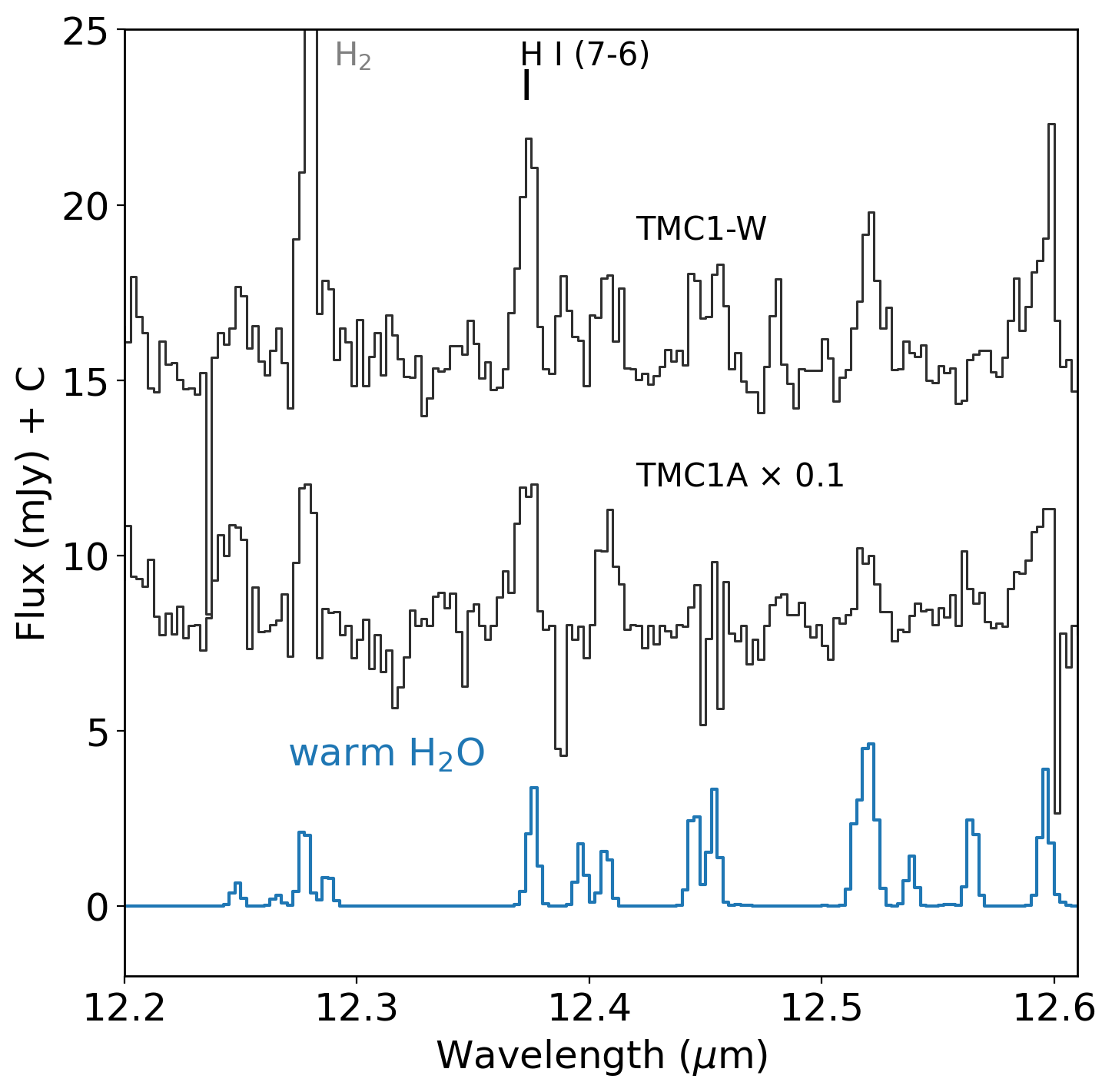}
\caption{Comparison of the H I 7-6 Humphreys $\alpha$ line detected in
  the JWST-MIRI spectrum of the Class I source TMC1A with that seen
  toward TMC1-W \citep{Tychoniec24}. A warm water spectrum at $T$=485
  K is also shown (best fit for TMC1-W, \citealt{vanGelder24overview})
  to illustrate that water emission can contribute to the H I 7-6
  feature.}
\label{fig:TMC1A-accretion}
\end{centering}
\end{figure}

Another well-studied nearby Class I source in Taurus within the JOYS
sample is TMC1A (IRAS 04365+2535). This $\sim$0.5 M$_\odot$ protostar
has a striking blue-sided molecular disk wind imaged with ALMA
\citep{Bjerkeli16} and has recently been studied with JWST-NIRSpec by
\citet{Harsono23} revealing a collimated jet in [Fe II] 1.644
$\mu$m. Using the NIRSpec Pa$\beta$ and Br$\gamma$ lines
and assuming that all the luminosity comes from protostellar
accretion, very low rates of $2\times 10^{-12}$ and $3\times 10^{-9}$
M$_\odot$ yr$^{-1}$ have been reported \citep{Harsono23}. We note that
neither value is corrected for extinction, which may also be
responsible for the differences between the two diagnostics.

We can revisit the TMC1A case with our MIRI data since the 7-6
Humphreys $\alpha$ line at 12.37 $\mu$m is detected and suffers much
less from extinction (Fig.~\ref{fig:TMC1A-accretion}). Moreover, the
extinction can be estimated, albeit with considerable uncertainty, to
be $A_V\approx 30$ mag from the silicate optical depth using
$A_V=18.5 \times \tau_{\rm 9.7}$.  Using the same relation of
\citet{Rigliaco15} and correcting for extinction and water, a mass
accretion rate of $6.6\times 10^{-7}$ M$_\odot$ yr$^{-1}$ is found for
TMC1A. This value is much higher than inferred from the uncorrected
NIRSpec data and consistent with those estimated from the bolometric
luminosity of the source and the need to build up a $\sim$1 M$_\odot$
star in less than 1 Myr. It agrees well with accretion rates found by
\citet{Fiorellino23} for a large sample of much less extincted Class I
sources based on Br$\gamma$ data. The case of SMM3 is discussed below
in Sect.~\ref{sec:SMM3}. We stress that all accretion rate
determinations depend strongly on the assumed extinction toward the
protostellar embryo; a detailed discussion of the different extinction
diagnostics is presented in Appendix~\ref{sec:app_extinction}.

The H I 7-6 line at 12.372 $\mu$m used to determine these accretion
rates is actually located near another recombination line, the H I
11-8 transition at 12.387 $\mu$m. The {\it Spitzer}-IRS could not
resolve these lines, but MIRI-MRS can \citep{Franceschi24}; the
empirical relation of \citet{Rigliaco15} is therefore taken to refer
to the sum of both components. Also, as
Figure~\ref{fig:TMC1A-accretion} shows, the H I flux can be
contaminated by warm water emission. However, the fluxes of TMC1-W and
TMC1A cannot be fully explained by water: H I still dominates the
12.37 $\mu$m emission. Any water contribution can be subtracted using
a fit to the H$_2$O emission over a larger wavelength range, as has
been done for the values listed in Table~\ref{tab:accretion}
\citep{Rigliaco15,vanGelder24overview}. An updated water-corrected
relation is presented in a recent JWST-based study by
\citet{Tofflemire25}, which gives consistent results for our sources.

Another complication is that not all H I recombination line emission
need to originate from accretion onto the protostars and that
scattering can complicate the extinction correction at shorter
wavelengths \citep{Delabrosse24}. Brackett $\alpha$ NIRSpec imaging
shows bright extended emission in some sources that is clearly
associated with shocks in protostellar jets
\citep{Federman24,Neufeld24}. VLTI-Gravity observations that resolve
the Brackett $\gamma$ emission in Class I/II T Tauri systems also
demonstrate that winds and outflows commonly contribute to the H I
emission at the scale of the inner disk
\citep{Gravity23Wojtczak}. Also, the \citet{Rigliaco15} relation was
derived from {\it Spitzer} data in large apertures that could have
included extended H I emission, thereby assigning all the H I to
accretion rather than a mix of accretion and ejection. Thus, the
inferred accretion rates from H I should be considered as upper
limits. A combination of spatially and spectrally resolved H I lines
observable with MIRI and NIRSpec tracing a range of temperatures and
densities, combined with H I excitation calculations \citep{Kwan11},
is needed to disentangle the components.

\section{Protostellar jets, winds, and outflows}
\label{sec:outflows}

\begin{figure*}[h!]
\begin{centering}
\includegraphics[width=16.5cm]{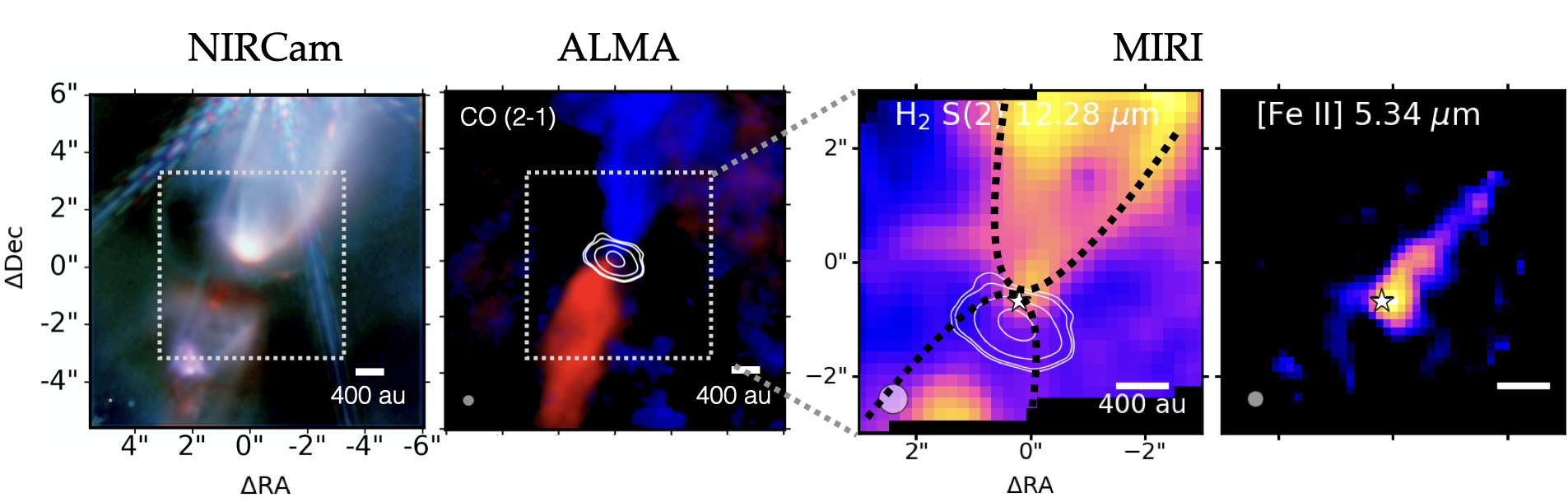}
\caption{Comparison of JWST and ALMA data for the Class 0 protostar
  Serpens SMM3. From left to right, (1) JWST NIRCam image of SMM3
  using a combination of the F210M (blue), F360M (green), and F480M
  (red) filters from archival data of PID 1611; (2) ALMA image of the
  CO $J$=2--1 red and blue outflow lobes in the inner $10"\times 10"$
  together with the dust disk at 1.3 millimeter continuum emission
  (contours) \citep{Tychoniec21}; (3) JOYS MIRI-MRS maps of the H$_2$
  S(2) line at 12.28 $\mu$m, with the dotted line outlining the
  scattered light cavity seen with NIRCam; and (4) the [Fe II] line
  5.34 $\mu$m, both in a zoomed-in region of $6"\times 6"$. Beam sizes are
  indicated in the lower-left corner of each panel.}
\label{fig:SMM3alma}
\end{centering}
\end{figure*}

Studies on the MIRI-MRS jet, wind, and outflow have been published within JOYS
for the high-mass protobinary IRAS 23385+6053
\citep{Beuther23,Gieser23} and for the low-mass protobinary system
TMC1 \citep{Tychoniec24}. Moreover, the entire blue outflow lobe and
part of the red lobe of HH 211, $1'$ in extent, has been mapped with
the MRS \citep{Caratti24}. The latter study thus includes also the bow
shocks at the tip of the outflow, in addition to the jets and winds
closer to the protostar. Figure~\ref{fig:protostar-cartoon},
Sect.~\ref{sec:components} and Appendix \ref{sec:back_outflows} provide
definitions of these terms as used in this paper and scientific
background.

\subsection{The Class 0 source Serpens SMM3 as an example}
\label{sec:SMM3}

Here we present the case of the Class 0 source Serpens SMM3 ($d$=439
pc, Table~\ref{tab:sources}) to illustrate typical findings in MRS
maps.  SMM3 is, with 28 L$_\odot$, among the most luminous of the
low-mass sources in the JOYS sample. It has been imaged by ALMA in
various molecular lines \citep[][and refs cited]{Tychoniec21} and has
a relatively narrow CO outflow opening angle of $< 20^o$. It is one of
the few Class 0 sources with clear ``knots'' seen in CO and SiO about
8$''$ offset from the source along the jet (also called extremely high
velocity (EHV) ``bullets''). It also shows hints of a large embedded
rotating disk, further discussed in Sect.~\ref{sec:disks}. Finally,
NIRSpec IFU data exist as part of JOYS+ (PID 1186) that can enhance
the analysis in future studies, especially by observing more accretion
tracers and ro-vibrational H$_2$ and CO lines
\citep{LeGouellec25S68N}.

\begin{figure}
\begin{centering}
\includegraphics[width=9cm]{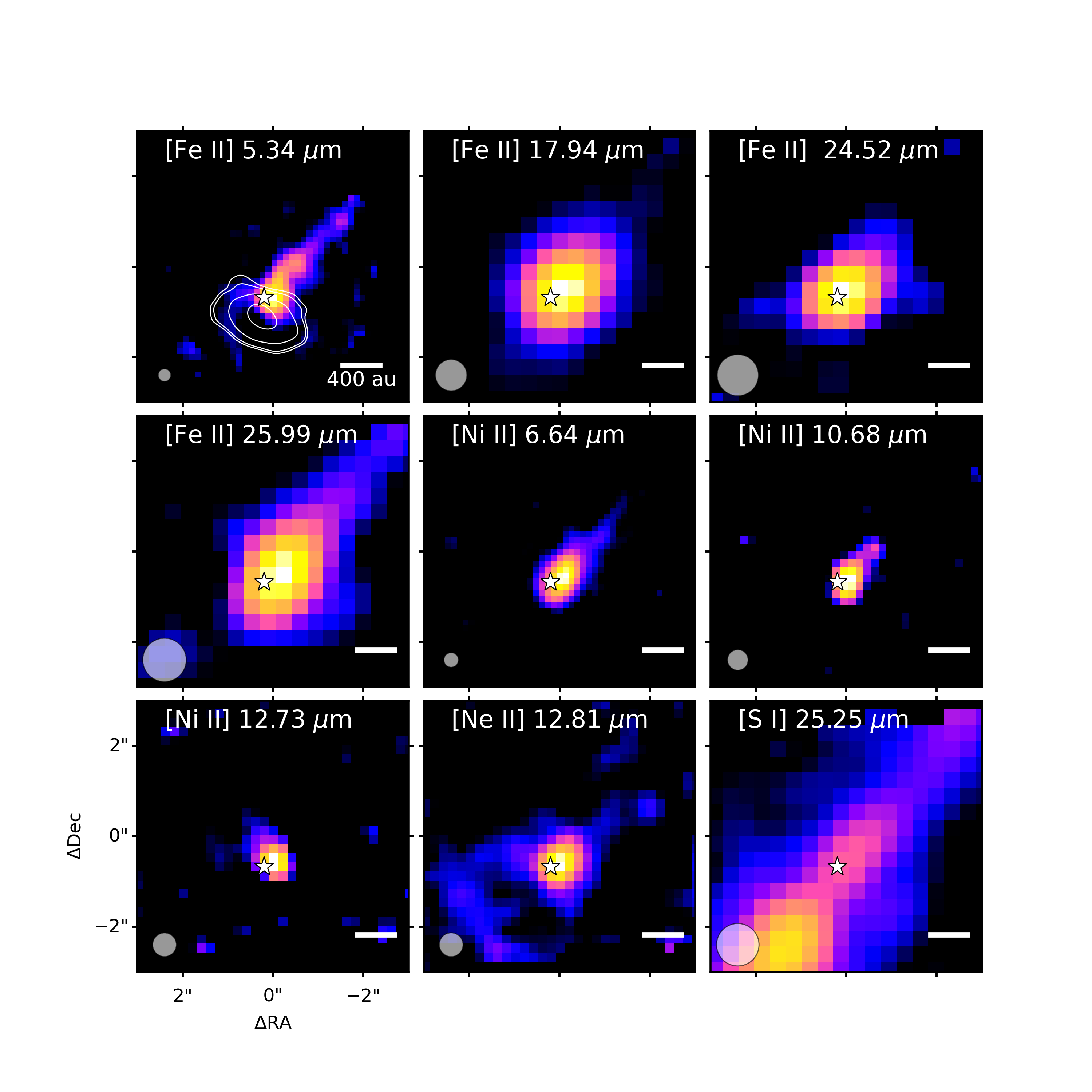}
\caption{Maps from MIRI-MRS of various atoms and ions toward the Class 0
  protostar Serpens SMM3. The white contours in the top-left panel
  outline the dust disk seen in millimeter continuum. Beam sizes are
  indicated in the lower left corner of each panel.  Each panel is
  scaled to the maximum emission of that species. The maximum and
  minimum colors are (in Jy km s$^{-1}$), from top left to bottom
  right: [Fe II] 5.34 (0.033, 0.003); [Fe II] 17.94 (0.84, 0.017); [Fe
  II] 24.52 (0.62, 0.025); [Fe II] 25.99 (1.86, 0.075); [Ni II] 6.64
  (0.05, 0.002); [Ni II] 10.68 (0.006, 0.001); [Ni II] 12.73 (0.005,
  0.001); [Ne II] 12.81 (0.017, 0.003); [S I] 25.25 (3.27, 0.033). }
\label{fig:SMM3atoms}
\end{centering}
\end{figure}

\begin{figure}
\begin{centering}
\includegraphics[width=9cm]{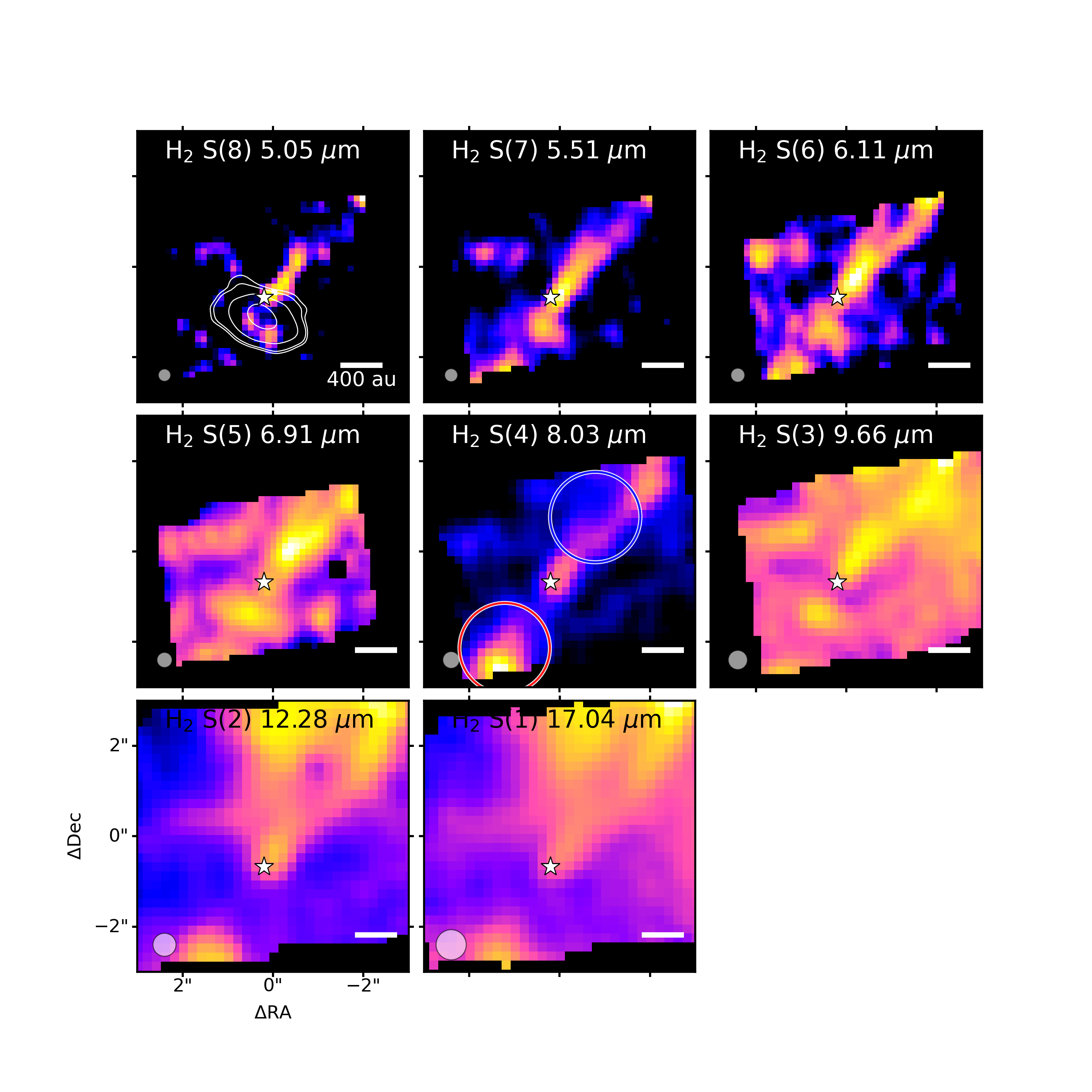}
\caption{Maps from MIRI-MRS of the various H$_2$ lines toward the Class 0
  protostar Serpens SMM3. The white contours in the top-left panel
  indicate the dust disk seen in millimeter continuum. The red and
  blue circles on the S(4) image indicate the positions where the
  H$_2$ spectra and rotational diagrams have been extracted. 
  The S(3) image at 9.66 $\mu$m is strongly affected by silicate
  extinction and the S(5) line at 6.9 $\mu$m by ice extinction; hence,
  their maps are more noisy. Beam sizes are indicated in the lower
  left corner of each panel. Each panel is scaled to the maximum
  emission of that species. The maximum and minimum colors are (in Jy
  km s$^{-1}$), from top left to bottom right: S(8) (0.008, 0.003);
  S(7) (0.033, 0.007); S(6) (0.013, 0.003); S(5) (0.026, 0.003); S(4)
  (0.089, 0.009); S(3) (0.035, 0.005); S(2) (0.090, 0.009); S(1)
  (0.084, 0.008). }
\label{fig:SMM3H2}
\end{centering}
\end{figure}

\paragraph{MRS maps.} Figure~\ref{fig:SMM3alma} (left) presents a
NIRCam image of SMM3 over the inner $10"\times 10"$ region together
with an ALMA image showing the swept-up CO outflow as well as the SMM3
dust disk in millimeter continuum. The two right-hand panels highlight
the inner $6"\times 6"$ in the H$_2$ S(2) line at 12.28 $\mu$m and the
[Fe II] line 5.34 $\mu$m. Only a single MRS pointing was obtained for
this source, thus covering only part of the outflow. The [Fe~II] image
clearly reveals the collimated jet, whereas the H$_2$ line shows the
wider-angle warm molecular gas inside the outflow cavity outlined by
the NIRCam scattered light image. Mid-infrared emission is stronger in
the blue lobe than in the red lobe due to enhanced extinction in the
latter close to the source (see also below).

The MIRI images of various other atomic lines and of H$_2$ covered in
the MRS data are presented in Figures~\ref{fig:SMM3atoms} and
\ref{fig:SMM3H2}, with the spatial resolution decreasing with
wavelength from $\sim 0.2''$ at 5 $\mu$m to $\sim 0.7''$ at 25
$\mu$m. All four [Fe II] lines trace the blue part of the jet, even at
the longest wavelength of 25.99 $\mu$m, but interestingly, the [S I]
25.25 $\mu$m line, which is close in wavelength, shows a more
prominent peak on the red side of the outflow. This demonstrates that
the lack of [Fe II] emission on the red side is real, and not just due
to extinction; it could be due to lower excitation and shock
conditions along the red-shifted jet. The intensity of all four [Fe
II] lines is sensitive to density and shock velocity, although to
different degrees \citep{Hartigan87,Hollenbach89}. In particular, the
[Fe II] 25.99 $\mu$m line is a fine-structure transition within the
ground electronic state with the lowest critical density, similar to
that of the observed [S I] 25.2 $\mu$m, yet this [Fe II] line is also
not detected. This suggests that in addition to excitation, abundance
differences also play a role: less Fe may have been returned to the
gas phase on the red side by the shocks.

The [Ni II] and [Ne II] lines peak close to the source in the blue
part of the jet (Fig.~\ref{fig:SMM3atoms}). For [Ni II] its limited
extent in SMM3 may be due to its lower $S/N$, since in other
still-unpublished JOYS sources it usually follows [Fe II].  In
contrast, [Ne II] often shows more central emission, although it has a
weak jet component as well for SMM3.

\begin{figure}
\begin{centering}
\includegraphics[width=4.3cm]{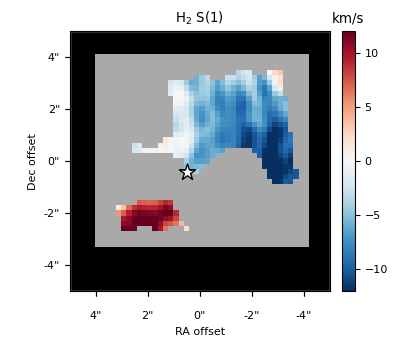}
\includegraphics[width=4.3cm]{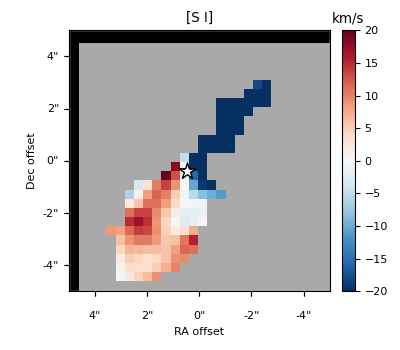}
\caption{MIRI-MRS moment-1 maps of the H$_2$ S(1) 17.0 $\mu$m and [S
  I] 25.25 $\mu$m lines toward Serpens SMM3 showing the ability of the MRS
  to distinguish the red- and blue-shifted gas. The velocities are
  relative to the source velocity of $V_{\rm LSR}=7.6$ km s$^{-1}$. }
\label{fig:SMM3velocity}
\end{centering}
\end{figure}

Figure~\ref{fig:SMM3H2} presents images of the various H$_2$ lines
covered by the MIRI-MRS. Whereas the low-lying S(1) and S(2) lines
trace the wide-angle wind, the images of the higher H$_2$ lines are
much narrower and clearly trace the jet, as seen most prominently in
the H$_2$ S(7) and S(8) lines at the shortest MIRI wavelengths of 5.51
and 5.05 $\mu$m. Here the images are sharpest and comparable in
resolution to those of the [Fe II] 5.34 $\mu$m. Moreover, as also seen
in Fig.~\ref{fig:SMM3alma}, H$_2$ is clearly detected in the red
outflow lobe as well, even at 5.5 $\mu$m. In fact, the H$_2$ S(4) line
reveals two new ``knots'' on opposite sides of the source.
Interestingly, there appears to be a slight tilt between the warm H$_2$
low-$J$ images and that of the hot H$_2$ and atomic lines. No HD is
detected at either position, consistent with the fact that the H$_2$
lines in SMM3 are not as bright as those in HH 211 where several
  HD lines are seen \citep[][see also
  Appendix~\ref{sec:app_background}]{Francis25HD}.
Figure~\ref{fig:SMM3velocity} presents the moment-1 map of the H$_2$
S(1) and [S I] 25 $\mu$m lines, clearly separating the red and blue
parts in [S I]. Radial velocities of $\sim$20 km s$^{-1}$ are seen for
H$_2$ and [S I].

Taken together, these data suggest that -- like for HH 211 -- the SMM3
jet has a ``nested'' structure with an ionized core surrounded by a
molecular layer. This ionized core, however, appears on just one side
of the outflow. Evidence of a wide-angle wind is found as well in
low-$J$ H$_2$ lines. This example also shows that the full suite of atomic,
ionic and molecular lines is needed to reveal the physical and
chemical structure of the outflow.

\begin{figure*}
\begin{centering}
\includegraphics[width=6.5cm]{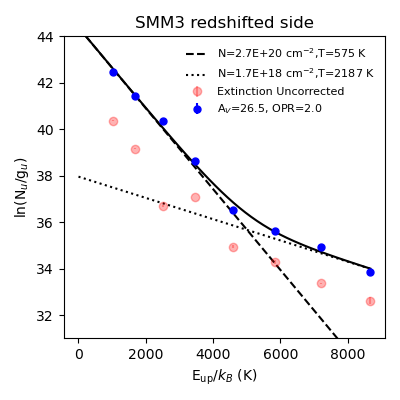}
\includegraphics[width=6.5cm]{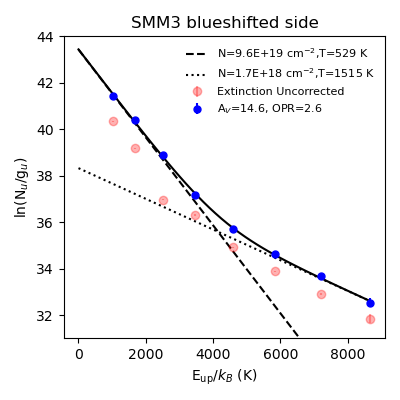}
\caption{Rotational diagrams of H$_2$ at the blue (left) and red
  (right) lobes using spectra extracted at the positions indicated in
  Fig.~\ref{fig:SMM3H2}. The light red points indicate the values
  before extinction correction, the full blue points after extinction
  correction. Two temperatures and an OPR have
  been fitted to the extinction-corrected fluxes.}
\label{fig:SMM3H2rot}
\end{centering}
\end{figure*}

\paragraph{H$_2$ excitation and extinction determination.} H$_2$ line fluxes
have been extracted near two ``knot'' positions in a $\sim$1$''$
aperture and put in a rotational diagram. The results for the blue and
red positions are presented in Figure~\ref{fig:SMM3H2rot}. As has been
commonly found, also from ISO and {\it Spitzer} data
\citep[e.g.,][]{vanDishoeck04,Neufeld06,Maret09,Giannini11,Gieser23},
the upper level column densities derived from the S(3) line clearly
fall below the trend of the other lines due to additional extinction
by the silicate feature in the envelope. This deficit can therefore be
used to infer the overall extinction in combination with an assumed
overall extinction curve. Here the KP5 extinction curve of
\citet{Pontoppidan24ext} has been used which has a somewhat stronger
silicate feature than the \citet{McClure09} extinction curve.

At least two temperature components are needed to fit the remainder of
the H$_2$ lines, consistent with previous findings (see above
references). Using a procedure to obtain a simultaneous best fit for
the two temperatures plus ortho-to-para ratio (OPR) and extinction
results in the values indicated in the panels \citep{Francis25HD}. The
warm temperature is typically $\sim$600 K, whereas the hot temperature
is $\sim$2000-3000 K. The OPR ratio is 2.0--2.5, close to the high
temperature value of three. Interestingly, the extinction is found to be
only slightly higher for the red lobe, $A_V\approx 25$ mag, versus 15
mag for the blue lobe (Fig.~\ref{fig:SMM3H2rot}).

The temperature structure as derived from H$_2$ can vary along the
outflow axis, but also across it. Indeed, the fact that the low-$J$
H$_2$ lines have a wider opening angle than the high-$J$ lines
indicates that gas temperatures decrease away from the jet axis
\citep{Tychoniec24,Caratti24,Delabrosse24}. For the case of Serpens
SMM3, the temperature of the warm component decreases from 600 K by about
100--150 K in an aperture centered close to the outflow
cavity wall. Ultimately, mapping of the temperature structure of the
wind may put constraints on the heating mechanisms and thereby also
its launching mechanisms.

\paragraph{Mass loss rates.} 
The H$_2$ column densities found from the rotational diagrams can be used
to calculate mass loss rates through

\begin{equation}
  {\dot M}_{\rm H_2}= 2\mu m_{\rm H} ({\overline N} A) (v_{\rm tg} l_{\rm tg}),
\end{equation}

\noindent
where ${\overline N}$ is the warm H$_2$ column density averaged over
the emission area $A$ (assumed to be circular and equal to the area
over which the spectrum has been extracted, Fig.~\ref{fig:SMM3H2} and
\ref{fig:SMM3H2rot}), $v_{tg}$ is the tangential velocity and
$l_{\rm tg}$ is the projected knot size.  The width of the
blue-shifted knot $l_{\rm tg}$ is measured to be 625 au at the
position of the extracted spectrum from the lowest $J$ lines, whereas
the line-of-sight velocity is --18 km s$^{-1}$ based on the average of
the velocity shifts from the H$_2$ S(1) to S(4) lines that best trace
the warm component (Fig.~\ref{fig:SMM3velocity}). Assuming an
inclination of 50$^o$ estimated from CO millimeter maps
\citep{Yildiz15} and taking the extinction corrected warm H$_2$ column
of $9.6 \times 10^{19}$ cm$^{-2}$ derived from the rotation diagram
(Fig.~\ref{fig:SMM3H2}), gives a mass loss rate from the warm H$_2$
component of ${\dot M}_{\rm H_2}=1.4 \times 10^{-7}$ M$_\odot$
yr$^{-1}$. The uncertainty is a factor of around two due to
uncertainties in inclination and extinction.  This value is a factor
of a few lower than the mass loss rate of $7\times 10^{-7}$ M$_\odot$
yr$^{-1}$ found for HH 211 from warm H$_2$.

The advantage of using H$_2$ is that the total column of hydrogen
nuclei is measured directly, as is its velocity, so no density or
velocity needs to be inferred indirectly from diagnostics and models.
Mass loss rates can also be derived from the atomic lines, as shown in
detail for the case of HH 211 by \citet{Caratti24}.
Appendix~\ref{sec:app_massloss} describes slightly different methods
used here to estimate ejection rates from the extinction-corrected [Fe
II] 26 $\mu$m line \citep{Watson16}, the [Ne II]+[S I] lines combined
with shock models from \citet{Hollenbach89} (Tychoniec et al., in
prep.) and from the [O I] 63 $\mu$m line \citep{Hollenbach85}.

These atomic diagnostics give values of $\dot{M}_{\rm jet}$ that are
slightly lower than derived from H$_2$ by a factor of two to three when
applied to one outflow lobe (Appendix~\ref{sec:app_massloss}).  Taking
into account the factor of roughly three higher velocities of the atomic
component compared with H$_2$, the atomic and molecular jets seem to
provide comparable thrust (momentum flux) $\dot M \times v_{\rm tot}$
to drive the material in the SMM3 system; the molecular
component is not as dominant here as it has been found to be for HH
211 \citep{Caratti24}.

When summed over the entire MRS channel 4 $7"\times 7"$ FoV, the [Fe
II] 26 $\mu$m mass loss rate is $\dot{M}_{\rm jet}$ = 4.0
$\times 10^{-7}$ M$_ {\odot}$ yr$^{-1}$. This value is within a factor
of two of that based on the {\it Herschel}-PACS [O I] 63 $\mu$m
luminosity in a $\sim 9"$ spaxel \citep{Mottram17}. We note that the [Fe
II] 26 $\mu$m relation of \citet{Watson16} has been calibrated against
[O I] 63 $\mu$m for a large sample of sources and that in general the
[O I] value should be considered as an upper limit (see
Appendix~\ref{sec:app_massloss}).

It is interesting to compare these mass outflow values with mass
accretion rates.  No hydrogen recombination lines are detected for
SMM3 but the upper limit on the H I 7--6 line constrains the mass
accretion rate to a low value of $< 5\times 10^{-9}$ M$_{\odot}$
yr$^{-1}$, assuming an extinction of 50 mag (see
Table~\ref{tab:accretion}).  An alternative rough estimate can be
obtained assuming that the observed bolometric luminosity is mainly
due to accretion for Class 0 sources like SMM3.  Using the relation
given in Sect.~\ref{sec:onset}, \citet{Mottram17} find
$1.2\times 10^{-5}$ M$_\odot$ yr$^{-1}$, close to our value in
Table~\ref{tab:accretion}.  The large difference with the upper limit
from H I is either caused by an underestimate of the H I extinction
due to the big massive disk of SMM3 making it more difficult to
peer close to the star, or by the magnetospheric accretion mode not
being applicable to Class 0 sources (and thus the H I relations).
Alternatively, not all luminosity needs to come from accretion. If it
does, the inferred
${\dot M}_{\rm jet} / {\dot M}_{\rm accretion} \approx$0.01. This
value is on the low side of previous determinations for other Class 0
and I sources \citep{Watson16}.

Taken together, this SMM3 example and the comparison of various
methods to derive mass outflow and accretion rates emphasize the need
for high quality spatially resolved observations of each of the
atomic, ionic and molecular components that are now available with
JWST, not only with MIRI but also with NIRSpec
\citep{LeGouellec25S68N}. They also indicate that earlier rates based on
large aperture data need to be used with caution, especially based on
[O I]. Similar comparisons for other Class 0 sources are needed to
assess the usefulness of each of the methods and calibrate them
better to trace mass loss and accretion.

\subsection{Morphologies: Which line traces what?}
\label{sec:morphologies}

\begin{figure*}
\begin{centering}
\includegraphics[width=13.5cm]{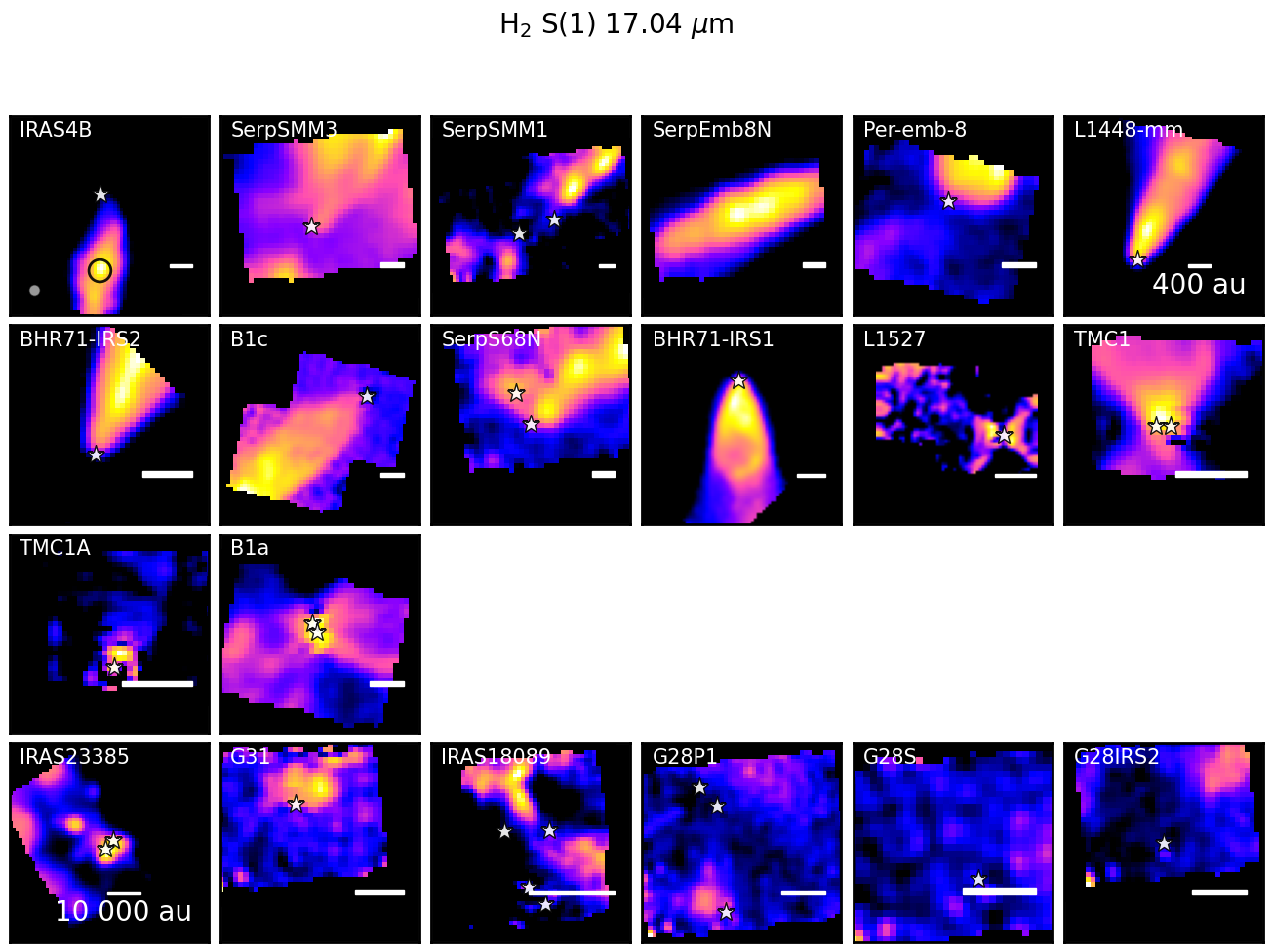}
\caption{Maps from MIRI-MRS of the H$_2$ S(1) line at 17.03 $\mu$m toward
  all JOYS sources. The top three rows show the low-mass protostars in
  order of increasing $T_{\rm bol}$, whereas the bottom row shows the
  high-mass protostars.  The white star(s) or arrow in each panel
  indicate the position of the protostar(s); for low-mass protostars
  those are derived from millimeter interferometry. A scale bar of 400
  au is shown in each low-mass protostar panel, and a bar of 10000 au
  in each high-mass protostar panel.  The beam size is indicated in
  the lower left corner of the first panel. The maximum colors in each
  panel (in Jy km s$^{-1}$) from the top left to bottom right
  for low mass are as follows: 0.59, 0.12, 0.03, 0.14, 0.08, 0.25, 0.08, 0.025,
  0.041, 0.11, 0.006, 0.18, 0.31, 0.064. For high mass, they are 0.21,
  0.02. 0.06, 0.021, 0.023, 0.027. }
\label{fig:JOYS-H2S4}

\includegraphics[width=13.5cm]{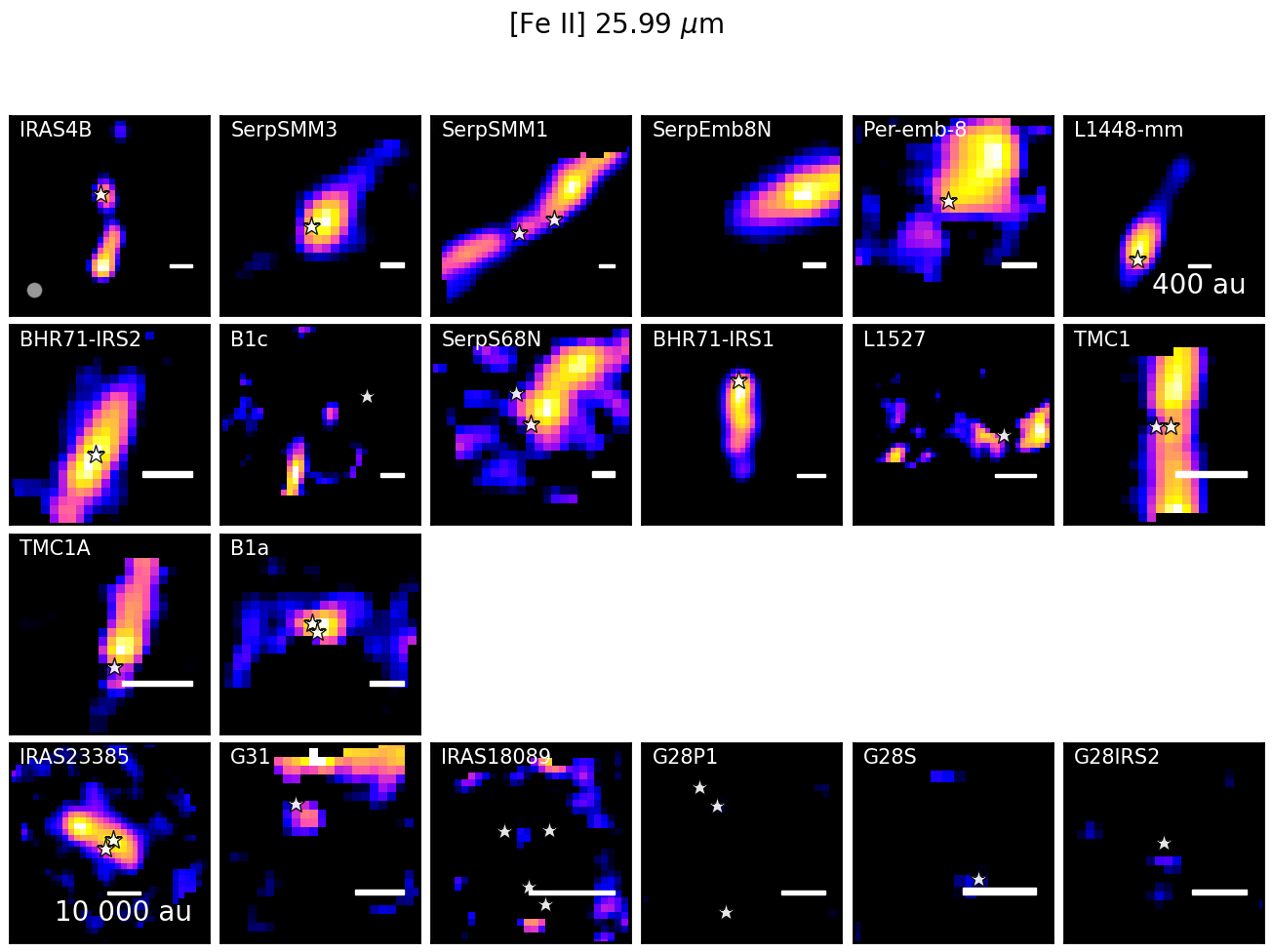}
\caption{Maps from MIRI-MRS of the [Fe II] $^4$F$_{7/2}-^4$F$_{9/2}$ line
  at 25.99 $\mu$m toward all JOYS sources. The top three rows show the
  low-mass protostars in order of increasing $T_{\rm bol}$, whereas
  the bottom row shows the high-mass protostars.  The white star(s) or
  arrow in each panel indicate the position of the protostar(s); for
  low-mass protostars those are derived from millimeter
  interferometry. A scale bar of 400 au is shown in each low-mass
  protostar panel, and a bar of 10000 au in each high-mass protostar
  panel.  The bright [Fe II] emission seen at the edges of some of the
  high-mass images are real. The beam size is indicated in the lower
  left corner of the first panel. The maximum colors in each panel (in
  Jy km s$^{-1}$) from top left to bottom right for low mass are as follows:
  1.18, 1.89, 15.65, 2.85, 0.62, 6.97, 1.41, 0.21, 0.56, 42.63, 0.17,
  2.00, 3.72, 1.26. For high mass, they are 0.79, 0.80, 0.80, 0.80, 0.80,
  0.80.}
  \label{fig:JOYS-Fe17}
\end{centering}
\end{figure*}

The next step is to compare the Serpens SMM3 results with those of
other sources studied with JWST.  Table~\ref{tab:lines} in
Appendix~\ref{sec:app_diagnostics} summarizes the main atomic and
molecular lines that are seen in the MIRI spectra of protostars,
together with their origin. It is well known that elements can be
divided into volatile (e.g., noble gases), moderately volatile or
semi-refractory (e.g., S, Cl), and refractory elements (e.g., Fe, Ni,
Co) depending on their condensation temperatures
\citep{Lodders03}. Observations of elemental abundances in diffuse
clouds using UV absorption lines show that the amount of depletion
strongly varies between these categories, with volatile elements like
Ne and Ar basically undepleted and refractory elements like Fe, Ni and
Co heavily depleted by up to two orders of magnitude due to their
incorporation into dust \citep{Savage96,Jenkins09}. The
semi-refractories are somewhere in between.  The MIRI-MRS line images
can indeed be divided along these different categories.

Figures~\ref{fig:JOYS-H2S4} and \ref{fig:JOYS-Fe17} provide an
overview of the H$_2$ S(1) line and one low-excitation fine-structure
line of a refractory element, [Fe II] 25.99 $\mu$m, for the entire
JOYS low- and high-mass protostar sample. In these figures, the
low-mass sources NGC 1333 IRAS4A, SVS4-5 and HH 211 have not been
included (see Sect.~\ref{sec:observations}); hence there are 14 rather
than 17 low-mass panels. The full set of all line images will be
presented in future papers, but the trends for low-mass sources are
similar to those presented for Serpens SMM3.

A common theme is that a narrow jet is seen prominently in the [Fe II]
lines at 5.34, 17.92, 24.52 and 25.99 $\mu$m, as well as in other
refractory species, such as [Ni II] 6.63 $\mu$m and [Co II] 10.52
$\mu$m, which have been returned to the gas phase by shocks.
Notably, [Fe I] 24.04 $\mu$m is exclusively detected in some
Class 0 sources, not in Class I sources, and always in the jet. This
finding is consistent with jets becoming more ionized with
protostellar evolution \citep{Nisini15}. The detection of both Fe and
Ni in jets points to the presence of iron/nickel grains which do
not have spectral features: this is the only way to discover the
nature of Ni-containing dust \citep{Henning10,Manfroid21}.

Among the semi-refractories, the [S I] 25.24 $\mu$m line is bright and
always elongated in the jet direction for Class 0 sources, where it
also shows up prominently in bow shocks, like in SMM3. As for the
refractories, it has been returned to the gas by shocks but it usually
follows H$_2$ more closely than [Fe II]. If detected in Class I
sources, [S I] emission is always compact on disk scales, and no
longer associated with any jet \citep[e.g.,][]{Tychoniec24}.  In this
context it has been suggested that sulfur-bearing species trace
accretion shocks onto the disk \citep[e.g.,][]{Sakai14}, but no
convincing indications have yet been found that this is the case for
[S I] (see Sect.~\ref{sec:disks}). A similar trend is seen with the other
moderately volatile element [Cl I] 11.33 $\mu$m, but this line is much
fainter so its detection is not clear for many targets.

Lines of the noble gases Ne and Ar show a mixed behavior: the strong
[Ne II] 12.81 $\mu$m line often follows the jet, but it may also have
a compact, spatially broader component centered close to the source,
likely tracing the X-ray or (E)UV-irradiated cavities and cavity walls
(see Figure~\ref{fig:SMM3atoms} and \citealt{Gieser23} for the
high-mass source IRAS 23385). The same holds for [Ar II] at 6.98
$\mu$m, but since its line is weaker than that of [Ne II] it is more
difficult to separate the components. In contrast, [Ne III] 15.56
$\mu$m and [Ar III] 8.99 $\mu$m are primarily detected in Class I
sources where they show only a moderately extended component,
plausibly tracing the ionizing radiation \citep[see
also][]{Gieser23,Tychoniec24}. The images of the noble gases
demonstrate that not just elemental depletion but also local
conditions (e.g., what is required to ionize Ne and Ar) determine the
emission pattern.

Heated dust is detected through 20 $\mu$m continuum emission in some
Class 0 sources such as HH 211, indicating that dust is launched in
the jet \citep{Caratti24}.  The question whether that dust is
subsequently destroyed or not by shocks requires comparison of one of
the refractory elements with a non-refractory one. Typically, sulfur
lines are used for the latter \citep{Nisini02}. However, the depletion
behaviors of S and Cl are not yet well understood (see also above
  discussion): both species are undepleted in diffuse clouds with
$A_V<$1 mag \citep{Jenkins09}, but in dense clouds the bulk of S has
been transformed into a more refractory component
\citep[e.g.,][]{Kama19}. Depletion of chlorine appears to be more
modest in dense clouds based on observations of HCl \citep{Blake86}.
Thus, it is not clear a priori which element to take as the undepleted
reference in determining the amount of depletion of elements in
shocks. One of the noble gases (e.g., neon) that does not suffer any
depletion could be a cleaner reference probe for abundances, as long
as their line flux does not depend sensitively on other physical
parameters. Another possibility for the case of S is to take H$_2$
itself \citep{Anderson13}.

In the youngest Class 0 protostars, H$_2$ shows a mix of wide-angle
wind and jet emission. The narrow jet is seen in the higher excitation
H$_2$ lines such as S(5) and S(7), whereas emission from the
lower-lying H$_2$ lines is dominated by the wide-angle wind
(Fig.~\ref{fig:SMM3H2}). For Class I sources, all H$_2$ lines fill the
outflow cavity with the opening angle of the disk wind increasing with
decreasing excitation (i.e., $J$ level) \citep{Tychoniec24}. Molecular
lines from CO, OH, HCO$^+$ and CO$_2$ are also detected associated
with the outflow \citep{Tychoniec24,Narang24,vanGelder24overview} (see
also Sect.~\ref{sec:hotcore}).

In conclusion, warm molecular gas is clearly present within the
outflow cavity, not just as part of the cavity wall, and its velocity
structure indicates that the gas is moving in the red- or blue-shifted
directions. Whether H$_2$ lines are indeed tracing disk winds close to
the source or simply gas inside the cavity that is heated and put in
motion otherwise (e.g., by internal jet shocks or shocks with ambient
material) would require velocity resolved line profiles and spatially
resolved H$_2$ temperature and column density maps. Farther removed
from the source, H$_2$ may also be mixed in by turbulence from the
entrained outflow gas, and that gas may also be heated by UV radiation
along the outflow cavity walls \citep[e.g.,][]{Yildiz15}.

The atomic and molecular line images for the high-mass source IRAS
23385 +6053 show similar trends, both in refractories, volatiles and
H$_2$ \citep{Beuther23,Gieser23}. The patterns for the other high-mass
sources are less clear from the images presented here: IRAS 18089
-1732 has a cluster of high-mass sources resulting in bright H$_2$
emission throughout the region, whereas the IRDC G28 positions show
much weaker emission. The high-mass outflows and accretion will be
studied in future papers.

\subsection{Comparison with other JWST studies}
\label{sec:otherJWST}

The JOYS findings are consistent with results of low-mass Class 0
protostars from a number of other JWST programs. Early MIRI-MRS
examples are the low-mass Class 0 sources IRAS 15398-3359
\citep[CORINOS program;][]{Yang22,Salyk24} (1.5 L$_\odot$) and IRAS
16253-2429 (0.2 L$_\odot$) \citep[IPA program;][]{Narang24}. Also,
NIRSpec images of [Fe II] and H$_2$ \citep{Federman24} as well as CO
\citep{Rubinstein24} ro-vibrational lines have been presented for all
five IPA sources from low to high mass, for the Class 0 binary Serpens
S68N \citep{LeGouellec24} and for the Class I source TMC1A
\citep{Harsono23}.
MIRI and NIRSpec images of the Class I source HH 46 IRS (16 L$_\odot$)
have been published by \citet{Nisini24} as part of PROJECT-J; their
Table 1 provides a useful overview of all atomic lines detected in HH
46.

A common finding of all of these early MIRI and NIRSpec results across
low-mass Class 0 and I sources as well as high-mass protostars, is the
presence of highly collimated jets, most notably seen in the various
[Fe II] lines and in the highest excitation H$_2$ pure rotational and
ro-vibrational lines.  Wider angle winds are traced by the lowest
H$_2$ pure rotational lines, as found within JOYS
\citep{Tychoniec24}. Lines of other refractory species as well as
noble gas (Ne, Ar) lines are also commonly detected, with the latter
species often showing a different distribution from either [Fe II] or
H$_2$, consistent with what is found in JOYS.  Analyses of the nested
physical structure of the Class I source DG Tau B and a small sample
of young Class II jets and winds are presented in
\citet{Delabrosse24,Pascucci25}.  Some H~I emission is seen also off
source in both NIRSpec and MIRI data \citep[e.g.,][]{Neufeld24},
demonstrating that not all the H~I recombination line luminosity is
due to the accretion column onto the star.

\begin{figure}
\begin{centering}
\includegraphics[width=9cm]{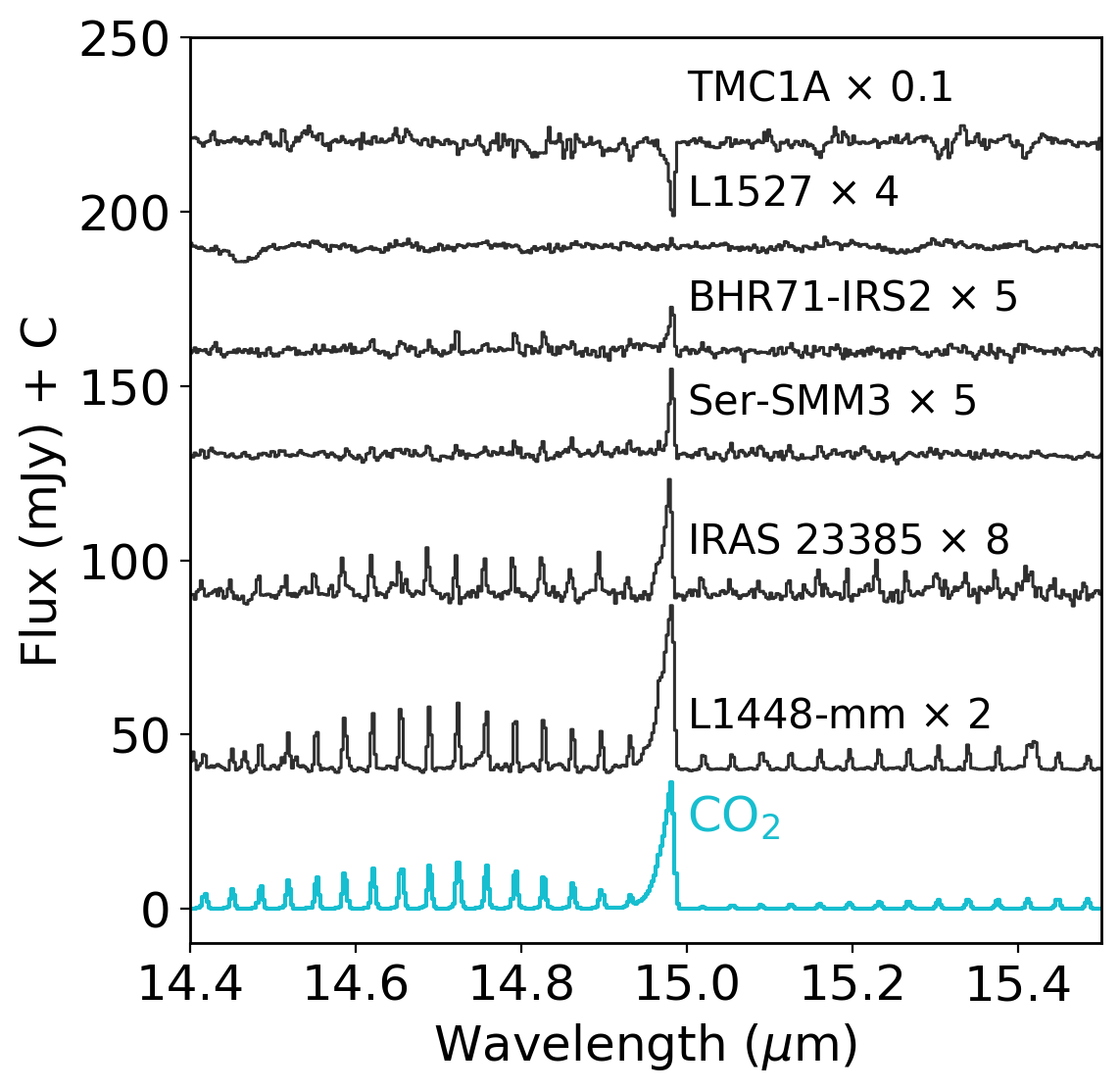}
\caption{Comparison of CO$_2$ 15 $\mu$m emission and absorption toward
  several low- and high-mass protostars. The bottom panel shows a
  model spectrum at $T$=120 K (best fit for L1448-mm), with the
  $P$-branch extincted by CO$_2$ ice.}
\label{fig:CO2overview}
\end{centering}
\end{figure}

\section{Hot cores and dense molecular shocks: Chemistry}
\label{sec:hotcore}

MIRI-MRS spectra of the first observed JOYS target, the high-mass
binary protostar IRAS 23385+6053, demonstrated that many gas-phase
molecular lines can be detected at the source positions and that not
just $Q$ branches but also individual $P$- and $R$- branches (see
Figs.~\ref{fig:molecules1} and \ref{fig:molecules2}) can be seen
\citep{Francis24}.  The MIRI-MRS spectra at central positions of the
low-mass JOYS sources have been analyzed in a recent overview paper of
molecular lines by \citet{vanGelder24overview}. Rich spectra
containing lines of up to 13 different molecules (including
isotopologs) are seen toward three out of 16 sources with mid-infrared
continuum, two in emission (L1448-mm and BHR71-IRS1) and one in
absorption (B1-c).  Other sources show primarily H$_2$O, CO, CO$_2$
and/or OH, but not C$_2$H$_2$, HCN, SO$_2$ or SiO.  The Serpens SMM3
source, discussed extensively in Sect.~\ref{sec:outflows} as an
example protostellar source, shows only CO$_2$ emission
(Fig.~\ref{fig:CO2overview}) but no H$_2$O. For most sources and
molecules, no significant ($> 10$ km s$^{-1}$) velocity offsets are
seen, except for BHR71-IRS1 (all lines) and L1448-mm (H$_2$O and
SiO). We here highlight the main results for both low- and high-mass
sources and put them in the context of the various scenarios of hot
core chemistry, dense molecular shocks, and winds (see
Sect.~\ref{sec:background} and Sect.~\ref{sec:back_outflows} for
background information).

\subsection{From CO$_2$ to SO$_2$: Cool emission}
\label{sec:CO2}

\paragraph{CO$_2$.}  Gas-phase lines of CO$_2$ at 15 $\mu$m are seen in
emission toward IRAS 23385+6053 \citep{Beuther23}, rather than in
absorption as found for the sample of high-mass protostars surveyed
with ISO-SWS \citep{Boonman03co2}. This difference is likely due to
the much higher spatial resolution of JWST, being able to detect the
weak emission on source, whereas any emission in the large ISO-SWS
beam was overwhelmed by the larger scale absorption against the
extended infrared continuum. For IRAS 23385, the resolved $R-$ and
$Q-$ branches provide an accurate temperature of $\sim$120$\pm 10$ K
\citep{Francis24}. One early lesson is that gas-phase lines can be
differentially affected by foreground ice absorption; this holds in
particular for the $P-$branch lines of CO$_2$ being obscured by CO$_2$
ice, but the same effect is also seen for other species such as SO$_2$
and H$_2$O in the 6--8 $\mu$m range, in addition to the overall
extinction \citep{vanGelder24}. This finding also puts direct
constraints on the location of the emission: it must originate from
behind where the bulk of the ice is located.

Figure~\ref{fig:CO2overview} shows a blow-up of the gas-phase CO$_2$
detections in several low-mass sources at their central position in
comparison with IRAS 23385. CO$_2$ is seen in 11 out of the 16 sources
studied in \citet{vanGelder24overview}, with four out of 11 in absorption
\citep[see also][]{Tychoniec24}. Where detected, the CO$_2$ emission
is cool, as in IRAS 23385, with typical rotational temperatures of
100--200 K (compare with Fig.~\ref{fig:molecules2}). For CO$_2$ in
absorption, the temperatures are somewhat higher, up to 300 K. The
temperatures derived from CO$_2$ emission in protostars are lower than
the values up to $\sim$400 K found in protoplanetary disks
\citep[e.g.,][]{Grant23,Vlasblom24}. Also, the hot bands are never
detected in protostars, consistent with low temperatures and lower
densities (see Fig.~\ref{fig:molecules2}).

Since CO$_2$ does not have a permanent dipole moment, the rotational
populations in $v=0$ should be thermalized and reflect the kinetic
temperature of the gas.  The rotational populations in $v=1$, measured
with JWST, should be similar if infrared pumping is indeed the main
mechanism for exciting the molecule (see
Sect.~\ref{sec:gasanalysis}). Several findings therefore argue in favor
of a hot core origin of CO$_2$ rather than in shocks
\citep{vanGelder24overview}: (1) temperatures of 100--200 K consistent
with ice sublimation; (2) lack of a velocity shift compared with the
source velocity; (3) similar gas and ice CO$_2$/H$_2$O abundance
ratios, and (4) location of the emission behind the ice absorption,
i.e., closer to the protostar.

\begin{figure}
\begin{centering}
\includegraphics[width=7.5cm]{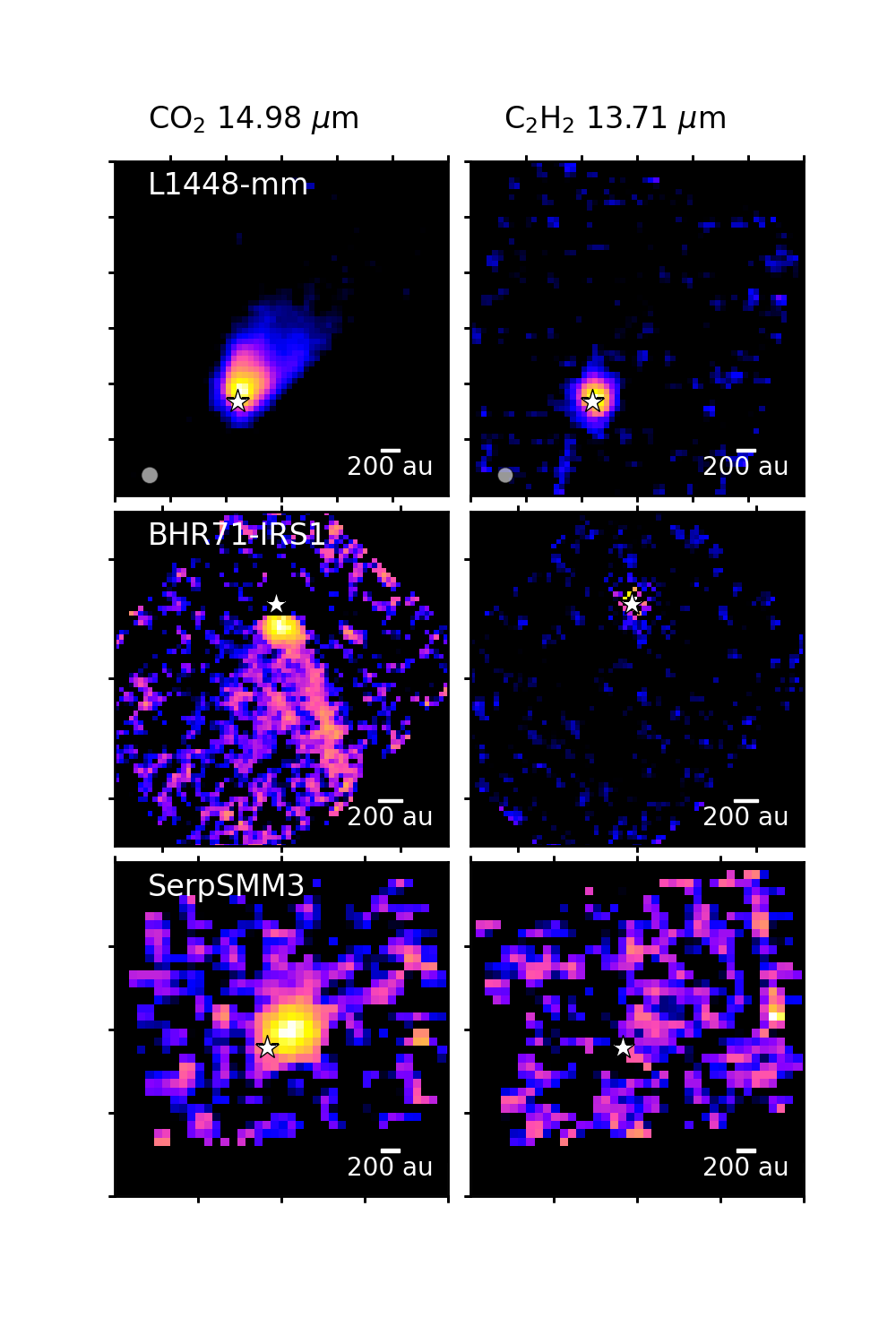}
\caption{Maps of CO$_2$ 15 $\mu$m and C$_2$H$_2$ 13.7 $\mu$m emission
  toward three low-mass sources. CO$_2$ is generally
  extended along the outflow, whereas C$_2$H$_2$ is centered on source,
  whose location is indicated by the white star. The beam size is
  indicated in the lower left corner of the top panels.}
\label{fig:CO2vsC2H2}
\end{centering}
\end{figure}

We note, however, that in some sources with prominent outflows CO$_2$
emission is also found to be extended along the outflow over distances
up to 1000 au, as illustrated in Figure~\ref{fig:CO2vsC2H2} \citep[see
also][]{Boonman03orion,Sonnentrucker07}.  One likely scenario is that
the observed gaseous CO$_2$ away from the hot core is enhanced by ice
sputtering of the dense and cold material along the outflow cavity
walls or in a slow wind, and subsequently excited by infrared
radiation from the source escaping through the outflow cavity. Warm
gas-phase chemistry, especially through the CO + OH $\to$ CO$_2$ + H
reaction, which is particularly effective in the 100--300 K range, can contribute as well \citep[e.g.,][]{vanDishoeck23}. 

\paragraph{C$_2$H$_2$ and HCN.} Both molecules are detected toward
IRAS 23385 and other high-mass protostars, as well as toward a few
low-mass protostars \citep{vanGelder24overview}.  Excitation
temperatures from emission features are similarly low as for CO$_2$,
100--200 K, again lower than the values up to 600 K found for
C$_2$H$_2$ and up to 900 K for HCN in protoplanetary disks around T
Tauri stars
\citep[e.g,][]{Grant23,Temmink24H2O,Schwarz24,Colmenares24,Arulanantham25}. In
contrast with CO$_2$, neither of these two molecules are seen along
the outflow (Fig.~\ref{fig:CO2vsC2H2}), or toward outflow positions
well offset from the high mass protostellar sources as was the case in
Orion \citep{Boonman03orion}. In low-mass outflows, such as HH 211,
C$_2$H$_2$ and HCN are also not seen at any of the strong knot
positions.  There is one clear exception, the NGC 1333 IRAS4B shock,
discussed in Sect.~\ref{sec:IRAS4B}, and one other spot near Serpens
SMM1. The fact that C$_2$H$_2$ and HCN are generally not extended
makes a hot core origin most plausible for these molecules, with high
temperature gas-phase reactions rather than ice sublimation dominating
the formation of C$_2$H$_2$ and HCN \citep{Francis24,vanDishoeck23}
(see also models below).

\paragraph{CH$_4$.} Methane gas is not readily seen toward
protostars. Analysis of the JOYS low-mass sample reveals detections
only in the two line-rich sources (B1-c, L14448-mm), as well as in the
source SVS4-5, which is likely a disk \citep{vanGelder24overview}. In
the former two sources, CH$_4$ is again cool, with temperatures of 100--200 K, suggesting a similar hot core origin as the other organic molecules. In disks around T Tauri stars, CH$_4$ is also seldom seen, with only two tentative detections reported \citep{Colmenares24,Temmink25}. In contrast, rich CH$_4$ line forests are prolific in many disks around very low-mass stars and brown dwarfs, even if not always easily distinguished due to line overlaps with C$_2$H$_2$ \citep[e.g.,][]{Arabhavi24,Kanwar24}.

\begin{figure}
\begin{centering}
\includegraphics[width=9cm]{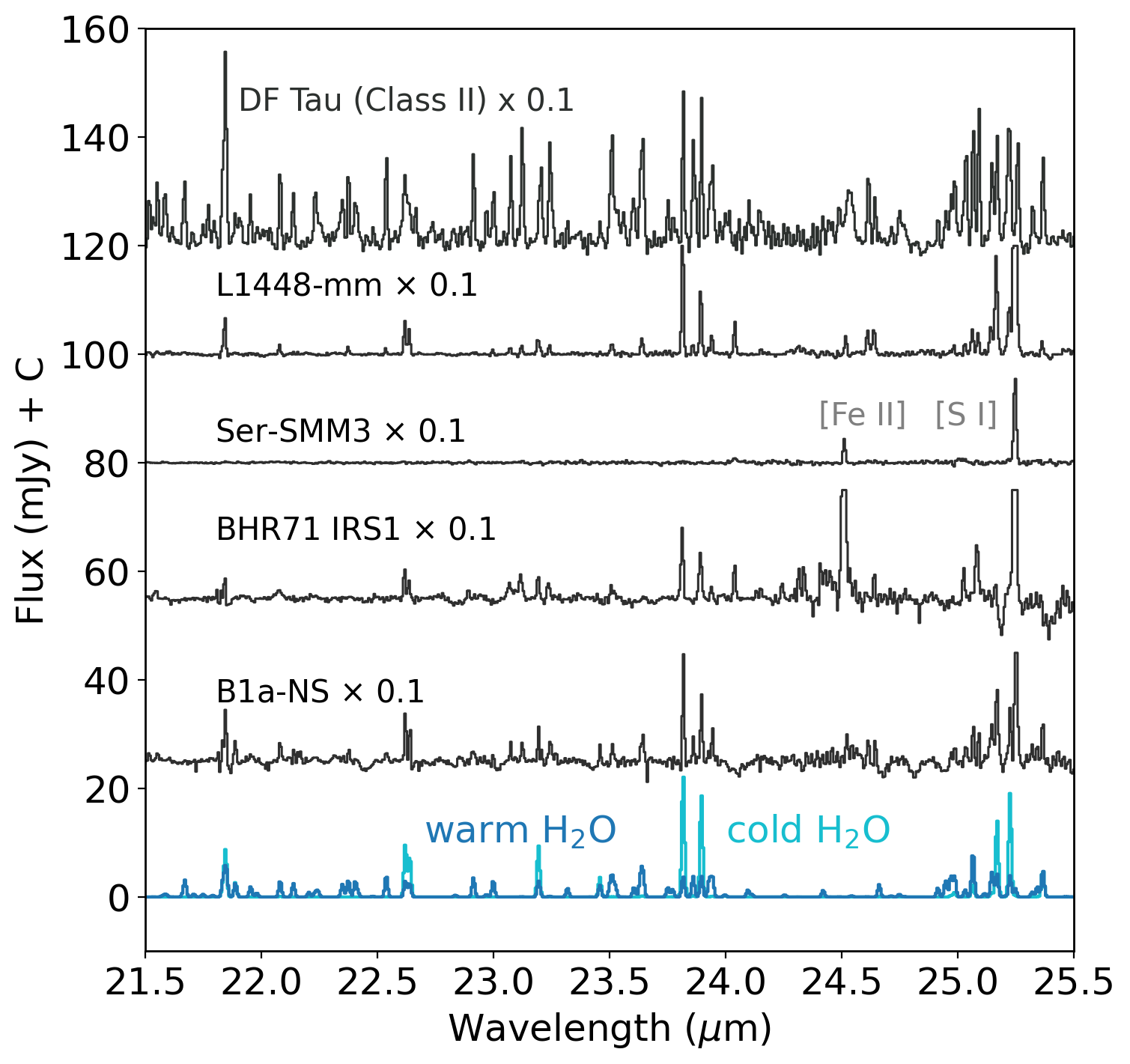}
\caption{Comparison of H$_2$O emission toward several low-mass
  protostars at long wavelengths in MRS channel 4. B1-a-NS is a Class
  I source, the other sources are Class 0 sources, including SMM3. The
  bottom panel shows warm and cold water model spectra at
  $T_{\rm rot}$=370 K (dark blue) and 160 K (light blue). The observed
  emission in protostars is dominated by the colder component. For
  comparison, the spectrum of a Class II disk, DF Tau, is shown at the
  top, illustrating that this disk has strong warm water emission
  \citep{Grant24}.}
\label{fig:H2Ooverview}
\end{centering}
\end{figure}

\paragraph{H$_2$O.} Water emission is surprisingly weak toward IRAS
23385, both through its vibration-rotation (5.5--7.5 $\mu$m) and pure
rotational ($>13$ $\mu$m) lines.  Toward one of the two sources in the
binary, H$_2$O is also seen in absorption at a higher temperature of
550 K, more characteristic of winds or the inner warm disk midplane
\citep{Barr20}.  H$_2$O emission and absorption is seen in other
high-mass protostars as well (see example of IRAS 18089 in
Sect.~\ref{sec:ices}).

The low-mass protostars do not show prominent water emission lines,
which is very different from the water emission forest seen in several
more evolved Class II disks
\citep[e.g.,][]{Gasman23,Pontoppidan24,Temmink24H2O,Banzatti25}. However,
upon close inspection, H$_2$O is detected in most low-mass protostars,
either in the ro-vibrational band or in the pure rotational lines or
in both, in absorption or emission \citep{vanGelder24overview}. These
data indicate the presence of water at two temperatures: a cool
temperature of 100--200 K, and a warmer component at 400--500 K,
likely reflecting both a hot core and shock component.
Figure~\ref{fig:H2Ooverview} shows prominent water lines at longer
wavelengths in the MIRI-MRS channel 4, comparing a few Class 0 and I
sources with the line-rich Class II disk DF Tau \citep{Grant24}. In
particular, the quartet of lines around 23.8 $\mu$m has been found to
be a good indicator of the presence of cold versus warm water as
  shown by \citet{Banzatti23,Banzatti25,Temmink24H2O}. The fact that
the warm water lines are clearly suppressed in the protostellar
sources indicates that the H$_2$O excitation temperature of water is
indeed low. In contrast, the spectrum of a water-rich Class II source,
DF Tau, has prominent warm and cold water lines in this wavelength
range: all four lines rather than just two lines of the 23.8 $\mu$m
quartet are seen.

We note that the two temperature components of water found in
protostellar sources are not necessarily the same as the hot, warm and
cold(er) components found in Class II disks where they have been
associated with temperature gradients in the disk surface in the inner
few au and water enhancement at the snowline
\citep{Banzatti23,Banzatti25,Gasman23,Temmink24H2O,Romero24}: while
there may be a similar water vapor enhancement in hot cores around 100 K,
H$_2$O is likely to be subthermally excited at the much lower
densities in protostellar environments, whereas the H$_2$O ro-vibrational
band can be pumped by infrared radiation.
Comparison with water in hot cores imaged with ALMA through lower
excitation pure rotational lines of H$_2^{18}$O and HDO
\citep[e.g.,][]{Persson16,Jensen19} will be done in future
studies.

\begin{figure}
\begin{centering}
\includegraphics[width=9cm]{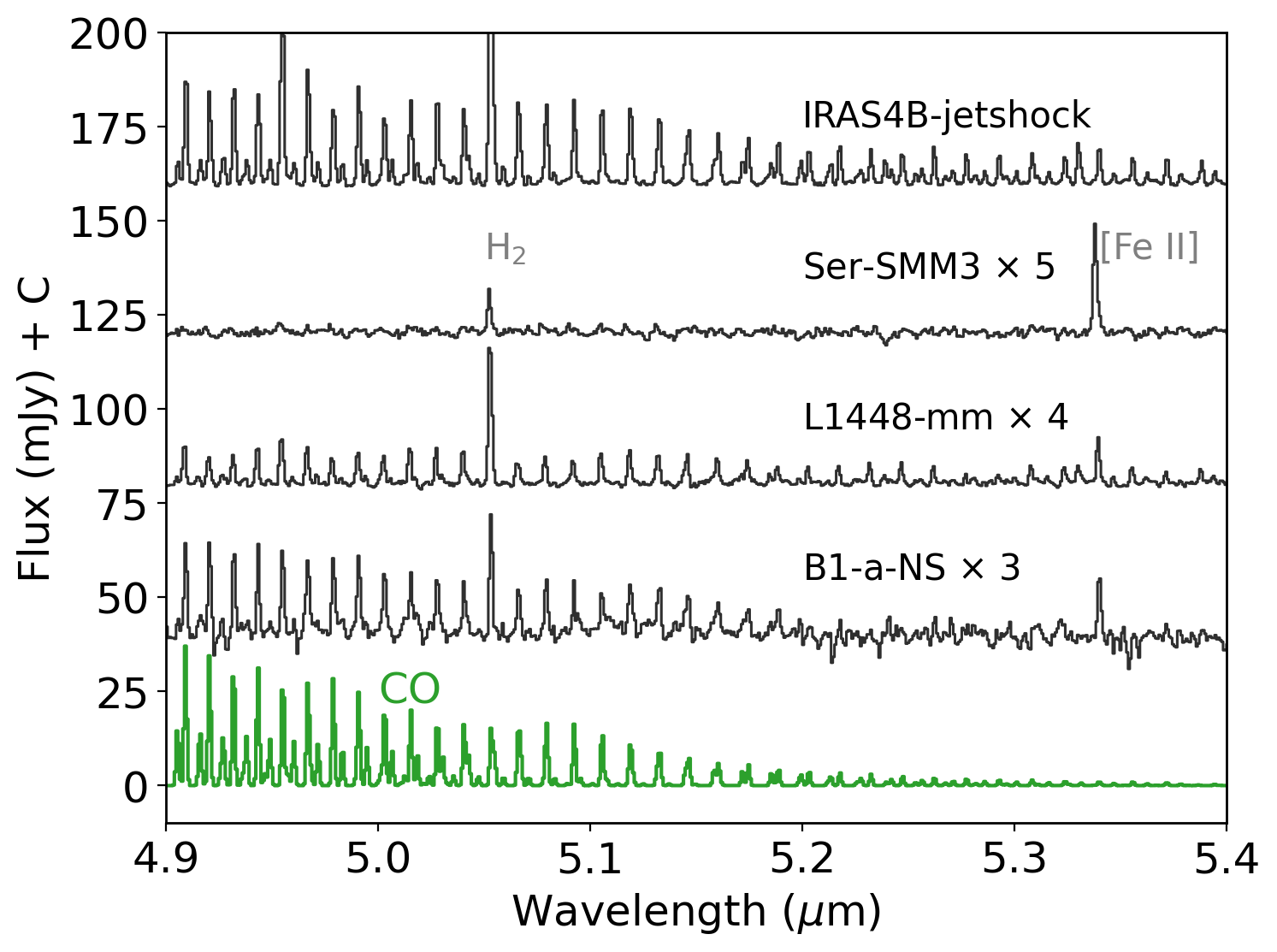}
\caption{Comparison of CO $v=1-0$ 5 $\mu$m emission toward several
  low-mass protostars. L1448 and SMM3 are Class 0 protostars, B1-a-NS
  is a Class I source, and the spectrum of NGC 1333 IRAS4B is taken at
  a shock knot position well off source (see
  Fig.~\ref{fig:IRAS4B}). The bottom panel shows an optically thick CO
  model spectrum at $T$=1500 K.}
\label{fig:COoverview}
\end{centering}
\end{figure}

\paragraph{CO.} The emission of CO is seen in about half of the MIRI spectra of
protostars in the 4.9--5.2 $\mu$m range, showing the tail of the
$v=$1--0 $P-$branch with $J_{\rm up}\geq 23$
(Fig.~\ref{fig:COoverview}) \citep[see also][]{Salyk24}. When
detected, it always indicates a high temperature
(Fig.~\ref{fig:molecules1}). A naive rotation diagram analysis
assuming optically thin lines would imply $\sim$2000 K, but the actual
kinetic temperature can be substantially lower, $\sim$1000 K if the CO
emission is optically thick \citep[see discussion
in][]{Herczeg11,Francis24,Rubinstein24,LeGouellec25S68N}. Such
emission could be consistent with the innermost disk region, as in
Class II sources \citep[e.g.,][]{Banzatti22}, but it can also be
associated with the disk surface \citep{Arulanantham24} or shock knots
(see Sect.~\ref{sec:outflows}) \citep{Federman24}. Indeed, spectrally
resolved ground-based CO data for Class I sources often show wind and
outflow absorption in $^{12}$CO $v$=1-0 profiles that could mask any
emission in spectrally unresolved data \citep{Herczeg11}. High
extinction by the inner envelope could also hide CO emission in
sources where it is not seen.

\paragraph{SO$_2$.} Mid-infrared emission of SO$_2$ at 7.3
$\mu$m has been detected for the first time by JWST in a low-mass protostar
\citep{vanGelder24}. It is seen toward the Class 0 protostar
NGC 1333 IRAS2A, a source that is part of JOYS+ in program 1236 (PI:
M.\ Ressler). Inspection of the full JOYS low-mass sample
shows that this band is not commonly seen, only toward L1448-mm in
emission and B1-c in absorption \citep{vanGelder24overview}.

Notably, SO$_2$ is one of the few molecules that can be observed with
both JWST and ALMA since it has a permanent dipole
moment. Interestingly, for IRAS2A the excitation temperature derived
from the mid-infrared data, $\sim 100$~K, is the same as that derived
from the wealth of SO$_2$ millimeter lines imaged with ALMA on scales
of 60 au radius as presented in \citet{vanGelder24} using data from
program 2021.1.01578 (PI: B. Tabone). Similarly low temperatures are
found for the two JOYS sources. A question then is to what extent 
this excitation temperature $T_{\rm rot}$ reflects the kinetic gas
temperature $T_{\rm kin}$.

\begin{figure}
\begin{centering}
\includegraphics[width=8cm]{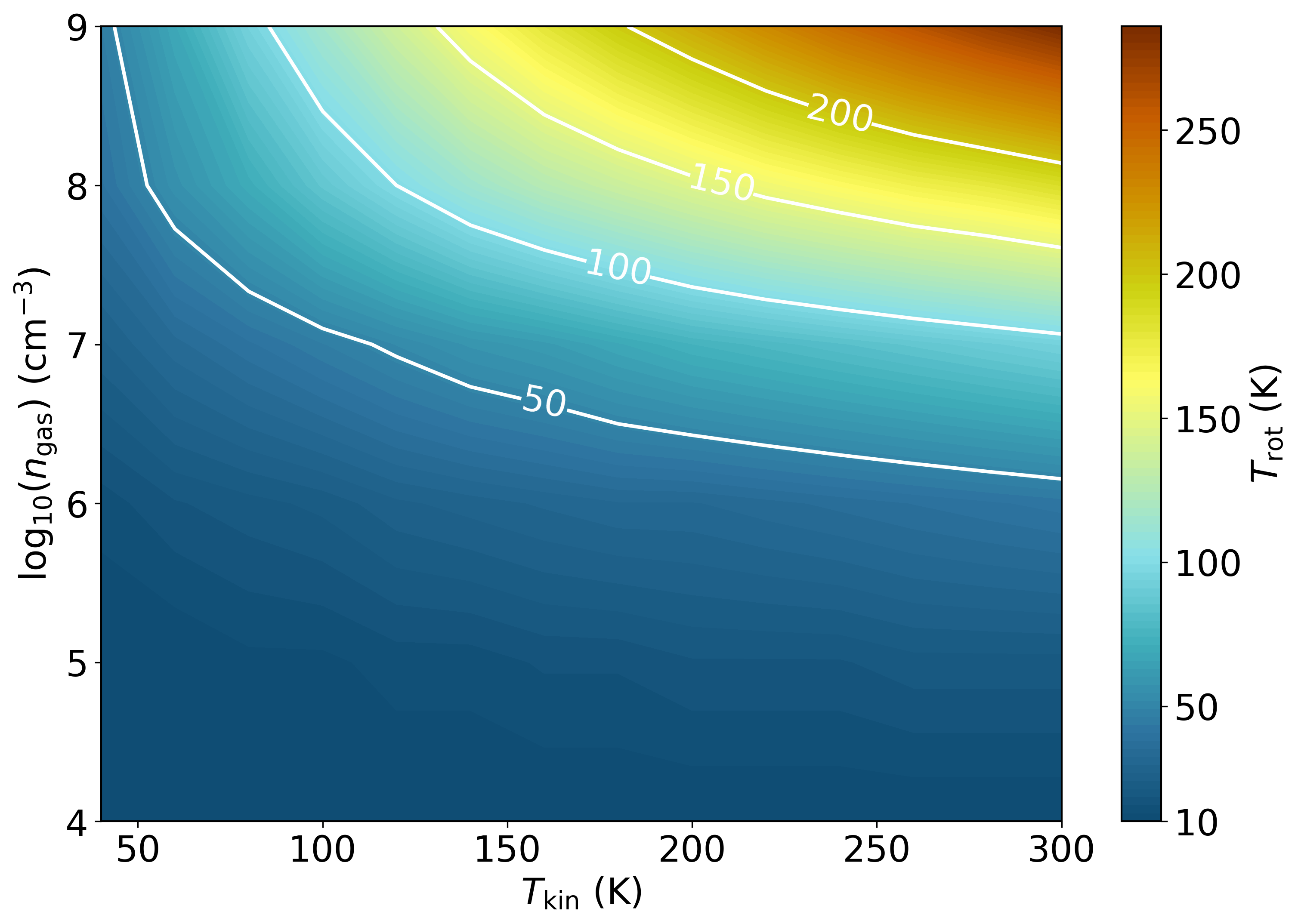}
\includegraphics[width=8cm]{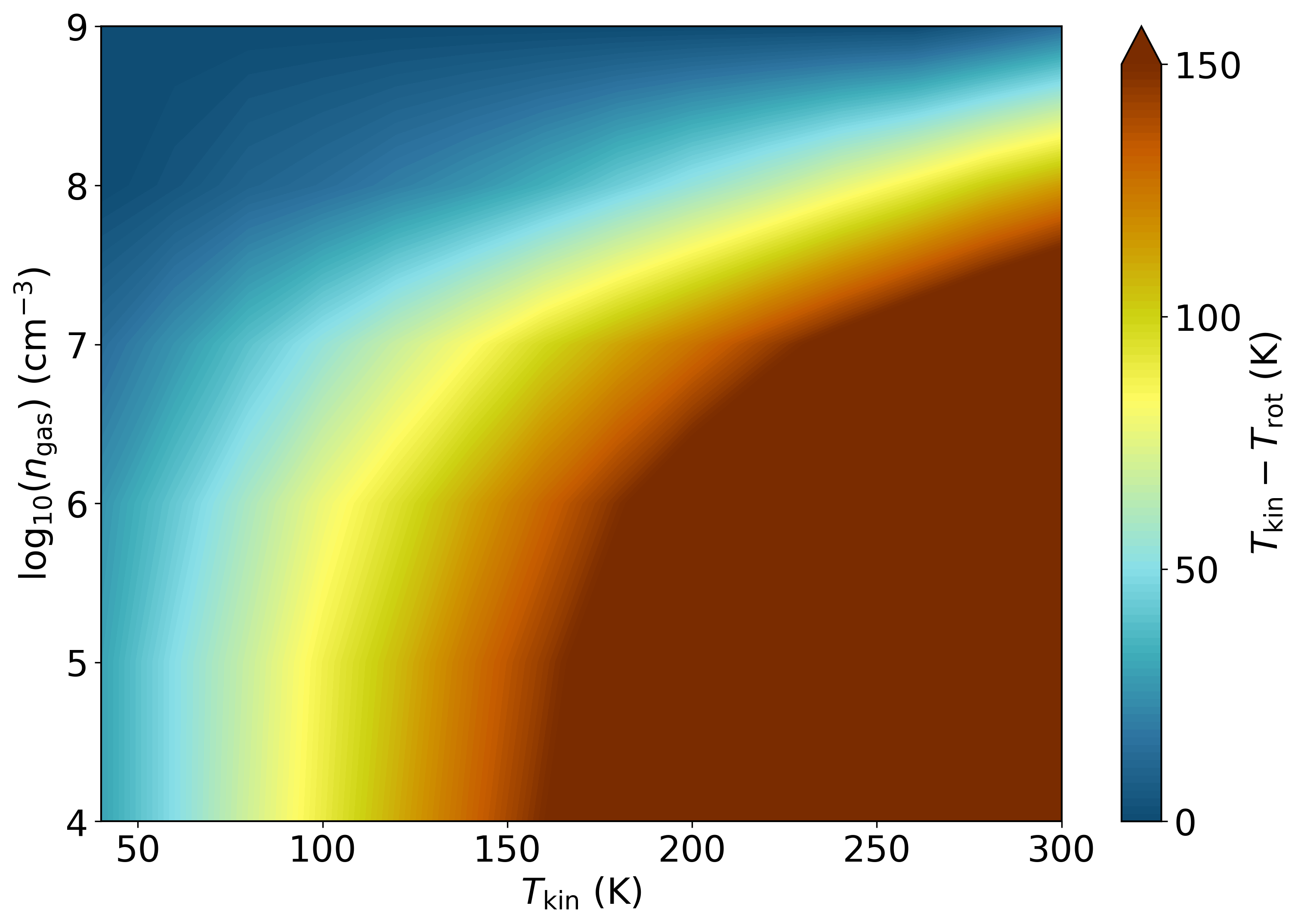}
\caption{Top: SO$_2$ rotational temperature $T_{\rm rot}$ within $v$=0
  as function of kinetic temperature $T_{\rm kin}$ and density
  $n$(H$_2$) obtained from non-LTE calculations. Bottom: Difference
  between $T_{\rm kin}$ and $T_{\rm rot}$: large deviations are apparent
  at low densities ($<10^7$ cm$^{-3}$). }
\label{fig:SO2Trot}
\end{centering}
\end{figure}

Within $v$=0, the level populations are determined by a mix of
collisional (de-)excitation and radiative decay. A non-LTE excitation
calculation using RADEX \citep{vanderTak07,vanderTak20} with
collisional rate coefficients from \citet{Balanca16} is adopted to
compute $T_{\rm rot}$ and the difference between $T_{\rm kin}$ and
$T_{\rm rot}$ as a function of kinetic temperature and H$_2$ density
$n$. $T_{\rm rot}$ is obtained by fitting a rotation diagram to the
model results for the same lines as observed with ALMA in IRAS2A.
Figure~\ref{fig:SO2Trot} shows that at low densities ($<10^7$
cm$^{-3}$) the excitation is subthermal, $T_{\rm rot}<<T_{\rm
  kin}$. However, typical envelope densities on scales of $<$100 au
are $10^7-10^9$ cm$^{-3}$ \citep{Jorgensen04i2,Notsu21} for which
differences $T_{\rm kin} - T_{\rm rot}$ are smaller.

The temperatures derived from the MIRI-MRS data should be
representative of the gas in which SO$_2$ resides. The number of
molecules inferred from LTE slab models, however, is not. It is
overestimated by a factor of 10$^4$ when compared with the ALMA
results. The mid-infrared lines are clearly pumped by scattered IR
radiation with a brightness temperature of $\sim$180 K, substantially
higher than the rotational temperature \citep{vanGelder24}. When the
infrared pumping is included, the number of molecules derived from the
JWST and ALMA data are consistent. Combined with the small millimeter
line width and lack of a velocity shift, a hot core rather than an
accretion shock origin seems most likely for SO$_2$.  This study
emphasizes the synergy between JWST and ALMA for molecules with a
dipole moment.

Although the importance of subthermal excitation and infrared pumping
is highlighted here for the case of SO$_2$ in IRAS2A,
Figure~\ref{fig:SO2Trot} is representative for other molecules with a
dipole moment such as HCN and H$_2$O. The analysis in
\citet{vanGelder24overview} also shows the
importance of infrared pumping for other molecules and sources.

\paragraph{OH.} As noted in Sect.~\ref{sec:outflows}, mid-infrared OH
emission has been detected with {\it Spitzer} in bow-shocks
\citep{Tappe08,Tappe12} and other environments with strong UV
irradiation such as the surface layers of protoplanetary disks
\citep{Carr14}. It is now also prominently seen with JWST in low-mass
protostars at the source position \citep{vanGelder24overview} and at
(bow) shock positions \citep{Neufeld24,Caratti24}, as well as in an
irradiated disk \citep{Zannese24}. Class II protoplanetary disks
commonly show OH
\citep[e.g.,][]{Gasman23,Temmink24H2O,Schwarz24,Vlasblom25} although
mostly at the longer wavelengths.  In the high-mass JOYS sample, OH
emission is only prominent in IRAS 23385 \citep{Francis24} although it
is detected in some other sources (Reyes et al.\ in prep.).

The OH lines originate from extremely high rotational states, with
rotational quantum numbers up to $N$=40 ($E_{\rm up}=40000$~K). This
phenomenon results from photodissociation of H$_2$O through its second
electronically excited state, the ${\tilde B}$ state, producing OH in
very high $N$ states which then cascade down through pure rotational
transitions to the ground state, a process that has been well studied
in the laboratory and through quantum chemical calculations
\citep[e.g.,][]{Harich01,vanHarrevelt00} (see summary in
\citealt{Tabone21}). This process requires UV photons at 1140 --1430
\AA\ and a wavelength range that includes Lyman $\alpha$ at 1216 \AA,
and it is thus a direct diagnostic of the strength of the UV radiation
multiplied by the number of H$_2$O molecules \citep{Tabone23}. The OH
``prompt emission'' occurs primarily at 9-11 $\mu$m with a
characteristic asymmetry in the emission of the $\Lambda$-doublet
lines. At longer wavelengths ($>13$ $\mu$m, depending on temperature),
chemical pumping through the O + H$_2$ $\to$ OH($v,J$) + H reaction
contributes. Most of the JOYS sources show only these longer
wavelength OH lines indicative of warm gas-phase chemistry
\citep{Francis24,vanGelder24overview}, although prompt emission is
seen in a few JOYS sources (TMC1, BHR71-IRS1) besides HH 211. The
implications of the presence or absence of 9--11 $\mu$m OH emission
for the amount of UV and water remain to be quantified.  Previously
inferred values of the enhancement factor $G_o$ compared with the
interstellar radiation field are of order $10^2-10^3$ on scales of
$\sim$1000 au using other tracers \citep{Yildiz15,Benz16}.

\begin{table*}[t]
  \caption{Column density ratios of key molecules observed toward the low-mass protostar B1-c.}
\begin{tabular}{lccccccc}
\hline
  \hline
  Species & Observed$^{a}$ & Model  & Model & Model & Model & Model & Model \\
          &              & 150 K   & 150 K  & 300 K & 300 K & 600 K & 600 K \\
          &    & no X-rays & $L_X=10^{29}$  & no X-rays & $L_X=10^{29}$
          & no X-rays & $L_X=10^{29}$ erg s$^{-1}$ \\
  \hline
  CO$_2$/H$_2$O    & 0.22 & 0.025 & 0.27 & 0.0016 & 0.0025 & 0.0012 & 0.0029\\
  HCN/H$_2$O        & 0.08 & 0.024 & 0.0001 & 0.037 & 0.037 & 0.063 & 0.086 \\
  C$_2$H$_2$/H$_2$O & 0.05 & $<$0.0001 & 0.0003 & $<$0.0001 & 0.0054 & 0.0001   & 0.0069 \\
  CH$_4$/H$_2$O     & 0.10 & 0.039 & 0.028 & 0.020 & $<$0.0001 & 0.048 & 0.013 \\
  NH$_3$/H$_2$O     & 0.04 & 0.05 & 0.0001 & 0.062 & 0.016 & 0.10 & 0.069 \\
  C$_2$H$_2$/HCN    & 0.64  & 0.0001 & 3.4 & 0.0004 & 0.15 & 0.002 & 0.08 \\
  C$_2$H$_2$/CO$_2$ & 0.21  & 0.0001 & 0.001 & 0.008 & 2.2 & 0.10 & 240 \\
  \hline
\end{tabular}

$^a$ Observed column densities taken from Table H.4 of
\citet{vanGelder24overview}, using the warm H$_2$O column of
$4.4\times 10^{18}$ cm$^{-2}$ as reference. Observational
uncertainties in the ratios are about 50\%.
\label{tab:B1c}
\end{table*}

\paragraph{Other molecules.} \citet{vanGelder24overview} report the first
mid-infrared detection of NH$_3$ in a protostar, seen in absorption
toward B1-c. The NH$_3$/H$_2$O column density ratio of 0.04 is again
consistent with ice sublimation in a hot core. Other first detections
toward a low-mass protostar include CS absorption (B1-c), and
C$_4$H$_2$ and SiO emission (L1448-mm). The $^{13}$C isotopologs of
CO$_2$ and C$_2$H$_2$ are also
seen for these two sources. \\

In summary, the relatively low temperatures of 100--200 K (somewhat
higher in absorption) coupled with the lack of velocity shifts point
toward a hot core origin for the bulk of the molecules. This is
further strengthened by the similarity of gas and ice abundances
relative to H$_2$O for molecules seen in ices: CO$_2$, CH$_4$ and
NH$_3$, especially in the case of B1-c \citep[see][and
below]{vanGelder24overview}. For L1448-mm and BHR71-IRS1, however,
some molecules like H$_2$O are clearly associated with the outflow
based on velocity offsets.

\subsection{Molecular abundances: Hot core chemistry}
\label{sec:modelchemistry}

The inferred molecular abundances can be used to test high temperature
chemical models. The cleanest case for studying hot core chemistry is
that of B1-c, for which the lines are in absorption and are therefore
not affected by radiative pumping. Also, isotopologs are detected to
correct for optical depth.  For emission lines, determination of
column densities and abundances is complicated by the fact that
infrared radiative pumping needs to be properly accounted for: column
densities can be too high by orders of magnitude if LTE excitation at
the same low temperature that characterizes the rotational
distribution is assumed for populating the vibrational levels (see
case of SO$_2$ above).

Table~\ref{tab:B1c} summarizes some key column density ratios of
molecules.  Determination of the fractional abundances with respect to
H$_2$ is complicated by the fact that there is no clean tracer of the
amount of 100--300 K molecular gas. Hence, warm H$_2$O is taken as a
reference. Values are taken from Table~H.4 of
\citet{vanGelder24overview} and have a typical uncertainty of
30\%. This can be higher if the optical depth of the lines is
underestimated. Specifically, for CO$_2$ and C$_2$H$_2$, the column
densities of $^{13}$CO$_2$ and $^{13}$C$_2$H$_2$ multiplied by 70
respectively 35 are used in Table~\ref{tab:B1c}. For CO$_2$ this value
is a factor of three higher than that found from $^{12}$CO$_2$. The
use of ratios implicitly assumes that the molecules are present in the
same gas.

\begin{figure*}[h!]
\begin{centering}
\includegraphics[width=7cm]{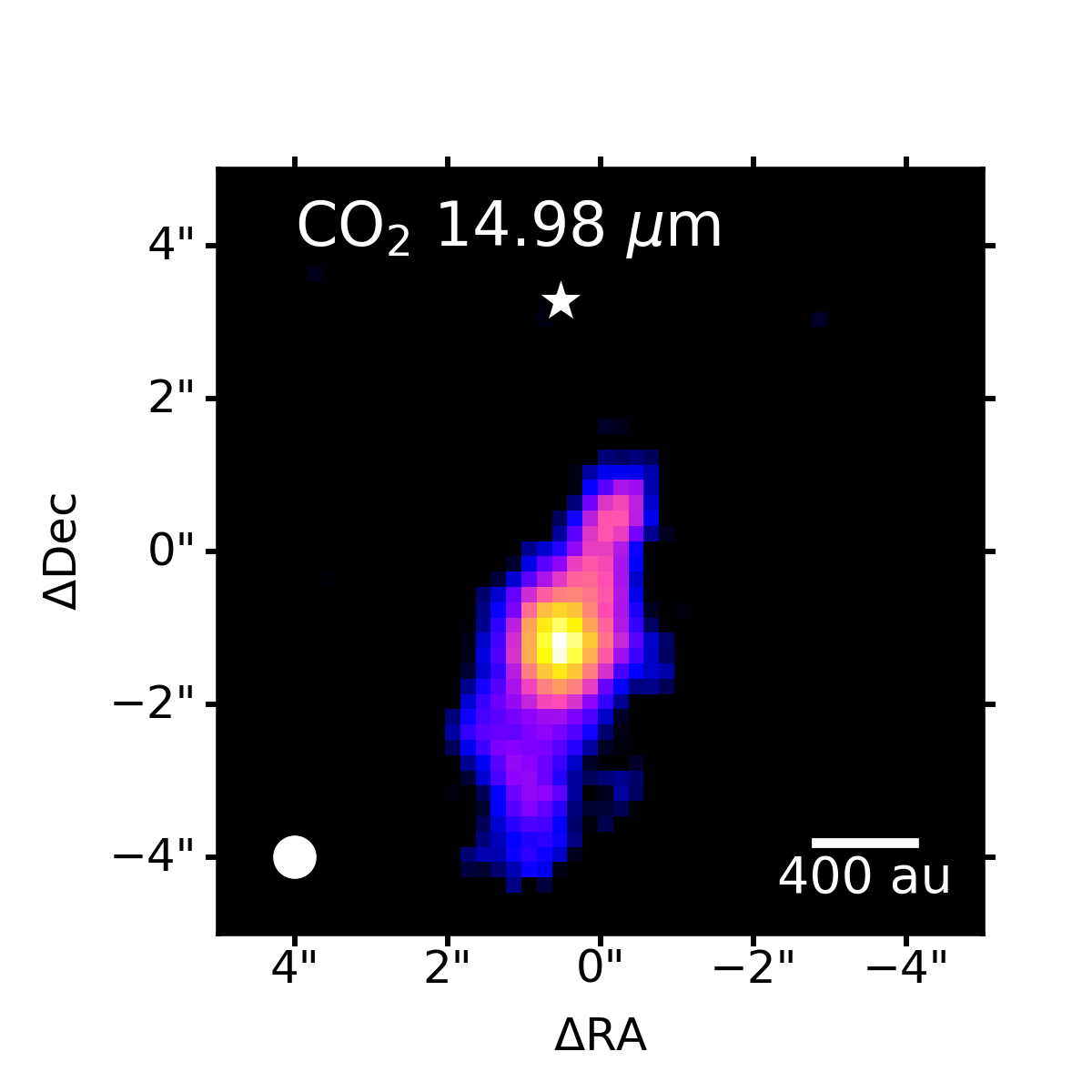}
\includegraphics[width=10cm]{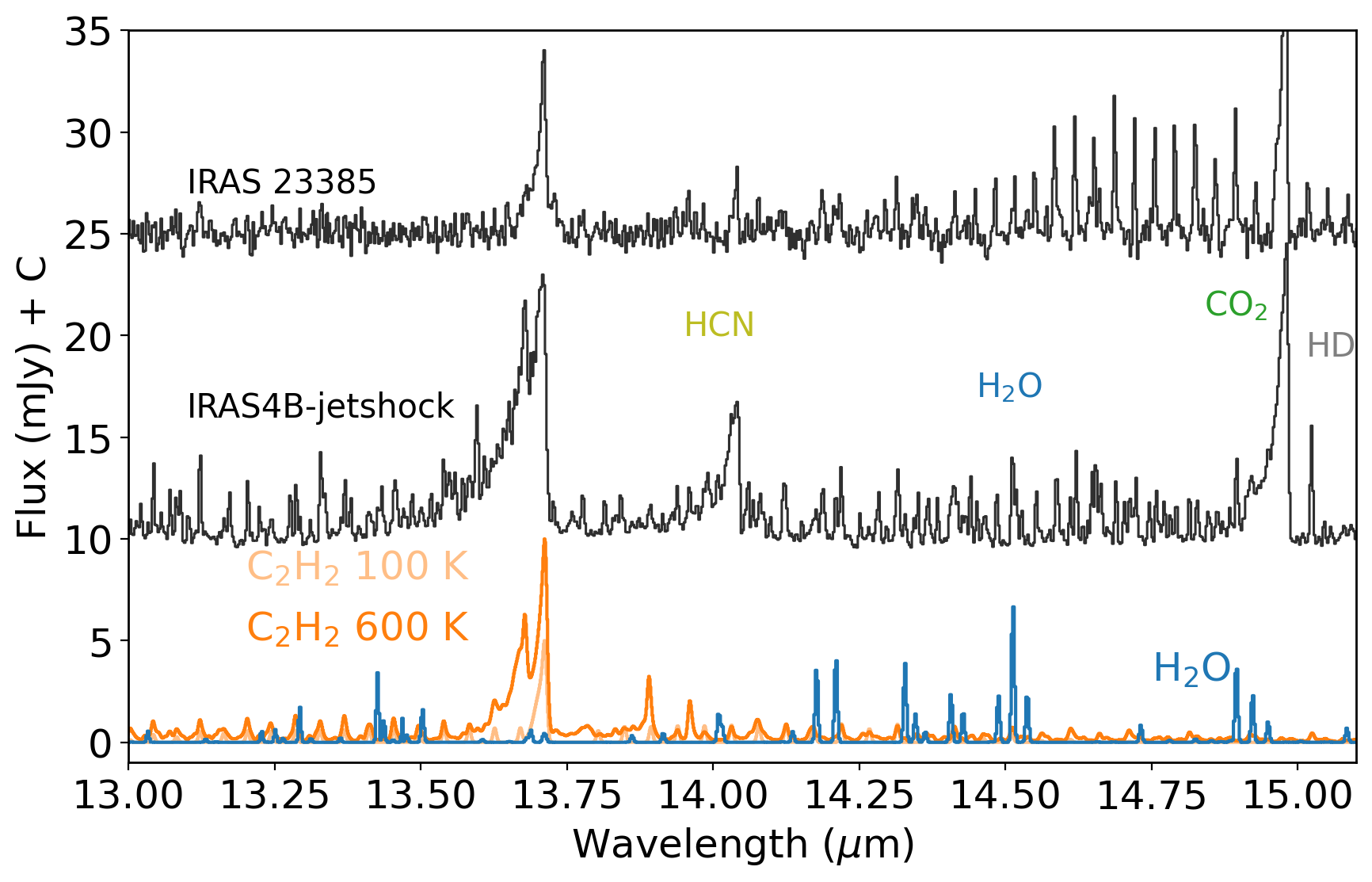}
\caption{Left: MRS image of the CO$_2$ 15 $\mu$m band toward NGC~1333
  IRAS4B showing the peak emission to be $\sim$1000 au ($\sim 4''$)
  offset from the source itself, whose position from millimeter
  interferometry is indicated by the white star. The beam size is
  indicated in the lower left corner. Right: MIRI-MRS spectrum of the
  NGC~1333 IRAS4B jet ``knot'' position in the 13--15 $\mu$m range
  highlighting its rich spectrum including hot C$_2$H$_2$ (600 K). For
  comparison, the spectrum of the high-mass protostar IRAS 23385,
  which shows only cold C$_2$H$_2$, is presented
  \citep{Francis24}. The model spectra at the bottom show the
  contributions from C$_2$H$_2$ (100 and 600 K) and H$_2$O (280 K) in
  this wavelength range. }
\label{fig:IRAS4B}
\end{centering}
\end{figure*}

The chemistry in hot cores has been studied in a number of models over
the past decades that start with the sublimation of ices at $t=0$
above their sublimation temperature (typically taken to be 100 K,
representative of H$_2$O) and then follow the gas-phase chemistry for
up to $10^5$ yr \citep[e.g.,][]{Charnley92,Doty04}.  It is well known
that the abundances of some molecules are particularly sensitive to
temperature, most notably HCN and C$_2$H$_2$ being significantly
enhanced in warmer gas \citep{Doty02}. In contrast, CO$_2$ production
through CO + OH $\to$ CO$_2$ + H is halted at $\sim$250~K when the OH
+ H$_2$ reaction takes over \citep[see discussion
in][]{vanDishoeck23}. Another parameter that plays a role in setting
the relative abundances is the X-ray luminosity. X-rays can both
destroy molecules (e.g., water) and enhance molecules (e.g.,
C$_2$H$_2$) \citep{Stauber05,Stauber07,Notsu21}.

Table~\ref{tab:B1c} lists the results of the gas-grain chemistry model
of \citet{Notsu21} using the envelope structure of NGC 1333 IRAS2A
derived by \citet{Kristensen12} at $r$=60 au. This position is just
inside the water snowline ($\sim$150 K) where the hot core starts and
the density is $2\times 10^8$ cm$^{-3}$. Such conditions are also
representative of the B1-c envelope inside its snowline, even though
it lies at somewhat smaller radii since B1-c is less luminous than
IRAS2A. Values without X-rays and for $L_X=10^{29}$ erg s$^{-1}$ are
tabulated. The model uses a detailed chemical network taken from
\citet{Walsh12,Walsh15}.

The adopted X-ray luminosity of $L_X=10^{29}$ erg s$^{-1}$ is
consistent with observations and chemical analyses of low-mass
protostars \citep{Benz16} and turns out to be a ``sweet spot'' where
the gas-phase CO$_2$ and C$_2$H$_2$ abundances are enhanced by one
respectively two orders of magnitude at 150 K. Thus, this choice
maximizes C$_2$H$_2$. H$_2$O and CH$_4$ are decreased by less than a
factor of two for this value, but HCN and NH$_3$ are lower by more
than two orders of magnitude, explaining the dramatic changes in the
abundance ratios of, for example, C$_2$H$_2$/HCN in models with and
without X-rays. For higher values of $L_X$, the abundances of all
these six molecules decrease sharply. The use of H$_2$O as a reference
introduces some uncertainties since its observed overall hot core
abundance with respect to H$_2$ has been found to be less than the
canonical value of $2\times 10^{-4}$ in models without X-rays
\citep{Persson14,Persson16}. This observed low water abundance has, in
fact, been one of the main motivations for introducing X-rays into the
models \citep{Stauber07,Notsu21}.

The overall conclusion of Table~\ref{tab:B1c} is that hot core models
of ice sublimation and gas-phase chemistry with modest X-rays
reproduce the observed abundance ratios reasonably well within the
model uncertainties of up to an order of magnitude. However,
C$_2$H$_2$ is consistently underproduced by orders of magnitude at
temperatures of only 150 K.

This underabundance of C$_2$H$_2$ has been noted before, also in the
context of explaining its observed high abundances in high-mass
protostars and even in deeply obscured (ultra)luminous infrared
galaxies \citep{Lahuis07AGN,Buiten25}. Temperature is clearly the one
parameter that affects C$_2$H$_2$ most. Table~\ref{tab:B1c} therefore
also includes model abundances for temperatures of 300 and 600 K at
the same density. These results show that the C$_2$H$_2$ abundance is
indeed boosted at higher temperatures, especially in the presence of
X-rays, although the observed high C$_2$H$_2$/H$_2$O ratios are not
yet reached. We note that the observed rotational temperature of
C$_2$H$_2$ in absorption toward B1-c is 285 $\pm$ 25 K, i.e., just
consistent with the temperature regime at which C$_2$H$_2$ is
significantly enhanced. HCN is increased as well with temperature, but
not as strongly as C$_2$H$_2$. As expected, the CO$_2$ abundance is
lower at $T>300$~K, since OH is now driven into H$_2$O. Taken
together, these results should be viewed as illustrative of the
parameter regime that can explain the observed column density
ratios. No single position model can explain all the column density
ratios and the hydrocarbon chemistry remains to be fully
understood. Explaining high C$_2$H$_2$ abundances derived from
emission lines in hot core regions, where this molecule has a low
rotational temperature of only $\sim$150 K, would be a significant
challenge if such temperatures were representative of the kinetic gas
temperature.  Source-specific depth-dependent models are needed to
test whether the type of models presented here can reproduce both
absolute and relative column densities. The models presented here do
not consider any contribution from shocks.  Sect.~\ref{sec:IRAS4B}
discusses the detection of strong and warm C$_2$H$_2$ emission in at
least one jet shock position that may have a different explanation.

\subsection{NGC 1333 IRAS4B: Dense shock}
\label{sec:IRAS4B}

Dense non-dissociative shocks can also emit copious molecular emission
of molecules other than H$_2$ \citep{Draine83,Kaufman96}. Indeed,
extended warm H$_2$O emission has been imaged with {\it Herschel}-PACS
in significant samples of low- and high-mass protostars showing both
emission on and off source on scales of several thousand au
\citep[e.g.,][]{Nisini10,Tafalla13,Karska13,Karska18,vanDishoeck21}. We
here investigate whether such molecular emission is seen in our data
at any shocked positions off source. The HH 211 MRS map revealed
bright OH prompt emission at the outer bow shock together with CO,
H$_2$O and weak HCO$^+$ and CO$_2$ (see Fig.~7 of
\citealt{Caratti24}), but no emission from other molecules. Also, in
contrast with the extended H$_2$O emission mapped by {\it
  Herschel}-PACS \citep{Dionatos18}, no H$_2$O emission is found with
MIRI at other HH 211 positions than the outer bow shock.

A special case within JOYS is formed by the Class 0 low-mass protostar
NGC 1333 IRAS4B in Perseus ($d$=293 pc, $L$=4.4 L$_\odot$,
Table~\ref{tab:sources}). This source attracted attention by {\it
  Spitzer} since its mid-infrared IRS spectrum shows a very rich
H$_2$O spectrum with an excitation temperature of $\sim$170 K. It was
speculated that these warm bright H$_2$O lines were due to a slow
accretion shock liberating water at the envelope-disk interface in
this young system \citep{Watson07}. The subsequent longer wavelengths
{\it Herschel}-PACS spectrum of this source taken with 9$''$ spaxels
was found to be equally rich in H$_2$O and other far-infrared
lines. However, the {\it Herschel} Nyquist-sampled spectral maps
suggested the bulk of the emission to come from a position $\sim 5''$
south of the central source position as measured by millimeter
interferometry, along the blue-shifted outflow lobe
\citep{Herczeg12}. The central source was concluded to be too heavily
extincted even at these longer wavelengths. Rather than high-density
($\geq 10^{11}$ cm$^{-3}$) thermalized emission suggested by
\citet{Watson07}, subthermal excitation of H$_2$O at
$T_{\rm kin}$=1500 K and $n\sim 3\times 10^6$ cm$^{-3}$ was inferred
by \citet{Herczeg12}.

The JWST MIRI-MRS spectral maps with their high spatial resolution can
test these two scenarios. Figure~\ref{fig:IRAS4B} (left) shows an
image in the CO$_2$ 15 $\mu$m $Q$-branch. The emission of all
molecules, including H$_2$O, as well as that of atoms is clearly seen
to be extended along the blue outflow lobe with the brightest ``knot''
about 4$''$ offset from the central source position (see also
Figures~\ref{fig:JOYS-H2S4} and \ref{fig:JOYS-Fe17}). No emission is
detected at the central source position, nor in the red-shifted
outflow lobe. This suggests that dense jet shocks are indeed
responsible for the bright H$_2$O mid-infrared lines originally
detected by {\it Spitzer} rather than an accretion shock onto the
disk.

Figure~\ref{fig:IRAS4B} (right) presents the 13--15 $\mu$m MIRI-MRS
spectrum extracted from the brightest knot position in a 1.4$''$
aperture. In contrast with most other jet and outflow positions in
other JOYS Class 0 sources such as BHR 71 IRS1
(Fig.~\ref{fig:CO2vsC2H2}) or HH 211 \citep{Caratti24}, this knot
shows very bright C$_2$H$_2$, HCN and CO$_2$ emission together with
the mid-infrared H$_2$O lines. The C$_2$H$_2$ is also clearly much warmer,
at least 600~K, compared with other low- and high-mass sources like
IRAS 23385 (Fig.~\ref{fig:IRAS4B}).

Although the spectrum can be well fitted with simple LTE slab models,
the derivation of column densities and abundance ratios requires a
detailed non-LTE excitation analysis since the density is well below
the critical density of $\sim 10^{12}$ cm$^{-3}$ for ro-vibrational
lines to be thermalized.  Infrared pumping by radiation escaping through the
outflow cavity can still contribute to the excitation of the
ro-vibrational bands, even though the shock knot is significantly
offset from the central source. The relatively high density of
$> 10^6$ cm$^{-3}$, likely higher than that of other jet knots,
clearly helps in producing bright lines compared with other
sources. The hot C$_2$H$_2$ emission points to high-temperature
carbon-rich gas-phase chemistry taking place in this knot, perhaps
enriched by the destruction or erosion of carbonaceous grains by
  processes such as sublimation \citep[e.g.,][]{Li21}, pyrolysis or
  oxidation \citep[e.g.,][]{Kress10,Lee10,Gail17} and reactions of
  carbon grains with atomic hydrogen, also called chemi-sputtering
  \citep[e.g.,][]{Lenzuni95}.  Recent protostar plus disk formation
  models by \citet{Borderies25} show that this latter process is
  particularly effective at temperatures below 1000 K. Also,
  liberation of PAHs that were locked up in ices by sputtering may
  contribute.  Taken together, this IRAS4B spectrum provides a
unique opportunity to test physical-chemical models of high density
shocks. A detailed analysis and comparison with the {\it Herschel}
results will be presented in a future paper.

\subsection{Chemistry in winds}
\label{sec:windchemistry}

The H$_2$ emission lines presented in Sect.~\ref{sec:outflows} show
evidence of wide-angle winds on scales of a few hundred au for
low-mass protostars \citep[see also][]{Arulanantham24,Salyk24}. The
question addressed here is whether there is evidence of other
molecules in these winds. In the edge-on case of the Tau 042021 Class
II disk, H$_2$O and CO are found to be spatially extended with
MIRI-MRS, but mostly along the disk surface and likely tracing
scattered inner disk emission.

Figure~\ref{fig:CO2vsC2H2} shows that CO$_2$ 15 $\mu$m emission is
extended along the outflow cavities for some sources, in contrast with
C$_2$H$_2$ and HCN. Therefore, it is unlikely to be tracing scattered
inner disk emission. Class I sources are good laboratories for
studying disk winds since they are not overwhelmed by emission from
hot cores or dense jet shocks. The Class I source TMC1, which
shows a prominent H$_2$ wide-angle wind, has CO$_2$, OH and HCO$^+$ 
detected off source, but not H$_2$O \citep[see Fig.~3
of][]{Tychoniec24}. Taken together, it appears that primarily CO$_2$
is seen in emission off source, which, as mentioned in
Sect.~\ref{sec:CO2}, could probe either sublimated ices or moderately
warm gas chemistry in a wind. Sputtered ices along outflow cavity
walls cannot be excluded, however. Since densities in winds are
expected to be significantly below those required for thermalization,
detailed non-LTE excitation models including infrared pumping are
needed to quantify abundances. This is left to a future study. The
other detected molecules, OH and HCO$^+$, point to a warm chemistry in
which UV photons play a significant role, both in photodissociating
H$_2$O and driving reactions of C$^+$ and O with excited H$_2^*$.

Winds can also be detected in absorption. Indeed, for the Class I
source Oph IRS 46, both CO$_2$, C$_2$H$_2$ and HCN absorption have
been seen with {\it Spitzer} and Keck \citep{Lahuis06} indicating
an origin either in the upper layers of the inclined disk or in a
wind/outflow. Within the JOYS sample, the Class I source TMC1A, known
to have a disk wind \citep{Bjerkeli16,Harsono23}, shows both CO$_2$
and ro-vibrational H$_2$O lines in absorption
(Fig.~\ref{fig:CO2overview}) but no HCN or C$_2$H$_2$, with the H$_2$O
line having a clear velocity shift \citep{vanGelder24overview}. This
makes it unlikely that the absorption arises in a hot core or inner
disk. Quantification of column densities from absorption lines is much
more robust than from emission lines; thus, TMC1A could serve as a
test case for studying chemistry in winds.

In summary, the cases of TMC1 and TMC1A indicate that the chemistry in
winds, showing CO$_2$, OH and HCO$^+$, is clearly different from
the high temperature dense chemistry that produces abundant C$_2$H$_2$
and HCN (Sect.~\ref{sec:modelchemistry}). Further development of,
  and quantitative comparison with, models of the chemistry in disk
  winds such as those of \citet{Panoglou12,Yvart16,Tabone20} that
  include OH and HCO$^+$ are warranted.

\begin{figure*}[tbh]
  \begin{centering}
\includegraphics[width=5.8cm]{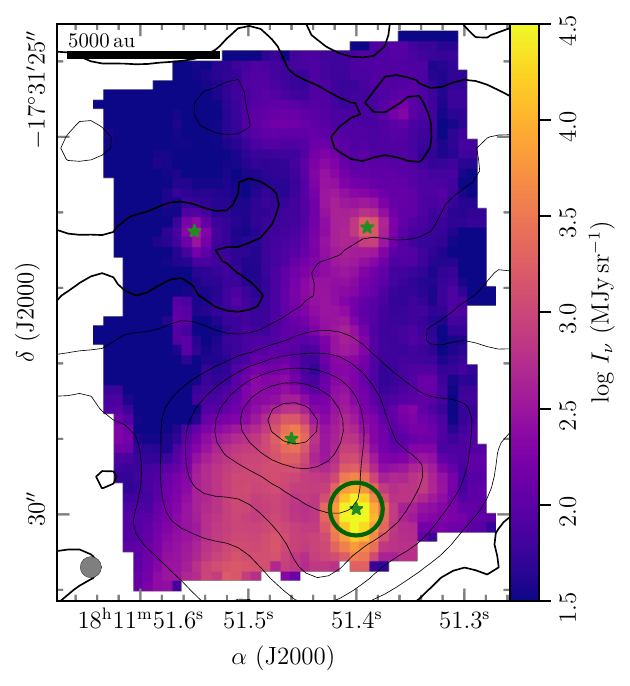}
\includegraphics[width=12cm]{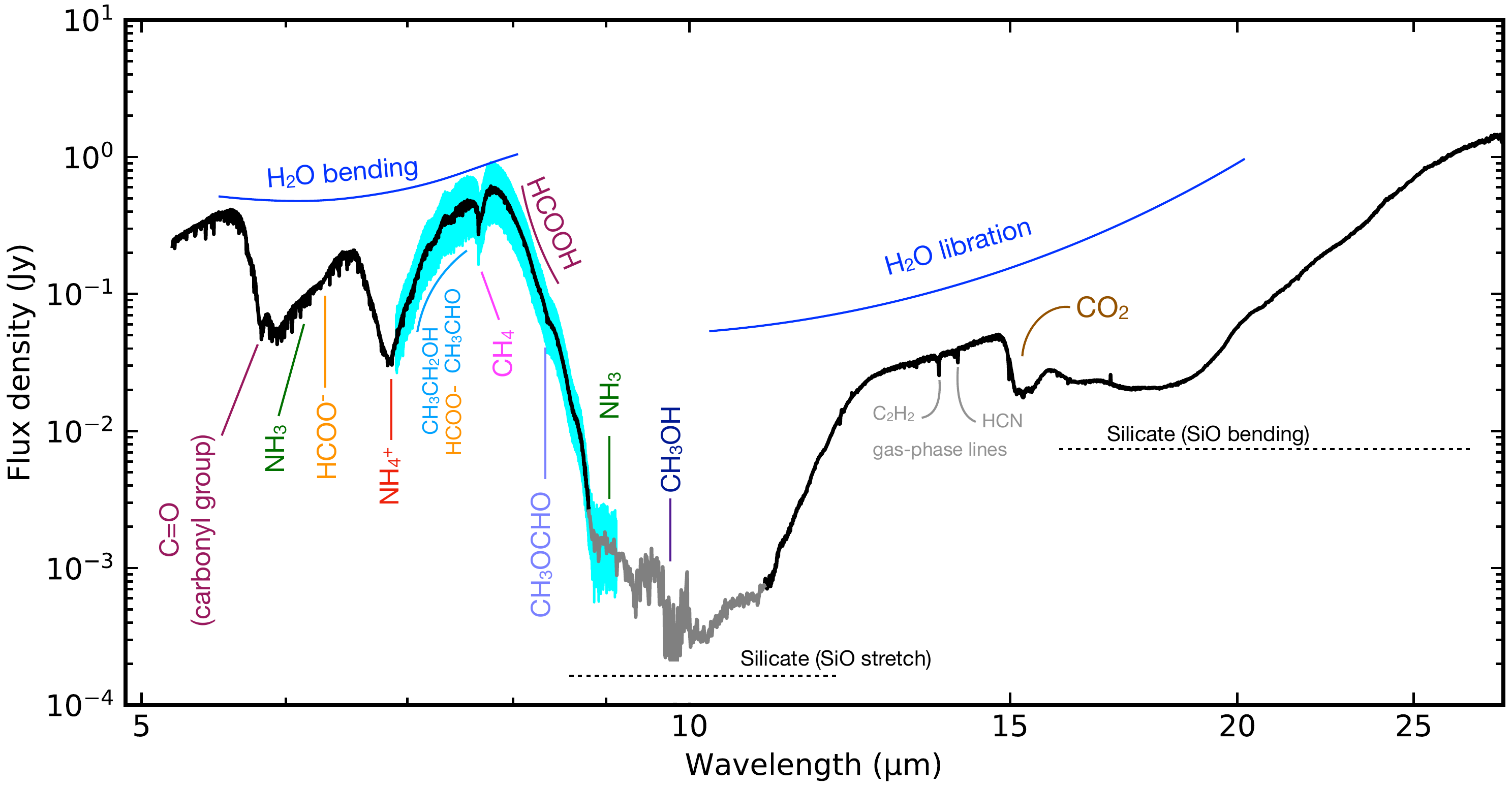}
\caption{Left: MIRI-MRS 5 $\mu$m continuum image of the IRAS
  18089-1732.  The four compact sources are marked by green stars. The
  0.7$''$ aperture used to extract the MIRI-MRS spectrum of the
  brightest mid-infrared source is indicated by a green circle. A
  scale bar of 5000\,au is shown in the top and the angular resolution
  at 5 $\mu$m is shown in the bottom-left corner.  Black contours are
  the ALMA 3\,mm continuum emission taken from \citet{Gieser23ALMA}
  with contour steps at 5, 10, 20, 40, 80, 160,
  320$\times\sigma_\mathrm{cont}
  (\sigma_\mathrm{cont}$=0.057\,mJy\,beam$^{-1}$).  Right: MIRI-MRS
  spectrum of IRAS 18089-1732. The main absorption features are
  annotated in the figure. Gas-phase lines are not labeled, except in
  the case of C$_2$H$_2$ and HCN at 13.7 and 14.0 $\mu$m. The cyan
  region between 7 and 9 $\mu$m is the spectral window analyzed in
  this work. Because of the low $S/N$ in the deep silicate absorption,
  the data between 8.9 and 10 $\mu$m are binned by a factor of five.  }
\label{fig:IRAS18089spectrum}
\end{centering}
\end{figure*}

\section{Cold envelope: Ices}
\label{sec:ices}

This section presents JOYS results on ices. As
Figure~\ref{fig:B1c-overview} shows, ice features are prominently seen
at 5--20 $\mu$m in MIRI-MRS spectra extracted from the bright
mid-infrared continuum. Detected ices range from simple molecules like
H$_2$O, CO$_2$ and CH$_4$ to CH$_3$OH as the most complex molecule
identified pre-JWST \citep{Boogert15}. Background information can be
found in Sect.~\ref{sec:backgroundices} and Sect.~\ref{sec:back_ices}.

\subsection{Detection of complex molecules in ices}

The high $S/N$ JWST spectra allow weaker features to be identified,
especially in the 5--10 $\mu$m range, where {\it Spitzer} only had
$R=50-100$.  One of the early highlights of JOYS is the detection of
complex molecules other than CH$_3$OH in ices toward the low-mass
protostars NGC 1333 IRAS2A and B1-c and the high-mass source IRAS
23385+6053 \citep{Rocha24,Chen24}. These features occur in the range
between 6.9-9.0 $\mu$m, also known as the icy COMs fingerprint region
\citep{Schutte99,Oberg11}. To investigate how similar these COMs features are
between low- and high-mass protostars, we analyze here a second
high-mass protostar, IRAS 18089-1732.

Figure~\ref{fig:IRAS18089spectrum} presents the spectrum of IRAS 18089
extracted from the peak mid-infrared continuum in a 0.7$''$ aperture
(RA=18h11m51.40s Dec=-17d31m29.93s). This aperture is optimized for
the 7--9 $\mu$m region and is taken to be small, also at longer
wavelengths, to avoid blending with another source. IRAS 18089 has a
luminosity of $\sim 10^4$ L$_\odot$ and envelope mass of $\sim$1000
M$_\odot$ based on single-dish observations
\citep{Sridharan02,Beuther02,Urquhart18}, at an estimated distance of
2.34 kpc \citep{Xu11}. The source is known to be rich in gas-phase
complex organic molecules (COMs)
\citep[e.g.,][]{Beuther04,Isokoski13}.

\citet{Rocha24} and \citet{Chen24} found signatures of several icy
COMs in their analyzed sources, including CH$_3$CHO (acetaldehyde),
CH$_3$CH$_2$OH (ethanol), CH$_3$OCHO (methylformate), and
CH$_3$OCH$_3$ (dimethylether). In addition, ice features of CH$_4$,
HCOOH (formic acid), SO$_2$ and the negative ions OCN$^-$ (cyanate
ion) and HCOO$^-$ (formate ion) are found in the 7--9 $\mu$m range.
Here, we adopt the same methodology as in \citet{Rocha24} to execute
the spectral fitting. The first step is to remove the continuum, the
deep silicate feature and the H$_2$O 12 $\mu$m ice libration band,
which is described in the Appendix~\ref{sec:app_IRAS18089}. This
procedure also provides the total H$_2$O ice column density, which is
found to be $(2.8 \pm 0.7) \times 10^{19}$ cm$^{-2}$.

\begin{figure}[tbh]
\begin{centering}
\includegraphics[width=9cm]{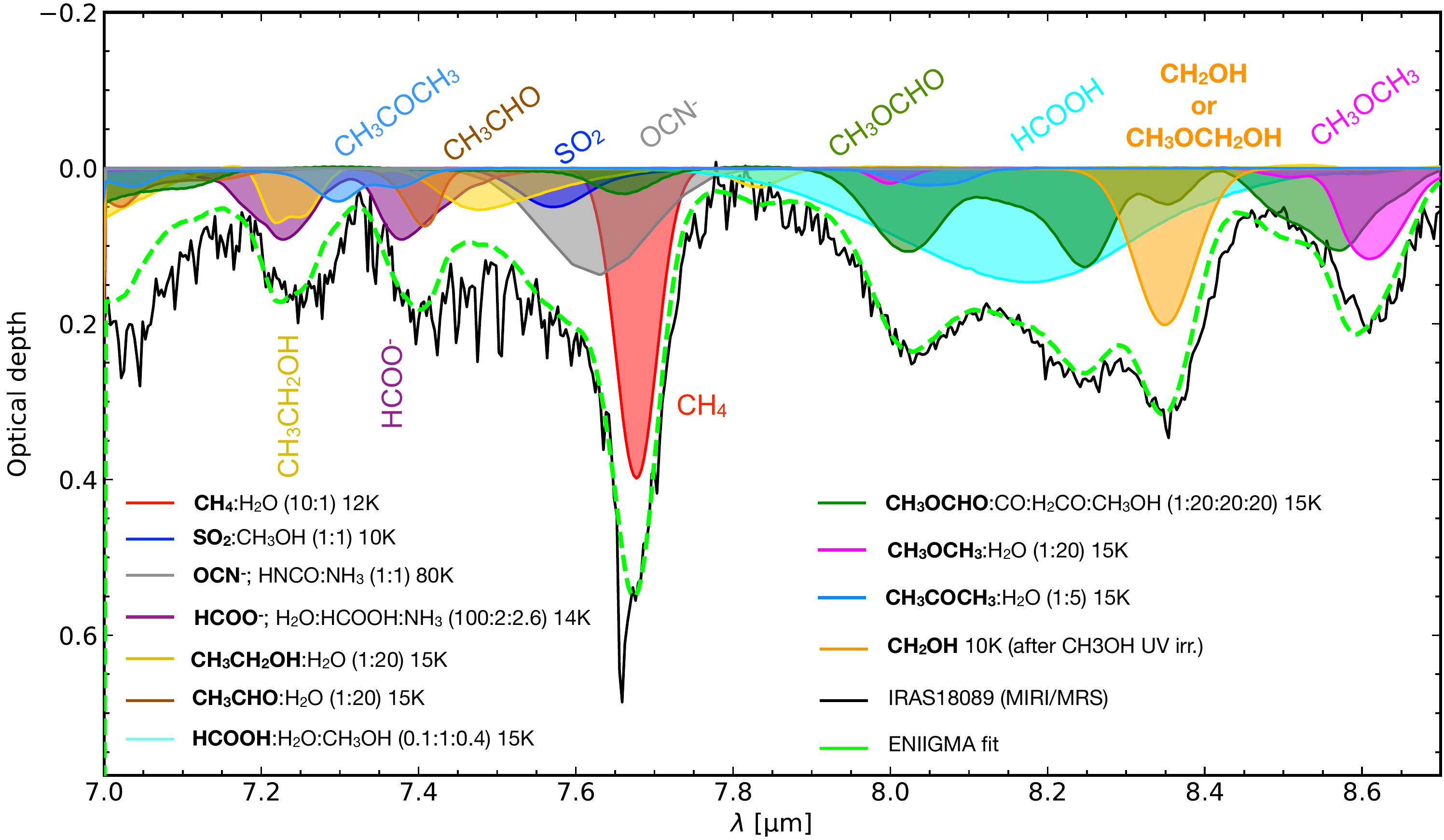}
\caption{Spectral fit between 7 and 8.7 $\mu$m of IRAS 18089 using the
  ENIIGMA fitting tool with laboratory ice mixtures from the LIDA
  database. Eleven ice components were used to obtain the best fit,
  which includes simple and complex molecules in ice mixtures. The
  boldface in the labels of the lab spectra indicates the main species that
  contributes in this spectral range. The narrow weak absorption
  features between 7 and 8 $\mu$m are primarily due to warm gas-phase
  H$_2$O; the narrow peak at 7.66 $\mu$m is due to gaseous CH$_4$
  absorption.}
\label{fig:IRAS18089COMfit}
\end{centering}
\end{figure}

The next step is to choose a local continuum for the 6.9--9.0 $\mu$m
region on the optical depth spectrum that has been corrected for the
global continuum, silicate feature and H$_2$O libration bands (see
Fig.~\ref{fig:IRAS18089H2Ofit}). This local continuum takes into account
absorption features from major species in this interval, such as the
H$_2$O ice 6 $\mu$m bending mode, the prominent NH$_4^+$ 6.8 $\mu$m
profile, and likely organic refractory material
\citep{Gibb04}. Figure~\ref{fig:IRAS18089localfit} in
Appendix~\ref{sec:app_IRAS18089} shows this local continuum profile,
which is obtained by fitting a fifth-order polynomial to guiding points
to avoid unrealistic inflections. We note that a small but important
excess is left around 7 $\mu$m since the CH$_3$ bending mode
of several COMs contributes to this range.

After this local continuum subtraction, the spectrum is fitted using
the infrared laboratory spectra of simple and complex molecules from the
LIDA database, following \citet{Rocha24} using the ENIIGMA fitting
tool \citep{Rocha21}.  This code combines a large number of laboratory
ice spectra in small groups of components aiming to find the best
solution given by the $\chi^2$. Starting from all possible species, a
solution was found for IRAS 18089 that combines 11 ice mixtures,
containing absorption features in the fitted wavelength range of
CH$_4$, SO$_2$, OCN$^-$, HCOO$^-$, CH$_3$CH$_2$OH, CH$_3$CHO,
CH$_3$OCHO, HCOOH, CH$_3$OCH$_3$ and CH$_3$COCH$_3$ (acetone), as
shown in Figure~\ref{fig:IRAS18089COMfit}.

Based on the statistical methods and criteria of \citet{Rocha24} and
as embedded in the ENIIGMA code, most of these identifications are
thought to be secure; acetone is considered tentative for the case of
IRAS 18089 (but see also discussion in \citealt{Chen24} for B1-c). The
presence of H$_2$CO is based only on its 8 $\mu$m band and would need
to be confirmed at shorter wavelengths. SO$_2$ ice is detected in this
high-mass source, in contrast to the case of IRAS 23385
\citep{Rocha24}.  CH$_3$COOH (acetic acid) is not detected in either
of the high-mass sources although its detection is sensitive to the
local continuum placement at 7.8--8.0 $\mu$m.

Indeed, more generally, the detections and abundances of icy COMs
remain sensitive to the details of the global and local continuum
placement and subtraction.  While the uncertainty in the global
continuum reflects the source structure and SED, the local continuum must reflect the contribution of
less understood species, such as salts and organic refractory
materials that show broader infrared absorption features from 6 to
8~$\mu$m. Laboratory spectra are often available for only a limited
range of ice mixtures and temperatures, although the current list
already reflects huge experimental efforts focusing on those icy COMs
that are also abundant in the gas-phase. Indeed, arguments in support
of their identification include the detection of the same COMs at
comparable abundances in the gas phase in the same source
\citep[see][and below]{Chen24,Nazari24}. New measurements to enlarge
the icy COMs inventory in the laboratory exploring new ice mixtures in
realistic abundances at a range of temperatures are encouraged, as
well as more studies on the chemical composition of the organic
refractory materials.

The IRAS 18089 spectrum contains a prominent 8.38 $\mu$m feature that
has been seen in only a few other JWST ice spectra analyzed so far
(B1-c, L1527, IRAS2A).  Possible identifications include the
hydroxymethyl radical (CH$_2$OH) or methoxymethanol (CH$_3$OCH$_2$OH).
It will be discussed in a future paper that will also include new
laboratory spectroscopy.

The derived ice column densities for the species fitted in this work
are presented in Table~\ref{tab:ices}. CH$_3$OH is extracted
from the bottom of the silicate feature for IRAS 18089, since no
NIRSpec data are available for this source. The adopted band strengths
are listed in Table~1 of \citet{Rocha24}. Uncertainties in inferred
column densities are included in the table and are typically
$\sim$30\%.

\begin{figure}[tbh]
\begin{centering}
\includegraphics[width=9cm]{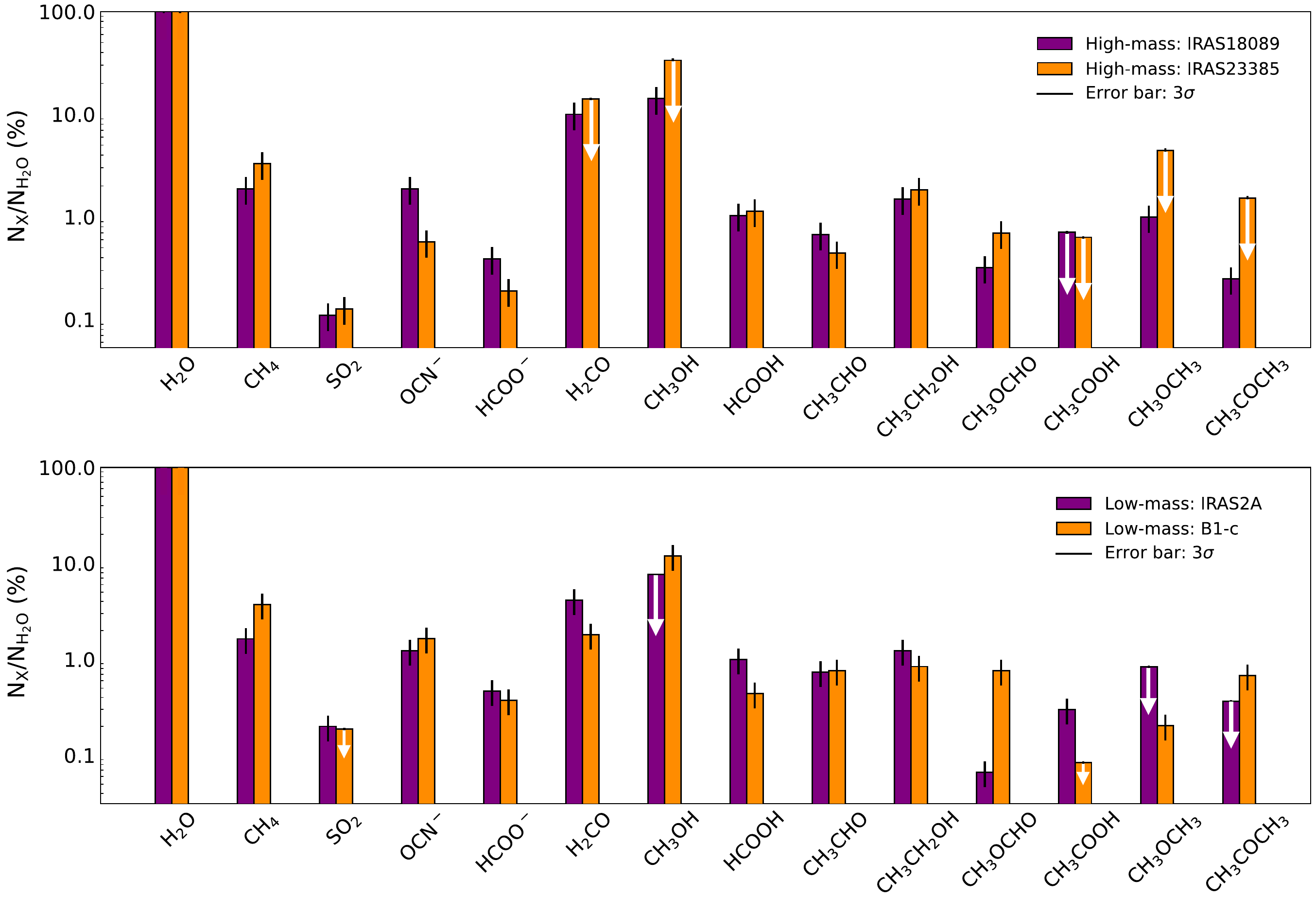}
\caption{Comparison of ice abundances relative to that of H$_2$O for
  the high-mass sources IRAS 18089 and IRAS 23385 (top) versus the
  low-mass protostars NGC 1333 IRAS2A and B1-c (bottom). White arrows
  indicate upper limits. The IRAS2A and IRAS 23385 results are taken
  from \citet{Rocha24} and B1-c from \citet{Chen24}.}
\label{fig:IRAS18089icecomp}
\end{centering}
\end{figure}

Figure~\ref{fig:IRAS18089icecomp} compares the ice columns relative to
that of H$_2$O for the high-mass sources IRAS 18089 and IRAS 23385
versus the low-mass sources IRAS2A and B1-c. Very similar patterns are
seen: not only are the same molecules detected in all four sources,
but their abundances with respect to H$_2$O are also found to be comparable
within the uncertainties for many species.
Comparing high- and low-mass protostars, interestingly, methyl formate
(CH$_3$OCHO) seems to show larger variations from source to source.

\citet{Chen24} went one step further and compared the results for
organic molecules in ices for the two low-mass protostars B1-c and NGC
1333 IRAS2A with those found in the gas phase in hot cores with ALMA
toward the same sources. Interestingly, the ratios of some of the most
abundant O-COMs with respect to methanol, CH$_3$OCHO/CH$_3$OH and
CH$_3$OCH$_3$/CH$_3$OH, are very similar in ice and gas, supporting
the picture of ice sublimation producing these species in their hot
cores. However, the gaseous ratios of CH$_3$CHO/CH$_3$OH and
C$_2$H$_5$OH/CH$_3$OH appear lower than those found in ices. Unless
these COM ice abundances are overestimated, this suggests that some
COMs may be destroyed quickly in the gas following sublimation (see
discussion in \citealt{Chen24} and in \citealt{Chen25} for the case of
acetone).

Besides quantification of ice column densities, comparisons between
laboratory and observed infrared spectra also reveal information on
ice temperatures and mixing conditions (i.e., mixing constituents and
mixing ratios) of the detected COM ices, because these factors vary the
peak position and the profile of their absorption
bands. \citet{Rocha24} and \citet{Chen24} found that most of their
detected COM ices are likely surrounded by a H$_2$O-rich environment,
with CH$_3$OCHO as an outlier that is suggested to be present in a CO-
and CH$_3$OH-rich layer. The inference on temperature is much more
degenerate since many COM ice bands do not change much below their
crystallization temperatures (usually $\sim$100 K in the laboratory,
20\%—40\% lower in space due to lower pressures). Investigation of a
larger sample is needed to solidify these results.

\subsection{Ammonium salts in ices}

The fit to the 7--9 $\mu$m range of the four sources for which the
COMs have been analyzed (Fig.~\ref{fig:IRAS18089icecomp}) also
indicate the presence of two negative ions in the ices: HCOO$^-$ and
OCN$^-$ (see Fig.~\ref{fig:IRAS18089COMfit}). The latter ion has been
identified previously through its strong band at 4.62 $\mu$m seen
along many lines of sight
\citep[e.g.,][]{Grim87,Novozamsky01,vanBroekhuizen04}, now also with
JWST-NIRSpec \citep{McClure23,Nazari24}. Together with the positive
ion NH$_4^+$ that is ubiquitously detected at 6.85 $\mu$m
(Fig.~\ref{fig:B1c-overview} and \ref{fig:IRAS18089spectrum}), they
make up ammonium salts that are readily produced by acid-base
reactions in ices \citep[][for review]{Knacke82,Boogert15}. Salts have
recently gained renewed interest since they are found in high
abundances in cometary material as one of the major nitrogen
reservoirs \citep{Altwegg19} and also in the gas-phase in some
high-mass protostars \citep{Ginsburg23}.

The JWST JOYS analysis confirms the identification of the OCN$^-$ ion
through its bending mode at 7.6 $\mu$m and strengthens the assignment
of the HCOO$^-$ ion as significant contributors to the 7.2 and 7.4
$\mu$m absorption bands, suggested by
\citet{Schutte99,Schutte03}. Moreover, they indicate the presence of
ammonium hydrosulfide NH$_4$SH as the dominant ammonium salt in ices
with SH$^-$ the main counterpart of NH$_4^+$ and locking up 18--19 \%
of the sulfur in dense clouds \citep{Slavicinska25sulfur}.
Figure~\ref{fig:IRAS18089NH4} shows a fit to the 6.85 $\mu$m band of
IRAS 18089. The ice anion abundances with respect to NH$_4^+$ are
similar to those in comet 67P, where NH$_4$SH is also the dominant
ammonium salt \citep{Altwegg22}. Together, the identified icy
N-species NH$_4^+$ and OCN$^-$ make up 19\% of the available N-budget
(see Appendix~\ref{sec:app_IRAS18089} for details). NH$_3$ ice is
difficult to measure in these data (Fig.~\ref{fig:IRAS18089spectrum}),
but, if its column were as high as 10\% of H$_2$O ice
\citep{Boogert15}, it would increase the N--budget to 31\%.  Future
studies using the larger JOYS+ sample and combining NIRSpec and MIRI
will allow these ions to be studied across the Class 0 and I stages.

\begin{figure}[tbh]
\begin{centering}
\includegraphics[width=8cm]{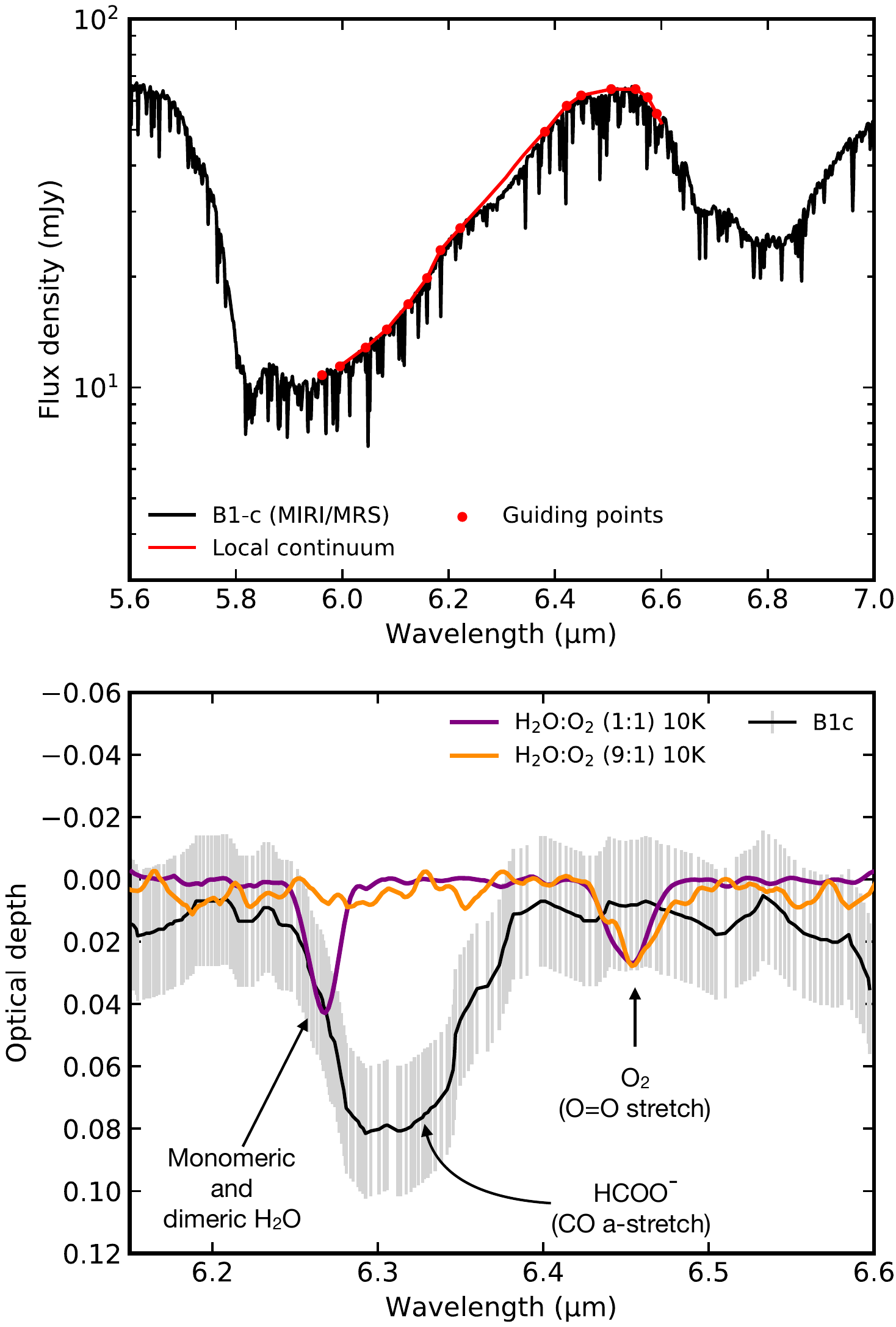}
\caption{Limits on O$_2$ ice in the spectrum of the low-mass protostar
  B1-c.  The top panel shows the local continuum used between 6 and
  6.6 $\mu$m together with the anchor points. The narrow absorption
  lines are all due to gas-phase water absorption
  \citep{vanGelder24overview}. The bottom panel presents the optical
  depth spectrum with 3$\sigma$ error bars compared with O$_2$:H$_2$O
  ice laboratory spectra \citep{Muller18}. }
\label{fig:O2}
\end{centering}
\end{figure}

\subsection{Limits on O$_2$ ice}

The most abundant molecule detected in ices is H$_2$O, but it contains
only a moderate fraction of the total oxygen budget, typically
$\lesssim 20\%$ \citep{Whittet10,vanDishoeck21}. A significant
fraction of oxygen must therefore be in other forms, and O$_2$ has
been suggested as an option. Observations to date however indicate
that it is not a major reservoir, neither in gas
\citep{Goldsmith11,Liseau12,Yildiz13o2} nor in ice
\citep{Vandenbussche99}. Sect.~\ref{sec:back_ices} provides more
background information.

New impetus for searching for interstellar O$_2$ ice has come from the
detection of abundant O$_2$ ice in comet 67P by the ROSINA instrument
on board of the Rosetta mission \citep{Bieler15,Rubin19}. The cometary
O$_2$ abundance is found to follow closely that of H$_2$O with
heliocentric distance at a mean ratio of
O$_2$/H$_2$O$\approx$0.03. While not a major oxygen reservoir, this
still makes O$_2$ a significant ice component. Moreover, it provides
insight into the major oxygen chemistry pathways: producing O$_2$ ice
at this level through gas-grain chemistry requires special conditions
(high density, low cosmic ray ionization rate) in the pre-stellar
cloud to suppress water ice formation \citep{Taquet16o2}.

Since O$_2$ is infrared inactive, it does not have a strong ice band
in its pure ice form. However, if mixed with H$_2$O ice as suggested
by the cometary results, its 6.45 $\mu$m band becomes weakly visible
\citep{Ehrenfreund92,Muller18}. ISO-SWS was able to put only weak
limits on the 6.45 $\mu$m feature in a few warm high-mass sources due
to its low $S/N\approx 10$ on the continuum \citep{Vandenbussche99}.

Here we used the deep MIRI spectrum of the most ice-rich source in the
JOYS sample, the low-mass Class 0 source B1-c, to search for solid
O$_2$. With an integration time of 1800 sec in channel 1C, this
spectrum achieves a $S/N>100$ on the continuum
(Fig.~\ref{fig:B1c-overview}). The data reduction details can be found
in \citet{Chen24}.  Figure~\ref{fig:O2} presents a blow up of the B1-c
MIRI-MRS spectrum extracted at the source position in the 6--7 $\mu$m
range. This region is dominated by the deep H$_2$O ice bending mode
absorption at 6.0 $\mu$m with NH$_3$ ice at 6.18 $\mu$m, and the deep
NH$_4^+$ band at 6.8 $\mu$m.  Moreover, the B1-c spectrum is rich in
narrow gas-phase absorption lines, most of them due to warm H$_2$O
vapor that are superposed on the ice features
\citep{vanGelder24overview}. A local continuum subtraction is
performed between 6 and 6.6 $\mu$m, where any contribution of O$_2$
ice is expected (Fig.~\ref{fig:O2}, top). Some anchor points are used
in this fit to leave some space for known absorption features and
ensure smooth curvature, and a fifth-order polynomial is then adopted to
trace the continuum.

Subtracting all of these features and overlaying laboratory spectra of
various O$_2$:H$_2$O mixtures provides a $3\sigma$ limit of O$_2$
(Fig.~\ref{fig:O2}, bottom). The inferred value is
$N_s$(O$_2$)$< 4.9 \times 10^{18}$ cm$^{-2}$, using a band strength
$A=3.9\times 10^{-20}$ cm molecule$^{-1}$ appropriate for 1:1 O$_2$
ice mixtures with H$_2$O \citep{Muller18}. Together with
$N_s$(H$_2$O)=$2.5 \times 10^{19}$ cm$^{-2}$ \citep{Chen24}, this
provides an O$_2$ limit of $\sim$20\% relative to H$_2$O ice. Due to the
strong dependence of the band strength on H$_2$O ice concentration,
this value may be increased by a factor of five. We note that layered ice
models indicate that O$_2$ is not uniformly mixed with H$_2$O ice
\citep[see Figure~5 in][]{Taquet16o2}; hence, the use of an O$_2$:H$_2$O
laboratory spectrum with a lower mixture than 1:30 could be
justified. Nevertheless, it is clear that even with the sensitivity of
JWST, it is not possible to put stringent limits on the O$_2$ ice
abundance \citep{Muller18} and test values of order 3\% as found in
cometary ices.

Future indirect constraints on the O$_2$ ice abundance may come from
the analysis of the $^{13}$CO ice profile at 4.8 $\mu$m and from
searches for photoproducts of O$_2$ ice such as O$_3$ and CO$_3$ ice
using both NIRSpec and MIRI data
\citep{Ehrenfreund98o2,Vandenbussche99,Pontoppidan03}.

\subsection{HDO/H$_2$O in protostellar ices}

While there have been multiple determinations of the gaseous
HDO/H$_2$O ratio in hot cores \citep{Persson14,Jensen19}, its value in
interstellar ices remained poorly constrained prior to JWST
\citep{Dartois03,Aikawa12akari}.  JWST-NIRSpec offers the opportunity
for deep searches for HDO ice through its O-D stretching mode at
$\sim$4.1 $\mu$m. Searches for this feature are complicated by the
nearby weak CH$_3$OH 3.8--3.9 $\mu$m combination mode
\citep{Dartois03}, and a potential S-H stretch from molecules like
H$_2$S or SH$^-$ at 3.9 $\mu$m. On the other side of the HDO ice band,
the deep $^{12}$CO$_2$ 4.3 $\mu$m ice absorption can distort the local
continuum due to scattered light off ices, a process that is sensitive
to grain shape effects \citep{Dartois22}.

\begin{figure}[tbh]
\begin{centering}
  \includegraphics[width=9cm]{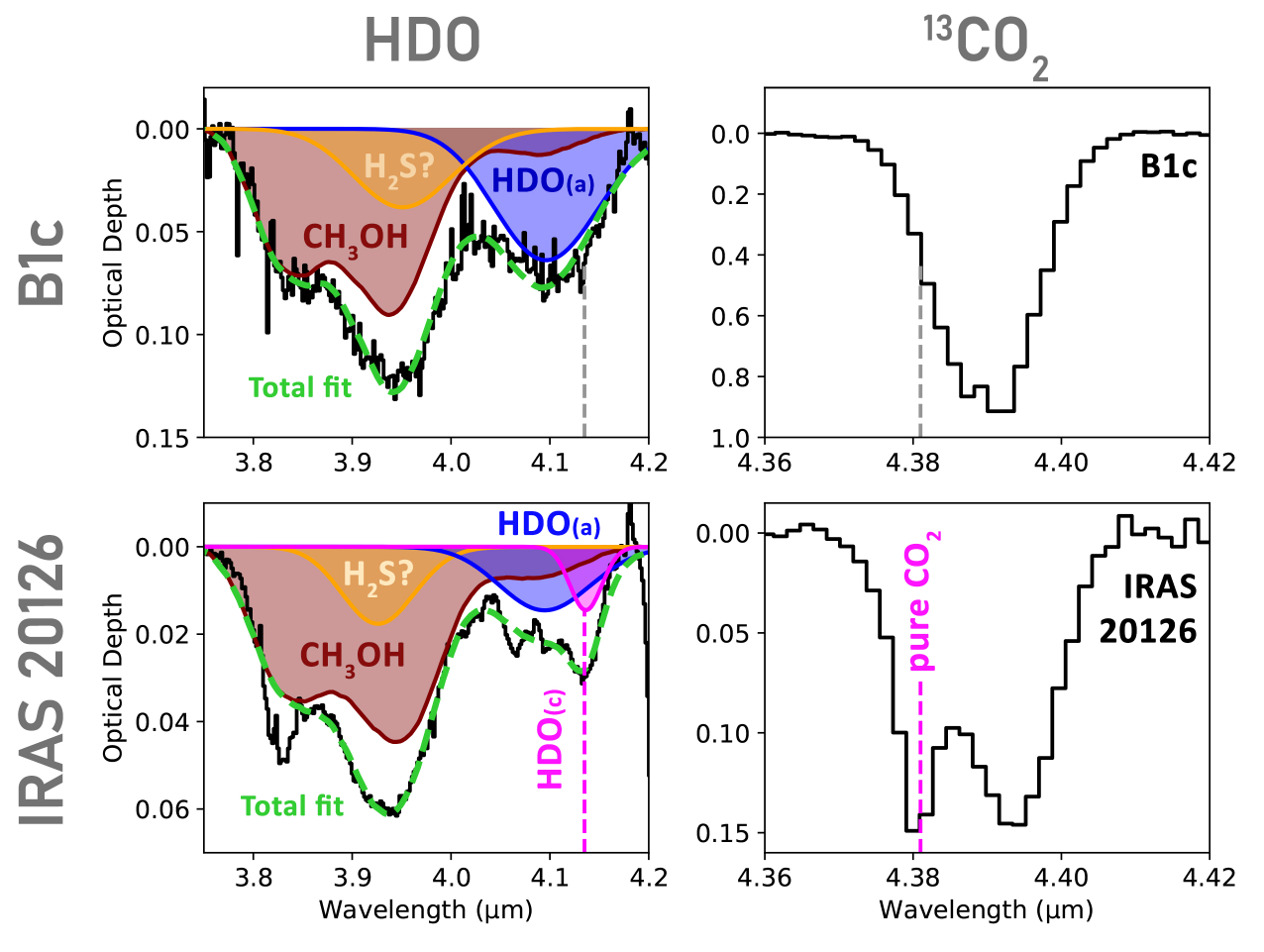}
\caption{Top-left panel: JWST NIRSpec spectrum of B1-c in the
  3.8--4.2 $\mu$m range highlighting the detection of HDO
  ice. Laboratory spectra of HDO amorphous ice (HDO(a)) as well as of
  CH$_3$OH ice are overlaid. A Gaussian has been added to represent an
  S-H ice feature. Top-right panel: Ice band of $^{13}$CO$_2$ 4.38
  $\mu$m of B1-c highlighting the lack of a heated ice
  component. Bottom-left panel: JWST NIRSpec spectrum of the high
  mass protostar IRAS 20126 taken from \citet{Slavicinska24HDO}
  showing the detection of both amorphous and crystalline HDO ice. The
  laboratory spectrum of crystalline HDO ice is shown in pink
  (HDO(c)). Bottom-right panel: Ice band of $^{13}$CO$_2$ 4.38
  $\mu$m of IRAS 20126 showing evidence of a heated ice
  component as analyzed by \citet{Brunken24}.}
\label{fig:HDO}
\end{centering}
\end{figure}

\citet{Slavicinska24HDO} recently presented the first robust
detections of crystalline and amorphous HDO ice toward the
intermediate- and high-mass protostars HOPS 370 (310 L$_\odot$) and
IRAS 20126 ($10^4$ L$_\odot$) as part of the IPA sample. A careful
decomposition of the 3.8--4.2 $\mu$m ice band was performed, using new
laboratory data and obtaining a consistent fit with the CH$_3$OH 3.54
$\mu$m band to assess the contribution of CH$_3$OH ice at 3.8--3.9
$\mu$m. Both crystalline and amorphous HDO ice were required to fit
the feature, consistent with the fact that both are warm sources as
deduced also from the shape of their 3 $\mu$m H$_2$O ice bands showing
crystalline H$_2$O ice and from the analysis of their $^{13}$CO$_2$ 4.38
$\mu$m ice bands \citep{Brunken24}. Both the HDO and $^{13}$CO$_2$ spectra
for IRAS 20126 are included in Figure~\ref{fig:HDO} (bottom). The
inferred HDO/H$_2$O ice ratios range from $(3-5)\times 10^{-3}$,
comparable with the highest reported warm gaseous HDO/H$_2$O ratios in
high-mass hot cores where ices have thermally sublimated.

The most prominent HDO ice detection to date is that toward the low
mass Class 0 protostar L1527 using the G395M NIRSpec spectrum from
program 1798 (PI: J.\ Tobin), which shows a significantly deeper
deuterated water ice feature \citep{Slavicinska25L1527}. This
detection is particularly exciting, since this low-mass source
provides a window into what the HDO/H$_2$O ice ratio in our own
protosolar system could have been. The inferred value of
$4.8\times 10^{-3}$ with an uncertainty of $\sim$60\% is at the upper
end of the gaseous HDO/H$_2$O ratios in other isolated low-mass
protostars \citep{Jensen19}.

To explore whether the HDO/H$_2$O ice ratio toward L1527 is
representative for solar-mass systems, we present here a second
detection of HDO ice toward a low-mass protostar, B1-c. A NIRSpec
G395M $R\sim1000$ spectrum was taken as part of JOYS (program
1290). We note that all other NIRSpec spectra in JOYS+ used the G395H
$R\sim 2700$ mode, which has a gap precisely at 4.1 $\mu$m and can
therefore not be used for HDO searches. Within the JOYS+ data set, the
low resolution PRISM mode could still be used for that purpose, albeit
with less fidelity at $R$ of only $\sim$200. Such a PRISM study is
left for a future paper.

Figure~\ref{fig:HDO} (top) presents a blow-up of the 3.8--4.2 $\mu$m
spectrum for B1-c over the same region as shown in
\citet{Slavicinska24HDO,Slavicinska25L1527} for the other sources. One
complication for B1-c is that the blue side of the deep CO$_2$ ice
feature shows clear signs of grain growth which affects the continuum
determination. This wing has been removed using a local continuum in a
similar manner as for L1527. Once removed, a clear HDO + CH$_3$OH
feature emerges. Using the same fitting procedure as in
\citet{Slavicinska24HDO,Slavicinska25L1527}, an HDO column of
$1.2\times 10^{17}$ cm$^{-2}$ is derived. The H$_2$O 3 $\mu$m band is
completely saturated and below the noise in this source but an H$_2$O
column density of $2.5\times 10^{19}$ cm$^{-2}$ has been derived by
\citet{Chen24} from the H$_2$O libration mode (see also above section
on O$_2$ ice limits for B1-c). Together this gives
HDO/H$_2$O=$(4.8 \pm 3.0)\times 10^{-3}$, consistent with that found
toward L1527. As for L1527, the uncertainties are dominated by the
placement of the continuum and variations in band strength with
temperature (see discussion in \citealt{Slavicinska25L1527}).

Both values are close to that recently found in warm gas in a young
protoplanetary disk where ice has sublimated \citep{Tobin23} (see
  Fig.~4 in \citealt{Slavicinska25L1527}).  They are somewhat higher
  than those measured in comets in our Solar system, suggesting that
  even though a large fraction of the water ice appears to be largely
  preserved in its journey from clouds to disks
  \citep{Visser09,Cleeves14,vanDishoeck21}, some modifications may
  occur along the way to the comet and planetesimal formation sites
  \citep{Furuya17}.

The HDO fit for B1-c requires only the presence of amorphous HDO
ice. Since the H$_2$O 3 $\mu$m band profile cannot be observed in B1-c
and the 12 $\mu$m band does not have a distinct crystalline feature,
it is not possible to check whether there is any crystalline H$_2$O
ice. However, the lack of crystalline HDO ice is consistent with the
fact that the $^{13}$CO$_2$ ice band, included in Figure~\ref{fig:HDO}
(top), has only a single broad feature that can be fitted with cold
CO$_2$:H$_2$O or CO$_2$:CH$_3$OH ice mixtures: no separate narrow
blue-shifted peak due to heated pure CO$_2$ ice is needed as found for
other sources \citep{Brunken24}. Similarly, the deep CO$_2$ 15 $\mu$m
bending mode of B1-c (Fig.~\ref{fig:B1c-overview}) shows no hint of a
double-peaked structure that is characteristic of heated ices. This
suggests that, in contrast with the other sources, B1-c has not
undergone episodes of intense accretion activity in the past that were
strong enough to crystallize a significant portion of the ices: once
crystallized, the process is irreversible.

\section{Embedded disks}
\label{sec:disks}

Disks are an integral part of protostellar systems, but little is
still known about their composition in the embedded stage. A related
question is to what extent these young disks are affected by accretion
shocks of material falling onto the disk. Background information on
this science case can be found in Sect.~\ref{sec:background} and
Sect.~\ref{sec:back_disks}.

\subsection{Low-mass protostars}

\subsubsection{JWST spectra of sources with embedded disks}

JWST MIRI-MRS data show rich spectra of Class II disks around pre-main
sequence stars pointing to a diverse chemistry in their inner few au
(see Sect.~\ref{sec:back_disks}).
Some disks are very rich in H$_2$O
\citep[e.g.,][]{Gasman23,Temmink24H2O,Romero24,Banzatti25}, some are
less H$_2$O rich \citep[e.g.,][]{Perotti23}, whereas other disks show
a prominent $^{(13)}$CO$_2$ feature
\citep[e.g.,][]{Grant23,Vlasblom25}. Disks around very low mass stars
often have strong emission from hydrocarbon molecules with only weak
water lines
\citep[e.g.,][]{Tabone23,Arabhavi24,Arabhavi25,Kanwar24}. Excitation
temperatures are high, from $\sim$300~K up to 900 K.  This diversity
is thought to be linked to icy pebbles drifting from the outer to the
inner disk, thereby enhancing molecules like H$_2$O and CO$_2$ when
they cross their snowlines on timescales of 1-2 Myr. Dust traps
locking up volatile elements in ices in the outer disks achieve the
opposite effect of lowering their abundances
\citep{Kalyaan23,Mah23,Mah24,Sellek25}.

\begin{figure}[tbh]
\begin{centering}
\includegraphics[width=9cm]{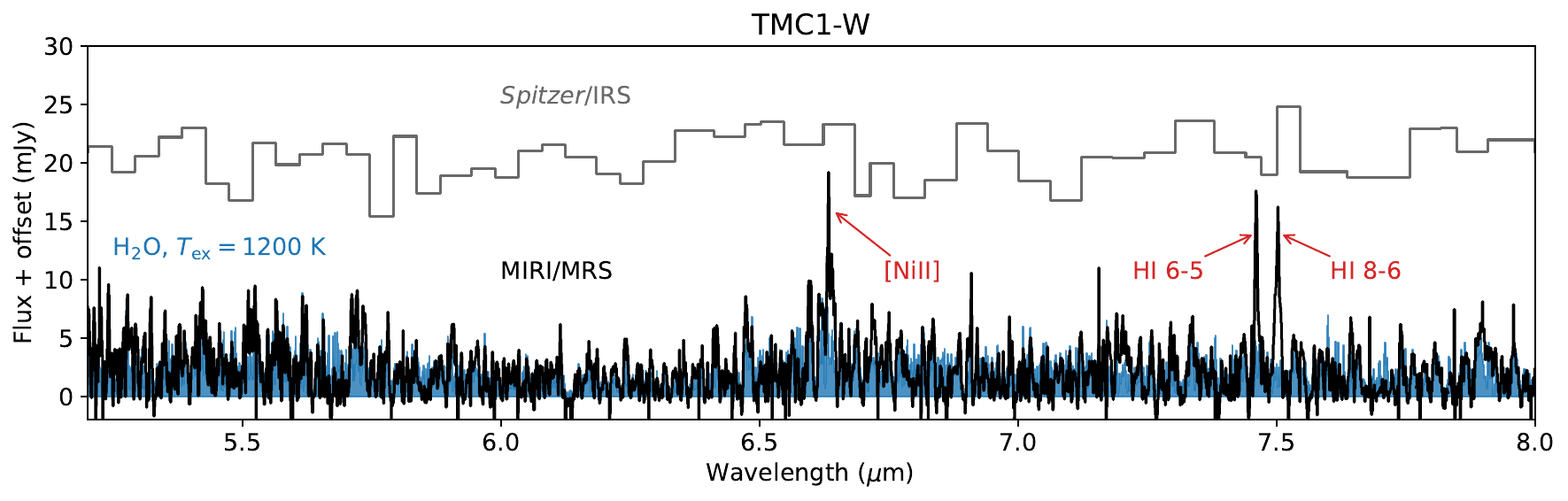}
\caption{JWST MIRI-MRS spectrum of the Class I TMC1-W source showing a
  forest of hot H$_2$O water lines. A LTE slab model at 1200 K is
  included for comparison in blue. The {\it Spitzer}-IRS spectrum of
  the binary TMC1 source is shown for comparison, illustrating that
  these types of detections were not possible prior to JWST. Figure
  adapted from \citet{vanGelder24overview}, which shows the model
  spectrum offset from the data.}
\label{fig:TMC1W}
\end{centering}
\end{figure}

Since disks already form early in the Class 0 phase and are well
established by the Class I phase \citep[e.g.,][]{Ohashi23,Tobin24}, a
prime question concerns the chemical composition of the inner regions
of embedded young disks, at a time of $< 0.5$ Myr when they are still
pristine and not yet affected by radial drift of icy grains from the
outer disk and/or when dust traps have not yet developed to interrupt
the pebble flux.

\begin{figure}[tbh]
\begin{centering}
\includegraphics[width=9.2cm]{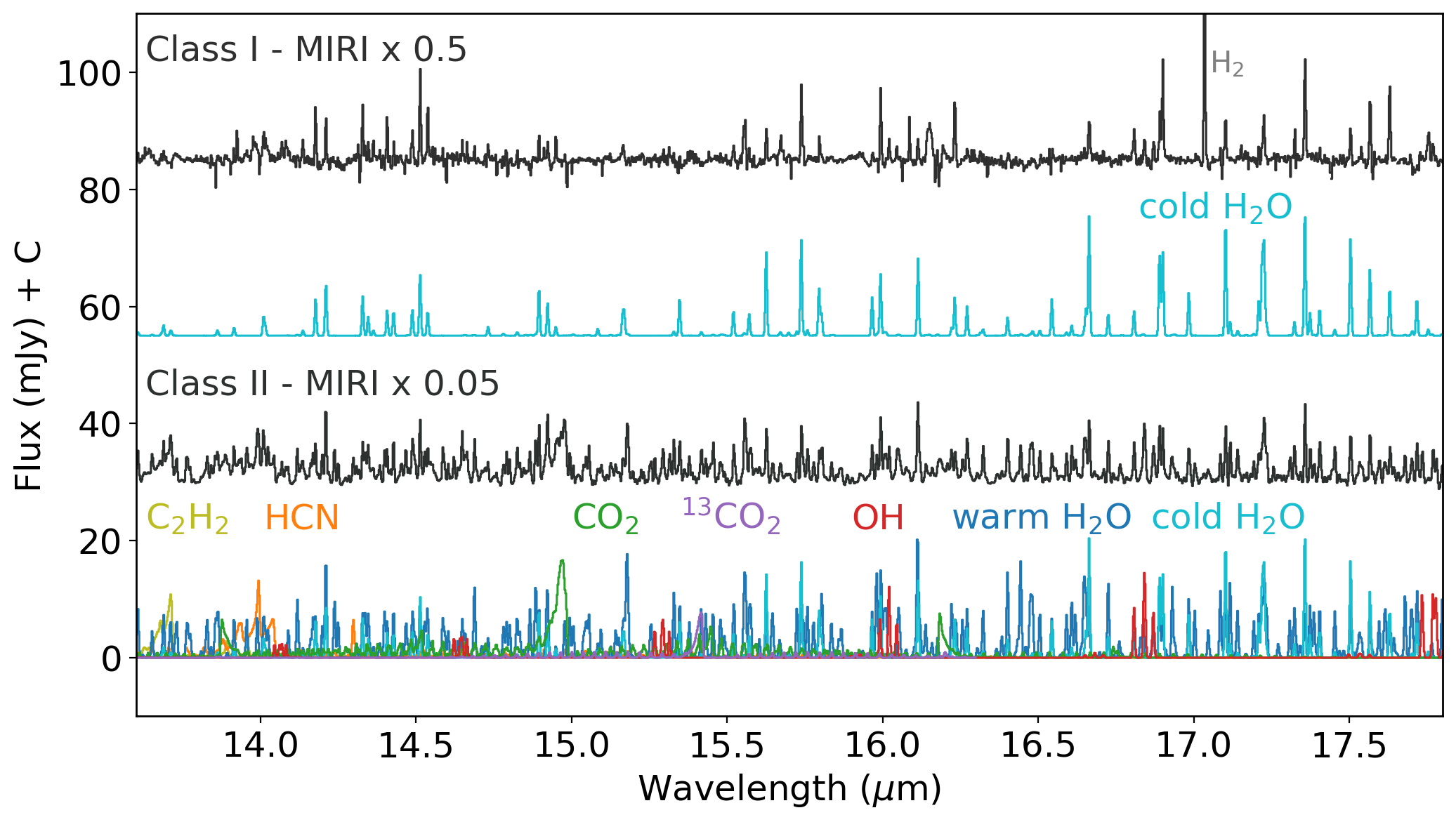}
\caption{Spectra from MIRI-MRS at 13.5--17.5 $\mu$m of the Class I source
  B1-a-NS showing only cold water \citep{vanGelder24overview} and of the
  Class II disk DF Tau showing it being rich in warm H$_2$O and other molecules
  \citep{Grant24}. The colored spectra show LTE slab models of
  individual molecules contributing to the observed spectra; they are
  meant to illustrate the positions of the lines, not as a fit to the
  data. The temperatures used on the H$_2$O slab models are 190 K
  (cold) and 490 K (warm)}
\label{fig:disks}
\end{centering}
\end{figure}

Class I disks should have warm gas in their inner regions, just as
Class II disks. Indeed, evidence of the presence of warm disk gas
down to $\sim$0.1 au is provided by VLT-CRIRES $^{12}$CO $v$=2--1
infrared line emission profiles at 4.6 $\mu$m \citep{Herczeg11}. It
therefore came as a surprise that the molecular line survey of the
JOYS sources by \citet{vanGelder24overview} found primarily low
excitation temperatures, not at all characteristic of inner disks
(Sect.~\ref{sec:hotcore}). Only one source with hot H$_2$O was uncovered,
TMC1-W, which is part of the $\sim$100 au separation binary source
TMC1 with both sources showing evidence of a disk in millimeter
continuum and line emission \citep{Harsono14,Tychoniec21,Tychoniec24}.
Figure~\ref{fig:TMC1W} presents the MIRI-MRS spectrum of the 5--8
$\mu$m range of this Class I protostar with the hot ($\sim$1200 K)
water model overlaid. At first glance, the MIRI spectrum looks like
pure noise, but blow-ups reveal that this is actually a forest of hot
water lines \citep[see
also][]{vanGelder24overview}. Figure~\ref{fig:TMC1W} includes the {\it
  Spitzer}-IRS spectrum for comparison to illustrate that such
detections were not possible prior to JWST.

Figure~\ref{fig:disks} presents the 13.5--17.5 $\mu$m MIRI-MRS
spectrum of one other Class I JOYS source, B1-a-NS. For comparison,
the line-rich Class II disk DF Tau is shown, taken from
\citet{Grant24}. This wavelength range contains several key molecules
that are readily detected in Class II disks due to their $Q-$branches:
CO$_2$, HCN and C$_2$H$_2$, in addition to the pure rotational lines
of H$_2$O. It is clear that these Class I and II sources are very
different, with B1-a showing only cold H$_2$O that is not even
necessarily coming from the inner disk.

Figures~\ref{fig:CO2overview} and \ref{fig:H2Ooverview} include the
spectrum of our example source, the Class 0 protostar Serpens SMM3,
which also shows evidence of a large embedded disk both in ALMA
continuum and CO lines \citep{Tychoniec21}
(Fig.~\ref{fig:SMM3alma}). Like B1-a-NS, SMM3 shows no hints for warm
molecular emission in the 13.5--17.5 $\mu$m range except for CO$_2$,
nor at 22--26 $\mu$m for cold H$_2$O emission. The CO$_2$ emission
appears extended along the outflow (Fig.~\ref{fig:CO2vsC2H2}) and may
therefore not be arising from the inner disk. This example illustrates
the importance of having spatial information at 30 au scales, as
offered by MIRI-MRS.

\subsubsection{Hiding warm molecular emission in young disks}

\begin{figure}[tbh]
\begin{centering}
\includegraphics[width=7.5cm]{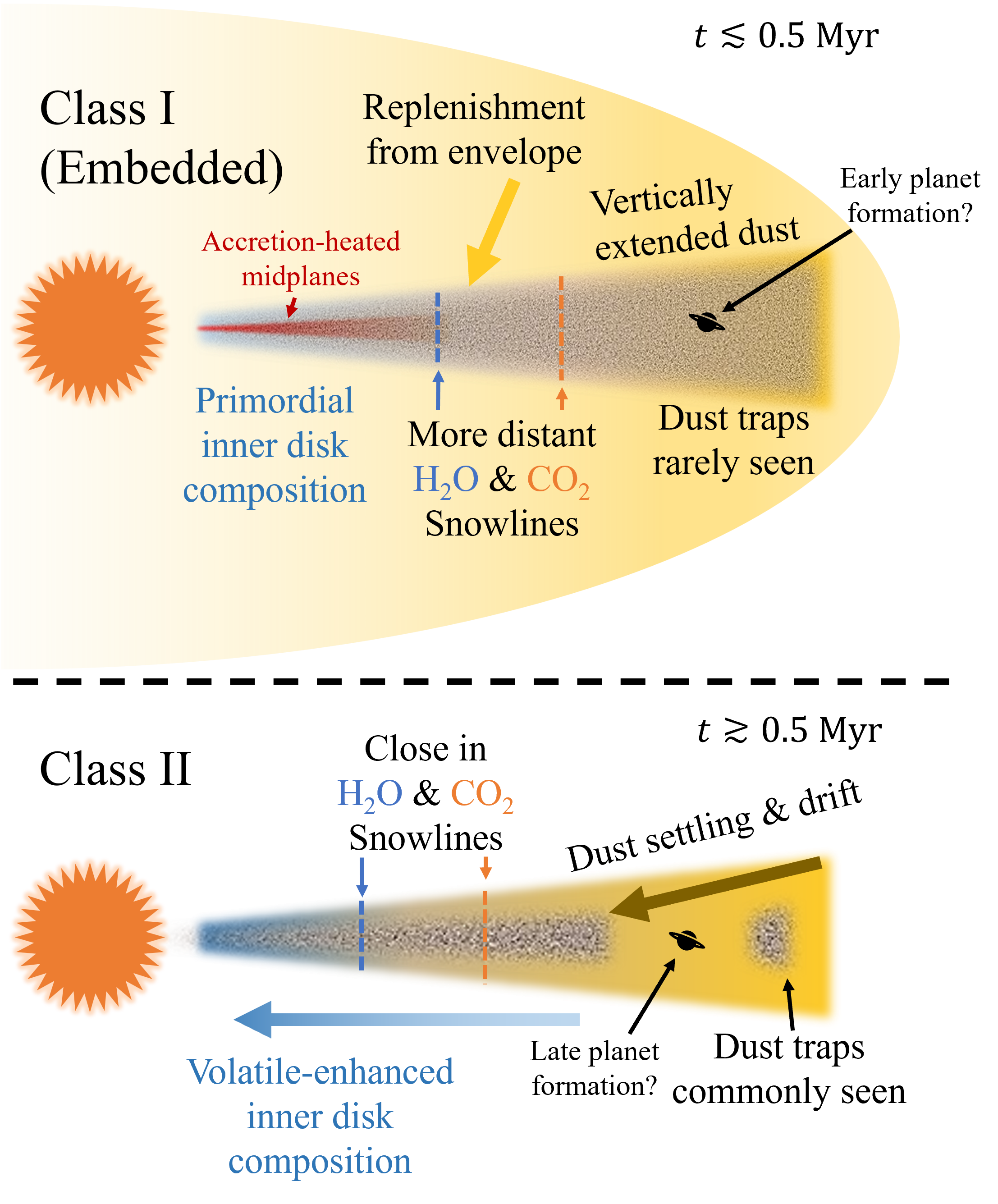}
\caption{Schematic illustrating differences in the physical and chemical
  structures of young Class I ($\sim$0.5 Myr) (top) and more mature
  Class II ($\sim$3 Myr) (bottom) disks.}
\label{fig:diskcartoon}
\end{centering}
\end{figure}

We here discuss possible reasons for the absence of warm molecular
disk emission in sources where a disk is clearly established.
Figure~\ref{fig:diskcartoon} illustrates some differences between
Class I and II disks. Some Class II disks also lack warm molecular
emission, and in those cases the presence of a large inner cavity has
been suggested \citep{Banzatti17,Vlasblom24}. Another related
possibility is that dust traps in the outer disk have locked up water
and other volatiles as ices, preventing the drifting icy pebbles from
reaching the inner disk \citep[e.g.,][]{Najita13,Banzatti20}. However,
the ALMA observations of a sample of representative Class 0 and I
disks, including TMC1A, show little evidence of dust traps down to 5
au resolution \citep{Ohashi23}. Also, even for Class II disks, gaps do
not fully prevent oxygen and water from getting to the inner disk
\citep[e.g.,][]{Perotti23,Schwarz24,Gasman25}.

A more likely explanation is that the inner disk emission is hidden by
small dust grains high up in the disk atmosphere, in any disk
  wind, and in the surrounding envelope. In Class II disks, the
grains have grown and settled to the midplane, as evidenced by the
fact that the analysis of their molecular emission generally invokes
gas/(small) dust ratios of 10000 rather than 100 in the line emitting
region to explain the line strengths and line/continuum ratios
\citep[e.g.,][]{Meijerink09,Bruderer15,Woitke18,Greenwood19,Bosman22H2O}.
Figure~11 of \citet{Vlasblom24} presents a series of thermochemical
disk model spectra of H$_2$O and CO$_2$ in the 13--17 $\mu$m range for
gas/dust ratios of 100, 1000 and 10000 respectively, illustrating that
the molecular emission is weakened by more than an order of magnitude
at normal gas/dust ratios of 100. ALMA observations show clear
evidence of highly settled large dust grains in the outer regions of
Class II disks \citep[e.g.,][]{Villenave20}, but much less in Class I
disks \citep{Villenave23}.

Assuming ISM dust-to-gas ratios, the time taken for dust grains to
grow large enough to settle is of the order of 1000 orbits at any
given radius, which is $\sim$100 kyr for the outer disk. Drift thus
likely begins sometime early on in the Class I phase. However, due to
(i) the initial enrichment of the inner disk with dust from the outer
disk, (ii) the timescale (100s of orbits) for the large grains to
settle and drift, and (iii) the resupply of dust from the envelope,
the dust-to-gas ratio can remain high up to $\sim$0.3--0.4 Myr
\citep[e.g.,][]{Cridland22,Appelgren23}, i.e., for most of the Class I
phase, likely even in the upper layers. Thus the levels of dust
depletion needed to reproduce the line strengths in Class II disks are
likely not achievable during the Class I phase even if drift is active
to supply volatiles.  Indeed, models of drifting icy grains find that
inner disks are not only enhanced in H$_2$O but also in small dust
that may mask the molecular features \citep{Sellek25,Houge25}.

While dust extinction could be an explanation for these MIRI
observations, it remains puzzling that CO infrared disk emission,
which should be similarly affected by dust extinction, is seen for
several Class I sources \citep{Herczeg11}, including TMC1A and TMC1
that are part of JOYS. This CO emission, particularly from the overtone bands, likely arises from the dust-free inner disk gas
\citep{Bosman19CO,Banzatti22}, as has been found from spatially
resolved near-infrared VLTI-Gravity observations of disks around more
massive stars \citep[e.g.,][]{Gravity21Kou}.  At least for the
TMC1A disk, grain growth and likely dust settling has been inferred
\citep{Harsono18}.

Disk wind absorption superposed on disk emission is another option to
minimize the molecular lines at the limited MIRI-MRS spectral
resolution \citep{Herczeg11}: high spectral resolution
($R\sim 100000$) VLT-CRIRES data reveal the presence of strong
blue-shifted disk winds in absorption in $^{12}$CO $v=$1--0 and
$^{13}$CO $v$=1--0 line profiles for many Class I sources. At MIRI-MRS
spectral resolution, the emission and absorption may well cancel out
\citep[see also][]{Banzatti22}. Such superposed absorption would be
particularly important for bands that connect to the vibrational
ground state. However it should not affect the highly excited pure
rotational H$_2$O lines. Also, no C$_2$H$_2$ or HCN has yet been found
associated with winds (see Sect.~\ref{sec:windchemistry}), except perhaps
in the case of Oph IRS 46 \citep{Lahuis06}.  Further observations and
analysis of a much larger sample of low-mass Class I disks, offered
partially by JOYS+ and by approved Cycle 4 programs, is needed to
determine whether the lack of warm molecular emission from embedded
disks is found more commonly.

\subsubsection{Evidence of accretion shocks?}

A final question is whether the JWST data provide any indication of
accretion shocks triggered by infall or streamers onto the disk. ALMA
observations have suggested that sulfur-bearing molecules, most
notably SO and SO$_2$, are good tracers of such shocks
\citep{Sakai14,Artur19,Liu25}. JWST MIRI has detected SO$_2$
ro-vibrational emission at 7.3 $\mu$m for the first time on disk
scales of a Class 0 source, but comparison with ALMA data suggests
that a hot core rather than accretion shock origin is more
likely. Also, SO$_2$ infrared emission is not commonly seen
\citep{vanGelder24overview}. The use of [S I] 25 $\mu$m emission as an
accretion shock tracer warrants further investigation, especially in
Class I sources where it is not associated with the outflow, but is
limited by the low spatial resolution in channel 4. Searching for
local H$_2$ emission hot spots near the disk surface is another
option.

\subsection{High-mass protostars}

Several high-mass protostars have been found to have Keplerian
disks based on high-resolution ALMA data (see
Sect.~\ref{sec:back_disks}), including IRAS 18089 \citep{Beuther07},
although this has not yet been established for all high-mass sources
included in JOYS. Due to their higher accretion rates and
luminosities, the midplanes of disks around high-mass protostars can
be heated to much higher temperatures than the upper disk layers
resulting in a temperature gradient that is decreasing rather than
increasing with height
\citep{DAlessio97,Harsono15,Nazari23HM}. \citet{Barr20} have proposed
that the warm H$_2$O absorption at 7 $\mu$m seen at high spectral
resolution data with SOFIA-EXES and ground-based instruments toward
high-mass protostars actually arises in the accretion-heated inner
midplane of disks. The warm H$_2$O absorption seen with MIRI-MRS
toward IRAS 23385 \citep{Francis24} and IRAS 18089
(Figures.~\ref{fig:IRAS18089spectrum} and \ref{fig:IRAS18089COMfit})
hints that this scenario could also hold for other high-mass
protostars. Inspired by this scenario and using a radiative transfer
model of \citet{Nazari23HM} to characterize the temperature structure
of an accreting massive disk, \citet{vanDishoeck23} suggest that the
cool CO$_2$ 15 $\mu$m emission observed toward IRAS 23385
(Fig.~\ref{fig:IRAS4B}) could arise from the colder surface layers of
the outer disk around one or both of the protostars. Alternative hot
core and outflow scenarios such as discussed in Sect.~\ref{sec:hotcore}
cannot be excluded, however. Combined JWST, ALMA and high-resolution
ground-based data are needed to disentangle the origin of the
mid-infrared emission or absorption. Because of the high extinction, a
pole-on geometry would be more favorable to peer into the inner
disk-like structures. If a disk origin can be established, it would
provide a unique probe of the physical and chemical structure of inner
disks of high-mass protostars.

\section{Conclusions}
\label{sec:conclusions}

The exquisite sensitivity of JWST combined with the much higher
spectral and spatial resolution offered by the MIRI-MRS IFU enable a
new view of low- and high-mass protostars in their most deeply
embedded stages. With this paper, we have presented a range of
illustrative results on a sample of 23 low- and high-mass targets (32
if binaries are included) while addressing their physical and chemical
evolution. A summary of published JOYS and JOYS+ studies to date is
listed in Sect.~\ref{sec:app_backgroundJOYS}. Our main findings are as
follows:

\begin{enumerate}

\item Mid-infrared H I recombination lines, in particular the H I 7--6
  Humphreys $\alpha$ line at 12.37 $\mu$m, provide a powerful tool to measure
  accretion rates onto protostars in the earliest highly extincted
  stages. Inferred accretion rates for low-mass sources show a wide
  range of values, suggesting some sources are in a low-accretion
  quiescent stage. Analysis of one high-mass source indicates
  accretion rates that are high enough to form high-mass stars.
     
\item Jets commonly show a nested physical structure with an ionized
  core traced best by [Fe~II] emission surrounded by a molecular layer
  traced by the higher-$J$ H$_2$ pure rotational lines. This structure
  is seen across the entire protostellar mass range. For some low-mass
  sources, the molecular gas provides the main thrust in the earliest
  Class 0 stages, but this phase appears to be short-lived. By the
  Class I stage, the outflows become fully atomic and ionized. Our
  results emphasize the need of high spatial resolution data to
  determine mass loss rates.

   \item Wide-angle winds traced by the lower-$J$ pure-rotational
     H$_2$ lines are commonly found inside the outflow cavity walls,
     which can be outlined by scattered light and by entrained cold gas in CO
     millimeter lines. Their temperatures of a few hundred to 2000 K
     are lower than those of the jets, with gradients seen
     perpendicular to the jet axis. The main heating mechanisms of the
     gas may vary from position to position and warrant further
     investigation. 
     
   \item A large variety of atomic and ionic lines from refractory,
     semi-refractory, and volatile elements have been detected and
     imaged in most sources. Their different morphologies are related
     to different levels of elemental depletions, local shock
     conditions, and (in the case of noble gases)
     photoionization. Also, 20 $\mu$m continuum emission due to heated
     dust has been found in a few very young Class 0 sources. Together
     with the atomic lines, this allows dust launching in jets and
     grain destruction in shocks to be quantified.

   \item Hot core and dense shock chemistry models can be tested with
     unprecedented detail through observations of simple key molecules
     such as H$_2$O, CO$_2$, HCN, C$_2$H$_2$, OH, and SO$_2$ in both low-
     and high-mass sources. Derivation of their abundances requires
     use of non-LTE excitation models including infrared pumping. Ice
     sublimation followed by high-temperature gas phase chemistry,
     perhaps stimulated by X-rays, explains the data for some sources,
     whereas dense jet shocks are more appropriate for others. No
     clear indication for accretion shocks onto low-mass disks has yet
     been found. The case of NGC 1333 IRAS4B provides a unique data
     set to test dense shock chemistry models, including the
     chemical processes that lead to bright hot C$_2$H$_2$ emission.

   \item Ice studies have made a significant jump forward thanks to
     the high-quality MIRI-MRS data in the 5--10 $\mu$m range. Several
     complex organic molecules and ions (salts) have been found with
     comparable abundances in low- and high-mass protostars. We report
     the second detection of HDO ice in a low-mass protostar, with
     HDO/H$_2$O ice ratios that point to inheritance from cloud to
     disk. Though O$_2$ ice remains elusive, JWST provides improved
     limits.

   \item Young disks in the embedded stage do not show the same forest
     of warm molecular lines as more evolved Class II disks do. A
     likely explanation is that small dust has not yet grown and
     settled and is being replenished by radial drift, thereby
     obscuring a clear view to their inner few au. Extinction from
     dust in a disk wind or envelope may also contribute to this.

\end{enumerate}

This paper provides a first comprehensive look into JWST
MIRI-MRS studies of a significant sample of protostars. So far, many
similarities between low- and high-mass sources have been found. Larger
samples across evolutionary stages and stellar mass ranges, such as
offered by JOYS+ and by other currently ongoing and future programs
such as the High angular-resolution observations of Emergence in
Filamentary Environments (HEFE; P.I. T.\ Megeath), will allow much more
robust trends to be identified. In particular, the low-mass sample in
JOYS contains only four (six with binaries) Class I sources and will be
more than doubled in JOYS+. HEFE will add many more Class 0 sources of
intermediate luminosity in a highly UV irradiated environment. Such large
samples will also help in interpreting similar spectra seen in
extragalactic sources.

More robust calibrations of mass loss and mass accretion rates need to
be developed and quantified. Studies of binaries and/or clustering will be
possible at 0.5--5$''$ separation with the JOYS+ sample. The combined
NIRSpec + MIRI study will increase the available diagnostics,
which includes providing higher excitation [Fe II] and H$_2$ lines that probe hotter
gas. The results from such studies ultimately need to be made
consistent with the physical components of warm and hot gas found in
CO and H$_2$O data with {\it Herschel}. The high spatial and spectral
resolution of NIRSpec also allows key ices to be detected and ice
mapping against the scattered light continuum to be conducted. The NIRCam
imaging and spectral mapping can do so as well. When combined with
ALMA and other instruments, JWST will truly help lift the veil of
key components of the youngest stages of stellar and planetary birth.

\begin{acknowledgements}
  The authors thank the referee for their constructive comments and
  the entire JOYS+ team, Adwin Boogert, John Black, Brunella Nisini,
  Teresa Giannini, and Christoffel Waelkens for useful
  discussions. Collaboration with Catherine Walsh on hot core chemical
  models is much appreciated.  This work is based on observations made
  with the NASA/ESA/CSA James Webb Space Telescope. The data were
  obtained from the Mikulski Archive for Space Telescopes at the Space
  Telescope Science Institute, which is operated by the Association of
  Universities for Research in Astronomy, Inc., under NASA contract
  NAS 5-03127 for JWST. These observations are associated with
  programs 1290 and 1257.  The following National and International
  Funding Agencies funded and supported the MIRI development: NASA;
  ESA; Belgian Science Policy Office (BELSPO); Centre Nationale
  d’\'Etudes Spatiales (CNES); Danish National Space Centre; Deutsches
  Zentrum fur Luft und Raumfahrt (DLR); Enterprise Ireland; Ministerio
  De Economi\'a y Competividad; The Netherlands Research School for
  Astronomy (NOVA); The Netherlands Organisation for Scientific
  Research (NWO); Science and Technology Facilities Council; Swiss
  Space Office; Swedish National Space Agency; and UK Space
  Agency. Astrochemistry in Leiden is supported by funding from the
  European Research Council (ERC) under the European Union’s Horizon
  2020 research and innovation program (grant agreement No. 291141
  MOLDISK), and by NOVA and NWO through TOP-1 grant 614.001.751 and
  its Dutch Astrochemistry Program (DANII). The present work is
  closely connected to ongoing research within InterCat, the Center
  for Interstellar Catalysis located in Aarhus, Denmark. TR
  acknowledges support from ERC grant no. 743029 EASY and TH from ERC
  grant no. 832428 Origins. ACG acknowledges support from PRIN-MUR
  2022 20228JPA3A “The path to star and planet formation in the JWST
  era (PATH)” funded by NextGeneration EU and by INAF-GoG 2022
  “NIR-dark Accretion Outbursts in Massive Young stellar objects
  (NAOMY)” and Large Grant INAF 2022 “YSOs Outflows, Disks and
  Accretion: toward a global framework for the evolution of planet
  forming systems (YODA)”. KJ and G\"O acknowledge support from the
  Swedish National Space Agency (SNSA), and LC from grant
  PIB2021-127718NB-100 from the Spanish Ministry of Science and
  Innovation/State Agency of Research
  MCIN/AEI/10.13039/50110001103. PJK acknowledges financial support
  from the Research Ireland Pathway programme under Grant Number
  21/PATH-S/9360. VJML’s research is supported by an appointment to
  the NASA Postdoctoral Program at the NASA Ames Research Center,
  administered by Oak Ridge Associated Universities under contract
  with NASA. PN acknowledges support from the ESO Fellowship and IAU
  Gruber Foundation Fellowship programs.  SN is grateful for support
  from Grants-in-Aid for JSPS (Japan Society for the Promotion of
  Science) Fellows Grant Number JP23KJ0329, MEXT/JSPS Grants-in-Aid
  for Scientific Research (KAKENHI) Grant Numbers JP23K13155 and
  JP24K00674, and Start-up Research Grant as one of the University of
  Tokyo Excellent Young Researcher 2024.  GP gratefully acknowledges
  support from the Max Planck Society and from the Carlsberg
  Foundation, grant CF23-0481. This paper makes use of the following
  ALMA data: ADS/JAO.ALMA$\#$ALMA 2015.1.00354.S, ADS/JAO.ALMA$\#$ALMA
  2017.1.01350.S, ADS/JAO.ALMA$\#$ALMA 2019.1.00931.S. ALMA is a
  partnership of ESO (representing its member states), NSF (USA) and
  NINS (Japan), together with NRC (Canada), NSTC and ASIAA (Taiwan),
  and KASI (Republic of Korea), in cooperation with the Republic of
  Chile. The Joint ALMA Observatory is operated by ESO, AUI/NRAO and
  NAOJ.
\end{acknowledgements}

\bibliographystyle{aa}

\bibliography{biblio_evd.bib}

\newpage

\begin{appendix}

\section{Source positions}
\label{sec:app_sources}

Table~\ref{tab:sources} contains the coordinates and properties of the
JOYS sources. See also Table A.1 in \citet{vanGelder24overview} for
more information on the low-mass sources. Table~\ref{tab:sources2}
contains more information on the observational set-up.

\begin{table*}[h!]
    \centering
    \caption{Main MIRI-MRS observational details of the sample of JOYS sources in the PID 1290 JWST GTO program.}
    \label{tab:sources2}
{\scriptsize
    \begin{tabular}{llllllllcc}
      \hline\hline
    ID & Archive name & Sources & RA & Dec & Dith & Exposure time & Background & Mosaic \\
     & & & J2000 & J2000 & &  (s)   & & \\ 
    \hline
    1 & B1b-S & -    &  03:33:17.953 & +31:09:31.49 & 2 & 1005, 1005, 1005 & - & 1$\times$1 \\
    2 & B1b   & B1-b & 03:33:20.341 & +31:07:21.36 & 2 & 200, 200, 200 & B1b-S & 1$\times$1 \\
    3 & B1c-1 & B1-c & 03:33:17.953 & +31:09:31.49 & 4 & 2009, 4029, 2009 & B1b-S & 1$\times$1 \\
    4 & B1c-2 & -    & 03:33:18.211 & +31:09:28.63 & 2 & 200, 200, 200 & B1b-S & 1$\times$1 \\
    5 & B1a & B1-a-1, 2 & 03:33:16.669 & +31:07:54.90 & 2 & 200, 200, 200 & B1b-S & 1$\times$1 \\
    6 & IRAS4B & IRAS4B & 03:29:12.012 & +31:13:07.07 & 2 & 200, 200, 200 & B1b-S & 1$\times$1 \\
    7 & IRAS4B-b1 & - & 03:29:12.040 & +31:13:01.35 & 2 & 200, 200, 200 & B1b-S & 1$\times$1 \\
    8 & IRAS4A-b1 & - & 03:29:10.595 & +31:13:19.40 & 2 & 200, 200, 200 & B1b-S & 1$\times$1  &  \\
    9 & L1448mm & L1448-mm &03:25:38.761 & +30:44:08.16 & 2 & 200, 200, 200  & B1b-S &  2$\times$3 \\
    10 & Emb-8 & Per-emb 8 & 03:44:43.982 & +32:01:35.21  & 2 & 200, 200, 200  & B1b-S &  1$\times$1 \\
    \hline
    11 & Dark-Taurus &  - &  04:39:32.290 & +26:05:14.50 & 1 & 200, 200, 200  & - & 1$\times$1 \\
    12 & L1527-1 & L1527 IRS &  04:39:53.955 & +26:03:09.56 & 2 &1000, 1000, 1000 & Dark-Taurus & 1$\times$1 \\
    13 & L1527-2 &  - &  04:39:54.245 & +26:03:11.67 & 2 &  200, 200, 200 & Dark-Taurus & 1$\times$1 \\
    14 & TMC1A &  TMC1-A &  04:39:35.203 & +25:41:45.17 & 2 &  216, 216, 216 & Dark-Taurus & 1$\times$1 \\
    15 & TMC1 &  TMC1-W,E &  04:41:12.700 & +25:46:34.80 & 2 &  200, 200, 200 & Dark-Taurus & 1$\times$1\\
    \hline
    16 & Dark-Serpens &  - & 18:29:48.250 & +01:14:26.20 & 2 & 200, 200, 200  & - & 1$\times$1 \\
    17 & SVS4-5 &  SVS 4-5 & 18:29:57.630 & +01:13:00.20 & 2 & 605, 605, 605  & Dark-Serpens & 1$\times$1 \\
    18 & SerpSMM3 &  Ser-SMM3 & 18:29:59.302 & +01:14:00.97 & 2 &  200, 200, 200  & Dark-Serpens & 1$\times$1 \\
    19 & SerpEmb8N &  - & 18:29:49.269 & +01:16:52.62 & 2 &  200, 200, 200  & Dark-Serpens & 1$\times$1 \\
    20 & SerpS68N &  Ser-S68N-N, S & 18:29:48.074 & +01:16:43.80 & 2 &  200, 200, 200  & Dark-Serpens & 1$\times$1 \\
    21 & SerpSMM1-2 & Ser-SMM1-b1, b2  & 18:29:49.602 & +01:15:21.90 & 2 &  200, 200, 200  & Dark-Serpens & 1$\times$1 \\
    22 & SerpSMM1-1 & Ser-SMM1-a & 18:29:49.874 & +01:15:19.79 & 4 &  200, 200, 200  & Dark-Serpens & 1$\times$1 \\
    \hline
    23 & Dark-BHR71 & - &12:01:30:720 & -65:08:44.60 & 2 & 200, 200, 200 & - & 1$\times$1 \\
    24 & BHR71-IRS2 & BHR71-IRS2 & 12:01:34.000 & -65:08:47.00 & 2 & 200, 200, 200 &  Dark-BHR71 & 1$\times$1 \\
      25 & BHR71-IRS1 & BHR71-IRS1 &12:01:36.548 & -65:08:53.59 & 2 & 200, 200, 200 &  Dark-BHR71 & 3$\times$3 \\
      \hline
  26     & G28IRS2 & G28IRS2 & 18:42:51.967 & -3:59:53.985 & 2 & 200, 200, 200
& - & 1$\times$1 \\
  27     & G28P1 & A, B & 18:42:50.679 & -4:03:14.087 & 2 & 799, 799, 799& - &
2$\times$2 \\
  28     & G28S & - & 18:42:46.414 & -4:04:15.404 & 2 & 200, 200, 200 & - &
1$\times$1 \\ \hline
  29     & Dark-IRAS23385 & - & 23:41:11.033 & 61:10:28.928 & 2 & 200, 200, 200
 & - & 1$\times$1 \\
  30     & IRAS23385& A, B & 23:40:54.503 & 61:10:27.846 & 2 & 800, 800, 800 &
  Dark-IRAS23385& 2$\times$2 \\ \hline
  31    & Dark-IRAS18089 & - & 18:11:44.112 & -17:31:03.047 & 2 & 200, 200, 200& - & 1$\times$1 \\
      32   & IRAS18089 & A, B, C & 18:11:51.447 & -17:31:27.285 & 2 & 400, 400,
400 & Dark-IRAS18089 & 2$\times$1 \\ \hline
  33     & Dark-G31 & - & 18:47:32.382 & -1:13:44.317 & 2 & 200, 200, 200 & -
& 1$\times$1 \\
  34     & G31 & G31 & 18:47:34.308 & -1:12:46.000 & 2 & 200, 200, 200 &
Dark-G31 & 1$\times$1 \\
    \hline
    \end{tabular}
    }
    \tablefoot{The coordinates are the center position of the
      observation, not the positions of the protostars. The exposure
      time is the exposure time divided over the A,B,C gratings. Dith
      stands for Dithers.}

  \end{table*}
\renewcommand{\arraystretch}{1.0}

\section{Scientific background}
\label{sec:app_background}

This section presents background information on the science cases
addressed in this paper as well as the motivation for using JWST-MIRI
in more detail than described in Sect.~\ref{sec:background}. To avoid
interrupting the flow of the results in Sections 4--8, this
information is contained here in a single section.  See also
Sect.~\ref{sec:components} for terminology.

\subsection{Protostellar accretion}
\label{sec:back_accretion}

Most of the protostellar accretion occurs in the earliest, deeply
embedded stages \citep[e.g.,][]{Lada84,Hartmann16} but there are very
few direct measurements of the rates at which this process occurs. The
H I recombination lines are thought to be good tracers if they are
produced primarily in the accretion column onto the forming star
(magnetospheric accretion).  Most notably, the H I 7--6 Humphries
$\alpha$ line at 12.37 $\mu$m has been proposed as an accretion tracer
for highly extincted sources for which more commonly used shorter
wavelength tracers like Brackett $\gamma$ cannot be observed
\citep{Rigliaco15}. However, H I recombination lines may also be
produced in jet/wind shocks or circumstellar photoionized gas. The
high spatial resolution imaging of the H I lines offered by MIRI is
important to disentangle these processes. \citet{Beuther23} provide
the first JOYS application of H I for the IRAS 23385+6053 high-mass
protostar. We note that these disk-to-protostar accretion rates should
not be confused with the cloud-to-envelope-to-disk gas infall rates
measured on larger scales \citep{Myers96,Mottram13,Wyrowski16} and are
not necessarily the same \citep{Dunham14}.

Accretion is known to be an episodic phenomenon, with accretion rates
varying by orders of magnitude during the main accretion phase
\citep{Hartmann96,Frank14,Fischer23}. Since ejection and accretion
processes are closely linked, these episodic events have also
historically been traced through ``bullets'' or ``knots'' in the jets
away from the central source, where they are more easily observed at
near-infrared wavelengths \citep{Reipurth97,Bally16,Lee20}.  Models of
outflow propagation show that many of these ``knots'' can naturally be
explained by variations in the jet ejection velocity, with high
velocity blobs within the jet producing shocks as they catch up with
the slower moving flow within the jet
\citep{Raga93,Bonito10,Rohde19,Shang23}.
This hypothesis of linked accretion and ejection is supported by a
handful of observed cases where an outburst in a protostellar system
has been followed by the appearance of a new knot
\citep{Ellerbroek14,Fedriani23,Cesaroni25}.  With its high angular
resolution over a wide area, JWST NIRCam offers new opportunities to
survey these knots closer to the protostar and in more extincted
regions than before.

Another opportunity for JWST is that of time domain studies. The JWST
NIRCam study of the HH 211 outflow by \citet{Ray23} has demonstrated
the use of accurate imaging over a 20 year period for determining the
proper motions and tangential velocity of the jet knots. Another
example is provided by the change in the mid-infrared SED of some protostars
like EC 53 during the outburst cycle as material builds up in the
inner disk \citep[e.g.,][]{Connelley18,Baek20,Francis22}. This could
cause molecular lines to shift from emission to absorption as the disk
midplane is heated by accretion and thus serve as a valuable
diagnostic \citep{McClure25}. Similarly, some mid-infrared ice bands are
excellent signatures of thermal processing, most notably the
$^{12}$CO$_2$ 15 $\mu$m band
\citep{Ehrenfreund98,Ehrenfreund99,Pontoppidan08,Kim12} and now also
the $^{13}$CO$_2$ 4.38 $\mu$m band observed with JWST as part of the
IPA program \citep{Brunken24}.

\subsection{Protostellar jets, winds, and outflows}
\label{sec:back_outflows}

Jets and outflows are seen from young stars up to $\sim$2 Myr age,
spanning all protostellar phases from Class 0 to Class II, and for
low- and high-mass protostars, and are one of the most striking
phenomena in astrophysics \citep[see reviews
by][]{Reipurth01,Ray07,Arce07,Frank14,Bally16,Lee20}. Not only are they
supersonic and highly collimated but they can stretch for several
parsecs, even beyond the boundaries of their molecular cloud.
Since mass loss and accretion rates are closely coupled
\citep{Beuther02}, jets and winds are most powerful in the earliest
evolutionary stage and can be used as a proxy for past activity (see
Sect.~\ref{sec:onset}). In fact, estimates suggest that $\sim$10\% of the
accreted mass is ultimately ejected, a number that may change with
evolutionary state and be as high as 100\% in some MHD disk-wind
models at the earliest stages \citep{Pascucci23}.
As the statistically estimated lifetime of the low-mass Class 0 phase
is only about $5\times 10^{\rm 4}$ years \citep{Evans09,Kristensen18},
associated outflows in the deeply embedded phase are not as big in
their extent as those from more evolved sources. They are, however, a
full trace of the activity since the birth of the protostar.

Despite their clear importance, there are still many questions about
the physics of the jets, winds, and outflows from the youngest
protostars. In large part this is because they are deeply embedded,
particularly in the earliest stages and in the region close to the
source.  Protostellar jets have so far mainly been probed in the
near-infrared at larger distances from the source via the vibrationally
excited H$_2$ lines tracing temperatures in excess of 1000\,K and the
near-infrared forbidden [Fe II] lines tracing hot ionized gas
\citep[e.g.,][]{Nisini05,Podio06,Giannini11}. In contrast, millimeter
images of CO pure rotational lines commonly trace the slower moving
entrained cold ambient gas, whereas SiO and H$_2$O lines reveal higher
velocity jets in some but not all sources
\citep[e.g.,][]{Guilloteau92,Bachiller99,Kristensen11,Podio21,Bally16}. In
a few cases, also disk winds, i.e., wider angle (poorly collimated)
and slower moving material that is launched from a larger range of
disk radii than the jets, are found
\citep{Lee17,Tabone17,Nazari24wind}. Figure~\ref{fig:protostar-cartoon}
illustrates these components.

  The near-infrared and millimeter are two extremes in wavelength with only
  limited information on the intermediate temperature regime.  To
  observe the much more highly collimated underlying jet propelling
  forward these outflows close to the source at velocities close to
  the gravitational escape velocity ($\approx$\ 100--300
  km s$^{-1}$ for low-mass sources), they have to be studied in the
  mid- and far-infrared.  The usefulness of mid-infrared observations of
  outflows from Class 0 sources has been demonstrated using {\it
    Spitzer-IRS}
  \citep[e.g.,][]{Neufeld06,Neufeld09,Melnick08,Maret09,Lahuis10}.  Although
  they are of poor spectral and spatial resolution compared to MIRI,
  such studies show not only that an underlying atomic component is
  present but suggest that the mass loss rate in such young jets is
  much higher than those from less embedded sources of equivalent mass
  \citep{Dionatos09}.

  Additional information on the physical and chemical structure of
  jets and outflows has been provided by {\it Herschel} far-infrared
  spectroscopic studies. Like {\it Spitzer}-IRS, the spatial
  resolution is poor, typically 10--20$''$, thus probing scales of a
  few 1000 au. Still, surveys of nearly 100 low- and high-mass
  protostars with the HIFI and PACS instruments have provided insight
  into the warm (few hundred K--1000 K) gas using H$_2$O, OH and CO
  pure rotational lines up to very high $J_u$=49 ($E_{\rm up}=6457$
  K). Two different physical components can be distinguished that are
  clearly different in temperature and in spatial and velocity
  distribution from the entrained outflow gas probed by low-$J$ CO
  lines
  \citep{Kristensen10,vanKempen10,Herczeg12,Karska13,Manoj13,Nisini13,Tafalla13,Green16,Mottram17,Kristensen17b,Karska18,Dionatos18,Dionatos20,vanDishoeck21}.

  The [O I] 63 $\mu$m line has also been surveyed with {\it Herschel},
  and is commonly associated with the jets powering the outflows at
  speeds of more than 90 km s$^{-1}$ with respect to that of the
  source \citep{vanKempen10,Nisini15,Dionatos17,Karska18}. As the main
  coolant of the gas \citep{Hollenbach85}, it can serve as a direct
  measure of mass loss rates. We note, however, that optical depth,
  self-absorption or absorption from cold foreground clouds can affect
  the results inferred from spectrally unresolved [O I] line profiles,
  as indicated by spectrally resolved [O I] data with SOFIA
  \citep[e.g.,][]{Schneider18,Guevara24,Lis24}.
  Comparison of mass flux rates from [O I] and CO suggests that this
  atomic gas is not the dominant driver of the entrained outflow gas
  in the earliest stages: the jets in the Class 0 phase are mostly
  molecular.  However, toward the Class I stage, the jets become
  mostly atomic \citep{Nisini15}. This scenario can now be tested with
  JWST using its diagnostic lines and spectral imaging to
  compare with the [O I] results both qualitatively and quantitatively.

Finally, the abundances of H$_2$O, OH and other hydrides and the
distribution of $^{13}$CO 6--5 emission point to UV fields up to
$10^2-10^3$ times the general interstellar radiation field in the
outflow cavity walls \citep{Spaans95,Yildiz12,Benz16,vanDishoeck21}.
In particular, the H$_2$O abundance in outflows is low due to UV
photodissociation, of order $10^{-7}-10^{-6}$
\citep{Tafalla13,Karska18}.
The limited spatial resolution of {\it Herschel} and {\it Spitzer},
however, prevented imaging any of these processes
at the base of the outflow in the inner few 100 au.

JWST now provides imaging of a wealth of mid-infrared lines at 10--100
times higher spatial resolution than was possible with {\it
  Spitzer}. Their diagnostic potential for studying outflows is
summarized in Table~\ref{tab:lines}. Of prime importance are the [Fe
II] and lines of other refractory elements tracing the jets, together
with the H$_2$ lines imaging both the jets and the wider-angle winds.
Some kinematic data down to $\sim$10 km s$^{-1}$ accuracy can be
obtained. [Ne II] has been found to trace both jets and photoionized
gas in embedded sources using {\it Spitzer} data
\citep{Lahuis10,Guedel10}, which can now be disentangled through
direct imaging.  [Ne II] versus [Ne III] can provide constraints on
the ionizing radiation field \citep[e.g.,][]{Espaillat13}.  Comparison
of all of these lines with shock models such as those by
\citet{Hollenbach89} and from the Paris-Durham code
\citep[e.g.,][]{Godard19,Lehmann22,Kristensen23} then provide
constraints on shock velocity and pre-shock densities. The OH
  mid-infrared `` prompt emission'' pure rotational lines are proven
  to be powerful tracers of UV irradiation in deeply embedded regions
  \citep[e.g.,][]{Neufeld24}.

Even though JWST will not be able to resolve the launching zone of the
protostellar jets (the innermost few au), it can investigate the
spatial structure and stratification of the jets in their close
environment and constrain the launching mechanisms.  Also, the
structures seen with JWST on small scales can ultimately be linked
with the physical components inferred from data at other wavelengths
on larger scales.  Initial JWST papers on outflows using JOYS data of
the HH 211 Class 0 outflow \citep{Caratti24} and the Class I TMC1
binary source \citep{Tychoniec24} indeed reveal a nested jet structure
consisting of an ionized core traced by [Fe II] with a molecular layer
traced by high-$J$ H$_2$ surrounding it, with a wider angle wind seen
in low-$J$ H$_2$ lines \citep[see also][]{Federman24,Delabrosse24}. By
studying the temperature and chemical structure of such winds,
ultimately constraints on their origin may be obtained, in particular
whether they are MHD winds \citep[e.g.,][]{Pudritz86,Romanova05} or
photoevaporative winds \citep[e.g.,][]{Ercolano17}.

Another question that MIRI can help answer is whether dust is
launched inside jets from Class 0 sources.  This, in turn, provides
constraints on the launching radius of the jet, in particular whether
it is inside or outside the dust sublimation radius ($\sim$1400 K,
typically located at $<$0.1 au for a low-mass protostar,
\citealt{DAlessio97}). The presence of dust in jets can either be
imaged directly through its mid-infrared continuum, or it can be
inferred more indirectly from line ratios of refractory to
non-refractory species to determine their depletion. The latter method
has been used for Class II sources using near-infrared lines
\citep[e.g.,][]{Nisini02,Podio06}.  While inferred depletion values
are not as high as in the general ISM, the gas in optically visible
jets has not attained solar abundances, at least close to Class II
sources \citep[e.g.,][]{Giannini15}.  Thus, this suggests that there
is dust in the jet and that it is only gradually destroyed via the
mild internal jet shocks, as expected from models of grain destruction
in dense gas \citep[e.g.,][]{Tielens94,Schilke97,Gusdorf08,Guillet09}.
This raises the obvious question whether dust is even more important
in jets from Class 0 sources than in Class I sources.  MIRI covers
many atomic lines of both refractory and less refractory elements,
most notably from [Fe I], [Fe II] and [S I], that can address this
issue \citep[e.g.,][]{Anderson13}. The HH 211 MRS map by
\citet{Caratti24} directly detects continuum dust emission associated
with the base of the jet directly suggesting that dust is indeed
launched in jets. It also indicates modest depletions of refractory
elements suggesting that not all grains are destroyed. A detailed JOYS
case is provided by the BHR 71 IRS1 jet and outflow (Tychoniec et al.,
in prep.).

Some evidence of dust in outflows can also be found on larger scales
from pre-JWST mid-infrared continuum ground-based and {\it Spitzer}
images \citep[e.g.,][]{Smith05,Velusamy07} and in absorption against a
background nebula \citep{Reiter17}. On these scales, dust can also be
mixed and entrained into the jet and outflow from the surrounding
cloud, especially in regions near the bow shock. This is shown for the
HH 211 case at larger distances from the source \citep{Caratti24}.

High-mass protostars have powerful outflows detected in CO millimeter
lines, with the CO momentum flux increasing strongly with source
luminosity and envelope mass
\citep{Bontemps96,Beuther02,Mottram17}. These high-mass flows have
been much less studied in other diagnostics because of their high
extinctions, although several jets that are likely driving such
outflows have been detected at near-infrared wavelengths (mostly in
H$_2$) \citep[e.g.,][]{Varricatt10,Lee13,Caratti15}.  Many of the IRDC
regions exhibit so-called green fuzzies or extended green objects
which are thought to be caused by shock-excited H$_2$ emission in the
4.6\,$\mu$m {\it Spitzer} band (e.g.,
\citealt{Noriega04,Beuther05,Cyganowski08}). JWST NIRCam and NIRSpec
data now demonstrate that CO ro-vibrational emission can dominate this
filter as well \citep{Ray23}. MIRI provides a wealth of mid-infrared
diagnostics to determine the physical structure of high-mass sources,
as demonstrated by initial JOYS results
\citep{Beuther23,Gieser23}. Moreover, such data allow one to quantify
the relative importance of shocks versus UV radiation in the
``feedback'' of a high-mass protostar on its immediate surroundings.

Finally, lines of HD starting at $R$(4) ($E_{\rm up}=1895$~K) are also
covered and provide together with H$_2$ constraints on [D]/[H]
\citep[e.g.,][]{Bertoldi99,Wright99,Yuan12}. One puzzle is the
depletion of deuterium in denser clouds \citep{Linsky06} compared with
that seen in the local diffuse interstellar medium. It has been
suggested that carbonaceous grains can lock up significant deuterium
\citep[e.g.,][]{Draine06D}; if so, then this deuterium should be
released by strong shocks that destroy grains. The power of JWST MIRI
to address this question has recently been demonstrated using JOYS
data of both low- and high-mass protostars by
\citet{Francis25HD}. Line images of HD in the nearby HH 211 outflow
show strong correlations with shock tracers such as [Fe II] and [S I],
but HD/H$_2$ varies by a factor of four without any clear trends with
shock conditions. This is consistent with the conclusion of
\citet{Caratti24} that not all grains have been destroyed. The
detection of HD in the high-mass IRAS23385+6035 source provides the
first measurement of HD/H$_2$ in the outer Galaxy, at a galactocentric
radius of 11 kpc. It falls significantly below the predicted trends from
galactic chemical evolution models with radius
\citep[e.g.,][]{vandeVoort18}, suggesting that there is a significant
reservoir of deuterium (such as that in grains) that is not measured
by HD.

\subsection{Hot cores and dense molecular shocks}
\label{sec:back_hotcores}

Protostars passively heat the gas in their surroundings through
gas-dust collisions to a range of temperatures depending on distance
from the source. The warm ($T>100$~K) inner parts of the collapsing
envelopes are called ``hot cores.'' The term ``passively heated''
distinguishes these hot cores from shock-heated or UV-heated gas.

Millimeter observations have found a rich chemistry to be associated
with the hot cores of deeply embedded protostellar sources, from
simple to rather complex organic molecules (COMs), i.e.,
carbon-containing molecules with at least 6 atoms \citep[see reviews
by][]{Herbst09,Jorgensen20,Ceccarelli23}.
This section is focused on simple molecules other than H$_2$ that are
detected by mid-infrared emission or absorption lines. This includes
molecules such as CO$_2$, CH$_4$ and C$_2$H$_2$ that cannot be studied
by millimeter techniques because they do not have a permanent dipole
moment, but that have strong dipole-allowed vibration-rotation
transitions. Some of them, most notably CO$_2$ and CH$_4$, are major
components of interstellar ices \citep{Boogert15}, but they cannot be formed
in large amounts in cold gas. Therefore, they are excellent probes of
ice sublimation in hot cores. Other molecules, most notably H$_2$O,
can have contributions from both ice sublimation or sputtering as well
as high temperature gas-phase chemistry in shocks \citep[see][for
overview]{vanDishoeck21}. Sect.~\ref{sec:irdiagnostics} summarizes
other simple molecules that can be studied at mid-infrared
wavelengths.

Several of these molecules were first detected at infrared wavelengths
by pioneering high spectral resolution ($R\sim 10^5$) ground-based
absorption line observations of the brightest high-mass protostars
\citep{Evans91,Lacy89,Lacy91,Knez09}.  The ISO-SWS and subsequently
{\it Spitzer} detected mid-infrared lines of gaseous CO, H$_2$O,
CO$_2$, C$_2$H$_2$, HCN, CH$_4$ and SO$_2$ in a variety of mostly
high-mass protostellar sources both in absorption and emission, albeit
at much lower spectral resolution than the ground-based data
\citep[e.g.,][]{Helmich96,vanDishoeck96,Lahuis00,Boogert04,Boonman03co2,Sonnentrucker07}
(see also review by \citealt{vanDishoeck04}). More recently, high
spectral resolution absorption line observations were obtained using
EXES on SOFIA and CRIRES on the ESO-VLT but are again limited to the
brightest high-mass sources
\citep[e.g.,][]{Dungee18,Barr20,Indriolo20,Barr22,Li23}. The CS
molecule was first detected at infrared wavelengths by
\citet{Barr18}. Based on their excitation temperatures and line
profiles or velocity offsets, the ground- and space-based studies have
suggested a variety of origins of these molecules: hot cores,
outflows, and shocks, and even the midplanes of disks around high-mass
protostars that are heated by the accretion luminosity.

Notably, JWST can contribute to answering these questions in a variety of
ways. First, it has the sensitivity to observe nearby low-mass Class 0
protostars. Its high spatial resolution allows
emission lines to be located and to be associated with different
physical components.  Second, its higher spectral resolution compared
with {\it Spitzer} boosts line/continuum ratios and enables many more
detections of molecules in emission, especially in the 5--10 $\mu$m
range. At $R\sim$3000, MIRI also has enough spectral resolution to
detect the gas-phase lines of several important molecules in
absorption (e.g., H$_2$O, CH$_4$, CO$_2$, C$_2$H$_2$, HCN)  and
distinguish them from the nearby much broader ice bands
\citep{Boogert98,Dartois98,vanDishoeck04}.  These pencil-beam
absorption line-of-sight data provide a unique method to peer deep
into the hot cores located close to the protostars compared with
beam-diluted emission data at millimeter wavelengths
\citep{Boonman01}.  Third, the observation of many individual $R$- and
$P$-branch lines as well as spectrally resolving the $Q$-branches,
plus detecting $^{13}$C isotopologs and hot bands (i.e., transitions
between vibrationally excited states), allows for more accurate
constraints on excitation temperatures and optical depth of the lines
\citep{vanGelder24overview}. Also, MIRI has the ability to
disentangle many of the H$_2$O pure rotational line blends
\citep{Banzatti23,Banzatti25}.

The JOYS overview of molecular lines in low-mass protostars at the
source position by \citet{vanGelder24overview} and that in the
high-mass protostar IRAS23385 by \citet{Francis24} highlight all of
these aspects.  They can test hot core models such as those by
\citet{Doty02,Notsu21} that include both ice sublimation as well as
high temperature gas-phase chemistry.  The SO$_2$ mid-infrared
emission detected by \citet{vanGelder24} underlines the importance of
taking infrared pumping into account in the excitation. With JWST,
also extragalactic hot cores can be studied \citep{Buiten25}, greatly
improving on early {\it Spitzer} infrared spectroscopic studies of AGN
\citep{Lahuis07AGN}.

At outflow positions offset from the source, the jet shocks plow into
dense gas that can produce not just atomic but also molecular emission
in the shocked and post-shock gas. Indeed, models of dense $J-$ and
$C-$type shocks produce strong H$_2$, CO, H$_2$O and/or OH mid- and
far-infrared emission
\citep[e.g.,][]{Draine83,Hollenbach89,Kaufman96,Flower10} that has
been observed with {\it Herschel}. {\it Spitzer}-IRS observations of
the HH 211 outflow have detected strong supra-thermal OH emission (see
Sect.~\ref{sec:irdiagnostics}) from its bow shock
\citep{Tappe08,Tappe12}, now also beautifully seen by JWST MIRI-MRS
\citep{Caratti24} in JOYS HH 211 data and in one of the HOPS 370 jet
knots in IPA data \citep{Neufeld24}. This emission is well modeled by
photodissociation of H$_2$O producing OH in very highly excited
rotational states \citep{Tabone21}. The mechanism is further confirmed
by the symmetric/antisymmetric line asymmetry, predicted by molecular
physics calculations \citep{Zhou15}, that is also observed in JWST
data \citep{Zannese24}. JWST NIRSpec spectra have confirmed that CO
ro-vibrational emission is strong and can outshine H$_2$ in broadband
emission at 4.6 $\mu$m \citep{Ray23}. Investigating at which knot and
bow-shock positions this molecular emission is seen, and where not,
offers opportunities to test the shock models further and ultimately
use the molecular emission as a diagnostic tool, in concert with the
results on larger scales from {\it Herschel} and {\it Spitzer}.

\subsection{Cold outer envelope: Ices}
\label{sec:back_ices}

Mid-infrared spectroscopy is unique compared with millimeter and near-infrared
data in its ability to probe solid-state features. The central IFU
spaxels centered on each source show a rising SED with broad bands of
cold ices and silicates superposed in absorption
(Fig.~\ref{fig:B1c-overview}). These ices are not a minor chemical
component; they are actually the main reservoir of the non-refractory
carbon and oxygen in cold regions \citep[see][for review]{Boogert15}.

MIRI improves significantly on ISO-SWS, {\it Spitzer} and ground-based
telescopes by having the sensitivity to observe the coldest, most
ice-rich low-mass Class 0 and the deeply embedded high-mass
sources. Also, late-type background stars behind dense clouds can be
targeted to reveal the ice content of pre-stellar cores prior to star
formation \citep[e.g.,][]{Knez05,Goto21,McClure23}. Equally important is
the fact that, compared with {\it Spitzer}-IRS, MIRI-MRS has much
improved spectral resolving power in the critical 5--10 $\mu$m range
($R\sim 3000$ vs $50-100$) which is key to resolve and identify weak
features.  Combined with complementary NIRSpec 3--5 $\mu$m data such
as part of JOYS+, this will allow for a complete ice inventory.

Pre-JWST mid-infrared spectra have identified primarily simple molecules in
ices up to CH$_3$OH.  Some of the weakest ice bands have been
suggested to be due to larger organic molecules such as CH$_3$CH$_2$OH
or CH$_3$CHO \citep{Schutte99,Oberg11}, but these molecules can only
be reliably assigned through a detailed comparison of laboratory ice
band profiles with the high-quality spectra that MIRI
provides. Recognizing the potential of MIRI in this area, the Leiden
Laboratory for Astrophysics has been collecting many laboratory
spectra of both simple ices and, recently, more complex molecules in
pure and in mixed ice form
\citep[e.g.,][]{Terwisscha18,Terwisscha21,Rachid21,Rachid22,Slavicinska23},
building on earlier efforts
\citep[e.g.,][]{Hudgins93,Ehrenfreund96,Boudin98,Mulas98,Oberg11,Ioppolo13}. Together,
they are collected and made publically available through the {\it
  Leiden Ice Database for Astrochemistry} (LIDA) \citep{Rocha22} at
{\tt icedb.strw.leidenuniv.nl}. Identification of larger ice species
would strongly support the view that many complex molecules are
actually formed in the ices during the cold pre-stellar and
protostellar warm-up phases, rather than solely by high temperature
gas-phase chemistry in the hot cores \citep[see reviews
by][]{Herbst09,Jorgensen20,Ceccarelli23}. Using JOYS(+) data for one
low- and one high-mass protostar, MIRI-MRS has indeed identified
several icy complex molecules demonstrating that they are made in
ices \citep{Rocha24}. Comparison of abundance ratios of complex
molecules in ice and gas in the same source for two low-mass JOYS
protostars indicates that some molecules are destroyed in the warm gas
following ice sublimation, whereas other molecules are not
\citep{Chen24}. Similar analyses using high quality MIRI-MRS data for
other low- and high-mass protostars such as provided by the CORINOS
and IPA programs \citep{Yang22,Nazari24,Tyagi25} are in progress.

Ions and salts have long been known to be a significant component of
ices \citep{Grim87,Grim89}, with the 4.62 $\mu$m band ascribed to
OCN$^-$ \citep{Novozamsky01} and a significant part of the 6.85 $\mu$m
band to NH$_4^+$ \citep{Schutte03,Boogert15}. These ions are naturally
produced through acid-base reactions between the acid NH$_3$ and bases
such as HNCO and HCOOH; laboratory experiments show that these are
thermal reactions that occur even at low temperatures in ices
\citep{Raunier03,vanBroekhuizen04}. The JWST-NIRSpec detection of the
OCN$^-$ ice band in a prestellar core toward a background star
confirms this low temperature formation route \citep{McClure23}. One
puzzle, however, is that not enough negative ions have been identified
to counterbalance the amount of NH$_4^+$ in ices implied by the 6.85
$\mu$m band. With MIRI, the prospects for identifying the weak and
broad features due to other negative ions such as HCOO$^-$
\citep{Schutte99} are much improved, and both OCN$^-$ and HCOO$^-$ are
indeed now detected in JOYS data \citep{Rocha24,Chen24}. In addition,
confirmation of the NH$_4^+$ identification and balancing the ice
charges is highlighted by the combined laboratory and observational
study of \citet{Slavicinska25sulfur}, including a likely JWST
detection of the main counter ion SH$^-$. This ion also locks up a
significant fraction, up to 20\%, of the sulfur budget in dense
clouds.

The ice band profiles give information on the ice environment (e.g.,
whether the molecule is in a water-rich or water-poor environment or
whether it has been heated) and thus whether the ice is layered or
segregated upon heating
\citep[e.g.,][]{Pontoppidan08,Boogert15}. Since crystallization
processes are irreversible, this also provides a record of the
temperature history of the sources, in particular whether they have
been heated in the past due to episodic accretion events. A well-known
case is that of the 15 $\mu$m bending mode of $^{12}$CO$_2$
\citep{Ehrenfreund99,Gerakines99,Kim12}, now also possible with JWST
NIRSpec for $^{13}$CO$_2$ \citep{Brunken24}.  Moreover, the gas/ice
ratio increases with overall envelope heating and thus with
evolutionary state \citep{vanDishoeck96,Dartois98,vanDishoeck98}. MIRI
can compare gas and ice column densities directly for sources where
the gas-phase lines are also in absorption, like the JOYS case of B1-c
\citep{vanGelder24overview}.

Isotopologs have long been known to be a powerful probe of the
physical and chemical structure of the region in which they are
residing \citep[see review by][]{Nomura23}. At low temperatures, the
abundances of the heavier isotopes can be enhanced by orders of
magnitude through gas-phase reactions. UV photons can also selectively
destroy isotopologs such as $^{13}$CO, C$^{18}$O and $^{14}$N$^{15}$N
\citep[e.g.,][]{vanDishoeck88,Heays14}. Now that these isotopes can
also be detected in exoplanetary atmospheres \citep{Zhang21,Snellen25}
they may contain information on their formation pathways that cannot
be inferred readily by other means \citep{Bergin24iso}. JWST-MIRI
builds on earlier work \citep{Boogert0013co2} by providing
high-quality spectra of $^{13}$CO and $^{13}$CO$_2$ that can be used
to determine $^{12}$C/$^{13}$C isotopolog ratios of ices in cold
clouds \citep{McClure23} and protostellar envelopes
\citep{Brunken24iso} before they enter the giant planet-forming zones
of disks.

For the specific case of water, the MIRI data will be highly
complementary to the gaseous water lines obtained from {\it Herschel}
in the WISH and related programs for the same sources
\citep{vanDishoeck21} (see Sect.~\ref{sec:back_hotcores}). They complement
ALMA and NOEMA data on warm gaseous H$_2^{18}$O emission in the inner
envelope \citep[e.g.,][]{Persson14,Jensen19}. Together, the data can
be used to put together the full picture of the evolution of water
(gas {\it and} ice) from pre-stellar cores to cometary objects, and
ultimately provide, through comparison of the deuterium fractionation,
further insight into the origin of water in comets and in oceans on
Earth. So far, this comparison has been based on gas-phase HDO/H$_2$O
ratios in hot cores assuming that they reflect those in ices, but this
does not need to be the case. Direct measurements of HDO/H$_2$O ice
are now possible with JWST. For this reason, a deep NIRSpec spectrum
covering the HDO 4.1 $\mu$m ice band has been obtained for the most
ice-rich solar-mass protostar in the JOYS sample, B1-c.
Prior to JWST, there had been upper limits and hints of HDO
ice detections \citep[e.g.,][]{Dartois03,Aikawa12akari}, but the first firm
detection of HDO ice, toward an intermediate and high-mass protostar,
has recently been reported by \citet{Slavicinska24HDO} as part of the
IPA program, demonstrating NIRSpec's ability for this purpose.

Related to the H$_2$O case is that of O$_2$ ice. Oxygen is the third
most abundant element in space, yet its major reservoirs are still
under debate. A long standing problem has been that some fraction of
oxygen (up to 50\%) may be unaccounted for if an interstellar
abundance of $5.8\times 10^{-4}$ \citep{Przybilla08} is assumed
\citep{Whittet10,Draine21} (see recent summary by
\citealt{vanDishoeck21}): H$_2$O ice only accounts for about 20\% of
oxygen in cold regions \citep{Boogert15}. Models of the chemistry in
dense clouds predict that O, O$_2$ and H$_2$O are the main volatile
oxygen carriers \citep{Tielens82}, with O and O$_2$ being converted to
H$_2$O ice through hydrogenation on the surfaces of grains
\citep{Bergin00}, processes that have been confirmed in the laboratory
\citep{Miyauchi08,Ioppolo08}. However, under certain conditions O$_2$
ice could become a significant reservoir. Indeed, cometary data of 67P
show a surprisingly high and constant O$_2$/H$_2$O ice ratio of a few
\% \citep{Bieler15,Rubin19}.

O$_2$ is notoriously difficult to observe both in the gas and in
ice. As a symmetric molecule, it does not have a permanent dipole
moment and can only be observed in the gas through its much weaker
magnetic dipole transitions from space or its much less abundant
$^{16}$O$^{18}$O isotopolog from the ground. While O$_2$ gas has been
detected at low abundances at a few PDR and shock positions
\citep{Goldsmith11,Liseau12}, {\it Herschel}-HIFI deep searches have
mostly obtained upper limits including a fractional abundance as low
as $6\times 10^{-9}$ for the NGC 1333 IRAS4A source that is part of
JOYS(+) \citep{Yildiz13o2}. Even in the inner hot core of the COM-rich
IRAS 16293-2422B source, the gaseous O$_2$ abundance appears to be
less than $10^{-6}$, thus containing only a small fraction of oxygen
\citep{Taquet18}. One possible explanation for its absence in hot
cores could be the presence of X-rays destroying O$_2$ following its
sublimation around 30 K \citep{Notsu21}. This suggests that O$_2$ ice
could still be present in the coldest parts of protostellar envelope
which JWST MIRI can search for.

\subsection{Embedded disks}
\label{sec:back_disks}

Disks around low-mass young T Tauri stars have been studied in great
detail over the last decades and much has been learned about their
physical and chemical structure \citep[see reviews
by][]{Dullemond07,Miotello23,Manara23,Oberg23}. In contrast, little is
still known about the physical and chemical properties of disks in the
embedded phase. It is well established theoretically that disks must
form by the Class 0 stage as the inevitable byproduct of the collapse
of a rotating core and observationally from the fact that all Class 0
sources drive collimated jets. Much of the observational difficulties
stem from the fact that the emission from the disk is readily
overwhelmed by that from the surrounding envelope if observed at too
low spatial resolution and sensitivity to image minor isotopologs.
With ALMA, this situation is rapidly changing and embedded disks can
now be detected through Keplerian rotation of CO isotopolog lines
\citep[e.g.,][]{Murillo13,Yen17}. An overview is presented in
\citet{Tobin24}. The recent eDisk ALMA survey of embedded disks at
0.04$''$ (down to 7 au) resolves also the millimeter dust emission showing
elongated disk images \citep{Ohashi23}.  Some of the eDisk sources are
part of JOYS.

Embedded disks are warmer than mature Class II disks
\citep{vantHoff20}, with fresh material that can be brought through
accretion streamers onto the disk
\citep[e.g.,][]{Pineda20,Murillo22,Hsieh23,Cacciapuoti24,Valdivia24,Podio24,Speedie25}.
One specific goal of the MIRI program will be to search for signs of
compact accretion shocks that could be revealed by H$_2$, atomic lines
(e.g., [S I]) and other shock tracers centered on the source. Indeed,
models of accretion shocks predict a rich spectrum of CO, H$_2$O and
OH lines \citep{Neufeld94}. Accretion shocks have been suggested based
on millimeter observations of SO and SO$_2$ molecules
\citep{Sakai14,Artur19,Artur22,Liu25}. The shock (its velocity) is a direct
measure of where most of the material enters the disk, although at
only a few km s$^{-1}$ this will be too low to measure with JWST. The
strength of the shock and inflow of envelope material, in turn, have
consequences for the thermal and chemical structure of the young disk,
in particular whether pristine pre-stellar core material can be
preserved during the transport from cloud to disk or whether the
chemistry is reset \citep[e.g.,][]{Visser09}. The mid-infrared
detection of SO$_2$ on disk scales with JWST as part of JOYS+
\citep{vanGelder24} could be a signature of an accretion shock,
although the line width and velocity measured with ALMA suggest that
thermal sublimation of SO$_2$ ice in a hot core is more likely.

ALMA studies provide insight into the outer parts of embedded disks,
but infrared studies are needed to probe the innermost regions ($<$1
au) that ALMA cannot reach.  Mature Class II disks show surprisingly
rich mid-infrared spectra with {\it Spitzer}
\citep{Carr08,Salyk08,Pontoppidan14} that may differ for very low mass
stars \citep{Pascucci13}. This is even more apparent now with JWST
\citep[e.g.,][]{Grant23,Kospal23,Banzatti23,Gasman23,Henning24,Pontoppidan24,Temmink24H2O,Romero24,Colmenares24},
with disks around very low mass stars showing a wide variety of
hydrocarbon molecules \citep{Tabone23,Xie23,Arabhavi24}.  Mid-infrared lines
from H$_2$O, HCN, C$_2$H$_2$, CO$_2$ and OH are commonly seen,
indicating hot molecular gas at temperatures of 300--900~K arising
from the inner few au, i.e., the terrestrial planet-forming zones of
the disks. This hot gas is also probed by pioneering velocity resolved
($R=20000-100000$) ground based CO infrared spectra
\citep[e.g.,][]{Najita03,Blake04,Salyk09,Brown13,Banzatti15,Banzatti22,Temmink24CO}.
Various atomic transitions, most prominently the [Ne II] 12.8 $\mu$m
line, have been studied as well pre-JWST
\citep[e.g.,][]{Pascucci07,Guedel10}.  An interesting question is
therefore whether young disks in the embedded phase have similarly
rich spectra as the more evolved disks that probe the inner few
au. Existing {\it Spitzer} data had too low feature/continuum and
spatial resolution to address this question \citep{Lahuis06,Lahuis10}
but VLT-CRIRES data show strong 4.7$\mu$m CO emission lines from
embedded Class I protostars that indicate Keplerian rotation and are
thus promising for JWST studies of other molecules
\citep{Herczeg11}. However, the JWST molecular line survey of the JOYS low-mass
protostars by \citet{vanGelder24overview} suggests that
emission from inner disks is not commonly seen.

The question whether high-mass protostars have disks has been debated
as well, although it is likely they do too, since they all show
collimated jets and outflows \citep{Beltran16,Beuther25}.  Their
characterization is even more difficult since they are all hidden by
huge amounts of extinction. Millimeter interferometer studies have
revealed elongated disk-like structures around the massive protostars
with signs of Keplerian motion
\citep[e.g.,][]{Cesaroni05,Ilee18,Maud19,Johnston20,Williams22}
\citep[see review by][]{Beltran16}.  Radiation transport studies
provide some information on the density structures of the underlying
disks \citep{Fallscheer11} and make predictions of expected
mid-infrared structures.
Obviously, the mid-infrared data trace completely different structures than
the millimeter images, and important additional parameters that could
be obtained in the mid-infrared regime include the vertical disk structure
or the disk flaring, if the emission were spatially resolved.

As for the low-mass disks, the spectroscopy will allow for
investigation into several other properties of the embedded disks, in
particular the temperature, ionization state, and chemical
composition.  Indeed, ground-based CO $v=2-0$ overtone emission at 2.3
$\mu$m has been detected with high resolution spectroscopy and
infrared interferometry.  The data are indicative of a disk origin
coming from inside the dust sublimation radius
\citep{Carr89,Bik04,Ilee14,Gravity21Kou}. Compared with previous
mid-infrared studies, the higher spatial resolution JWST MIRI spectra
will be much less beam diluted and could therefore pick up disk
emission. The higher spectral resolution of MIRI-MRS also helps to
boost the line-to-continuum ratio. Indeed, the relatively cool CO$_2$
emission seen toward the high mass protostar IRAS 23385+6053 in JOYS
data could arise from the cooler outer surface of a massive disk
\citep{vanDishoeck23} although other interpretations such as a hot
core are also possible \citep{Francis24}.

\subsection{JOYS studies}
\label{sec:app_backgroundJOYS}

Early JOYS publications on selected topics or individual sources are
discussed in Sections 4--8 and are briefly summarized here. They
include a study on the atomic and molecular lines in the high-mass
protostar IRAS 23385+6053 by \citet{Beuther23,Gieser23,Francis24}; on
the linked accretion and ejection in the Class I binary protostar
TMC-1 by \citet{Tychoniec24}; and on the analysis of ices including the
detection of complex molecules and sulfur-bearing salts by
\citet{Rocha24,Chen24,Slavicinska25sulfur}. Moreover,
\citet{vanGelder24overview} provides an overview of detected gaseous
molecular lines at the central position of all JOYS sources.  An early
comparison with Class II sources is provided by \citet{vanDishoeck23}.

The HH 211 outflow imaged with NIRCam has been highlighted by
\citet{Ray23}, whereas the MIRI-MRS large mosaic taken as part of JOYS
is analyzed in \citet{Caratti24}. A survey of HD and HD/H$_2$ in HH
211 and other sources with galactocentric radius is published by
\citet{Francis25HD}.

The L1527 outflow has been studied both within JOYS and in PID 1798
(PI J.\ Tobin).  This is a rare case in which the jets are
characterized by the presence of double peak emission in forbidden
emission lines with a low and high velocity component, but with no
such signature in molecular emission (Devaraj et al., in prep.;
Drechsler et al., in prep.). The low velocity components are thought
to be excited by shocks at the outflow cavity wall by interaction of
the disk wind with the ambient medium. The high velocity component is
excited by the fast moving jet inside the low-density cavity where no
molecular material is present anymore to be entrained by the fast
moving flow.

JOYS+ publications include the mid-infrared detection of gas-phase SO$_2$
\citep{vanGelder24}; the analysis of NIRSpec observations of the
Serpens S68N binary protostellar system \citep{LeGouellec25S68N}; a
combined JWST-MIRI + ALMA analysis of the fascinating WL20 triple
system with aligned disks and jets in Ophiuchus \citep{Barsony24}; and
a determination of the $^{12}$C/$^{13}$C ratio in ices
\citep{Brunken24iso}. Together, these JOYS(+) papers demonstrate the
rich and diverse science that can be done with a single protostellar
data set.

\section{Simultaneous MIRI imaging}
\label{sec:app_IRAS23385}

\subsection{Protostars in the outer Galaxy}

\begin{figure}
\begin{centering}
\includegraphics[width=9cm]{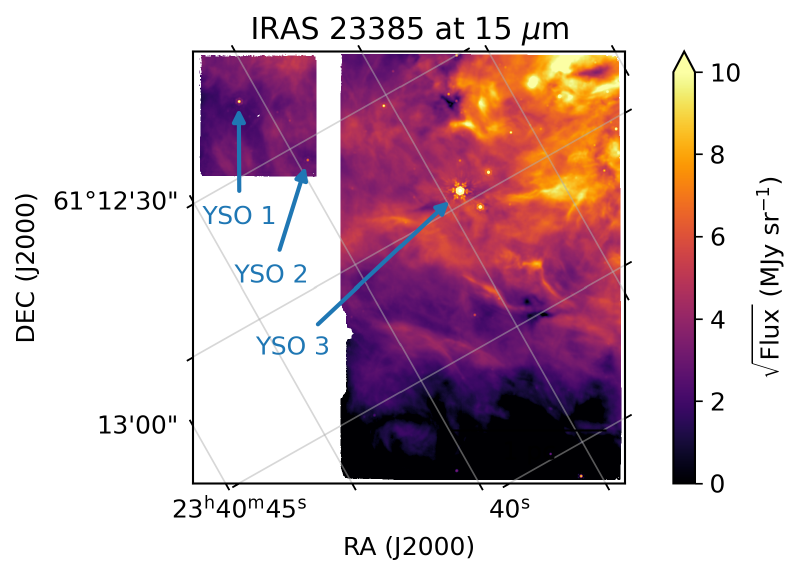}
\caption{Simultaneous MIRI 15 $\mu$m imaging near the young high-mass
  cluster IRAS 23385+6053 at a galactocentric distance of 11 kpc. Its
  coordinates are 23:40:49.18, +61:11:22.86.}
\label{fig:IRAS23385image}
\end{centering}
\end{figure}

The simultaneous off-source image at 15 $\mu$m obtained for IRAS
23385+6053 at a Galactocentric radius of 11 kpc is presented in
Figure~\ref{fig:IRAS23385image}. Since the MRS on-source observation
consisted of a $2 \times 2$ mosaic as well as a background pointing,
the simultaneous MIRI image is a composite of two off-source
pointings, a southern and northern region. The southern region has 3
exposures of 200 seconds each, and the northern region is a mosaic of
4 images, with each 3 exposures of 200 seconds. Thus, the northern
image is significantly deeper than the southern one. The resulting rms
values, after subtraction of a median background of 0.14 MJy sr$^{-1}$,
are 0.4 (north) and 3.8 (south) MJy sr$^{-1}$, respectively.

\begin{figure}
\begin{centering}
\includegraphics[width=8cm]{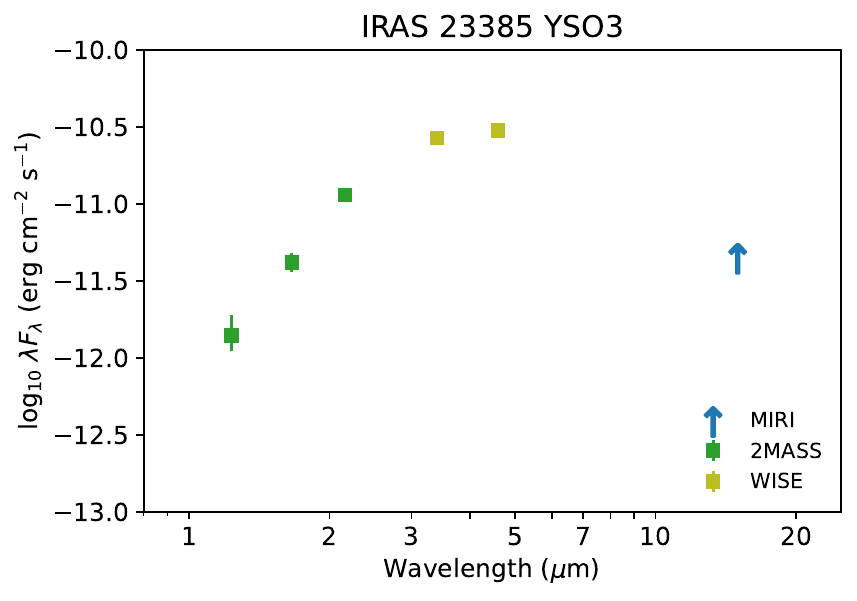}
\caption{Spectral energy distribution of a candidate embedded protostar in the outer Galaxy found by
  simultaneous MIRI imaging.}
\label{fig:IRAS23385SED}
\end{centering}
\end{figure}

Using the aperture photometry method, 47 point sources were found,
after discarding doubles, false positive detections or bad pixels. The
high spatial resolution of MIRI now allows point sources to be
distinguished from the extended bright background.  SEDs were made
using near-infrared observations of 2MASS and WISE at the shorter
wavelengths at the MIRI positions, see Figure~\ref{fig:IRAS23385SED}
for one example.  A rising SED between 2 and 15 $\mu$m was used as a
criterion to identify embedded protostars, as done in \citet{Greene94}
between 2 and 10 $\mu$m. For the source shown in
Figure~\ref{fig:IRAS23385SED}, the spectral slope $\alpha_{2-15}$ is
$>0.46$ taking into account that the 15 $\mu$m MIRI data point is a
lower limit due to saturation.  The integrated luminosity is at least
450 L$_{\odot}$, suggesting indeed an intermediate or high mass
source, but the lack of data longer than 15 $\mu$m prevents assessing
the total luminosity of these sources and quantities such as
$T_{\rm bol}$. This source was highlighted in the image accompanying
the press release of \citet{Rocha24} at

\noindent {\footnotesize \tt
  webbtelescope.org/contents/news-releases/2024/news-2024-111}.

Other MIRI point sources have $\alpha$ values closer to zero or
negative, suggesting that they are more evolved sources. This example
serves as an illustration of JWST-MIRI's capabilities to detect new
protostars in highly extincted distant and crowded regions.

\subsection{Role of PAH emission affecting ice bands}

The MIRI spectra and SEDs may also contain signatures of PAHs. MIRI
has the sensitivity to detect these features throughout a cloud. At
the position of deeply embedded sources, PAH features were largely
undetected with ISO for high-mass protostars \citep[e.g.,][]{Gibb04}
(unless associated with well-known PDRs such as Orion), and with {\it
  Spitzer} for low-mass protostars \citep{Geers09}.  Likely
explanations include freeze-out into ices and coagulation to larger
PAH systems that do not emit, rather than a lack of UV radiation to
excite them. With JWST's much higher sensitivity, PAH emission may be
seen both on and off source, however.

The case of IRAS 23385+6053 provides a particularly illustrative
example of how this emission can complicate the analysis of broad ice
features in the spectrum. Likely, the PAH emission, which is clearly
seen at 11.3 $\mu$m \citep{Beuther23}, comes from a background cloud
and is extincted by ices in the dense envelope surrounding the
high-mass protostellar cluster. This makes the 6.2 and 7.7 $\mu$m
features hard to detect and they can distort the ice bands. This case
serves as a warning for detailed analysis of ice bands in high-mass
sources.

\section{Mass loss rate determinations and shock model comparison}
\label{sec:app_massloss}

\begin{figure}[h!]
\begin{centering}
\includegraphics[width=7cm]{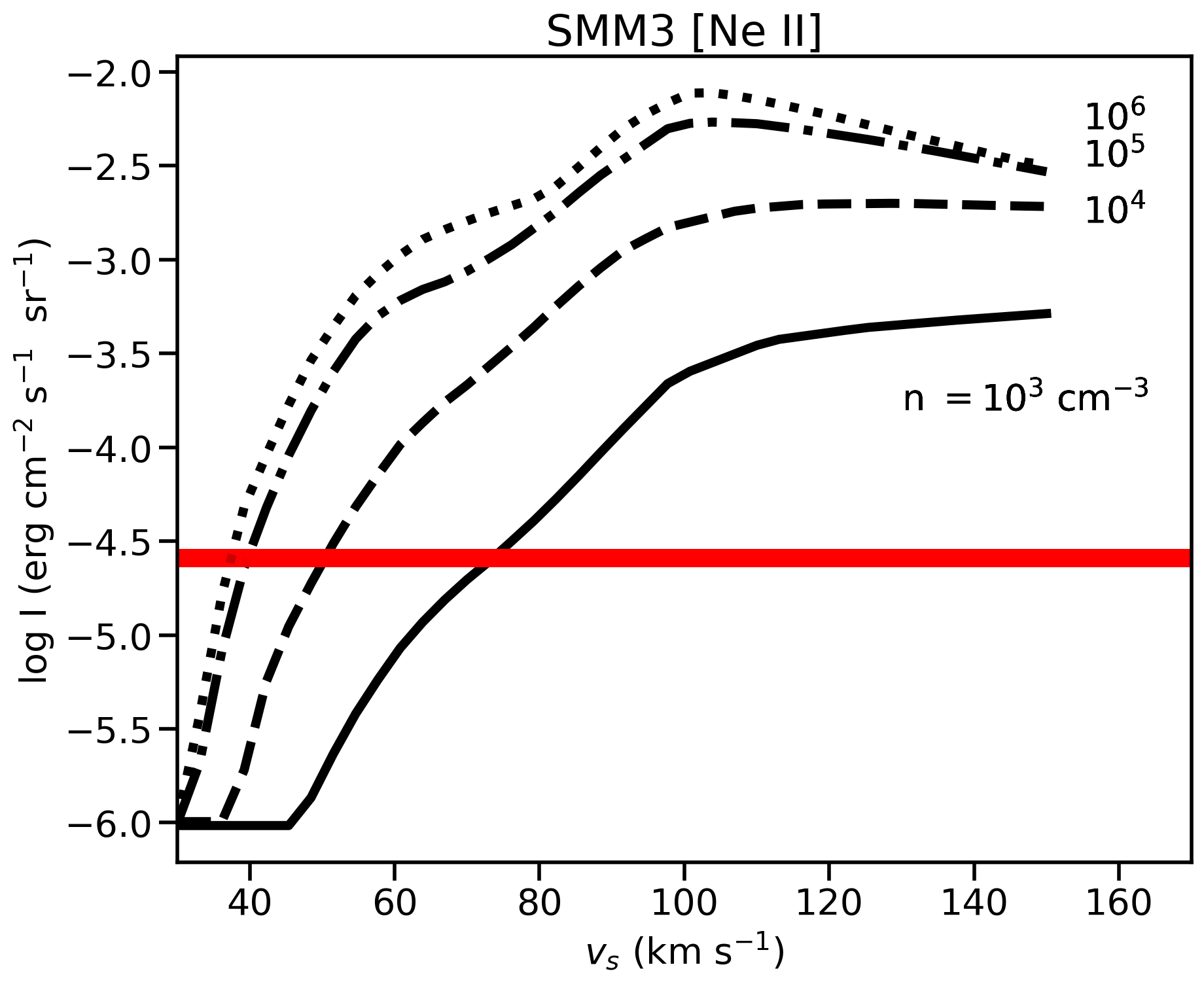} \\
\includegraphics[width=7cm]{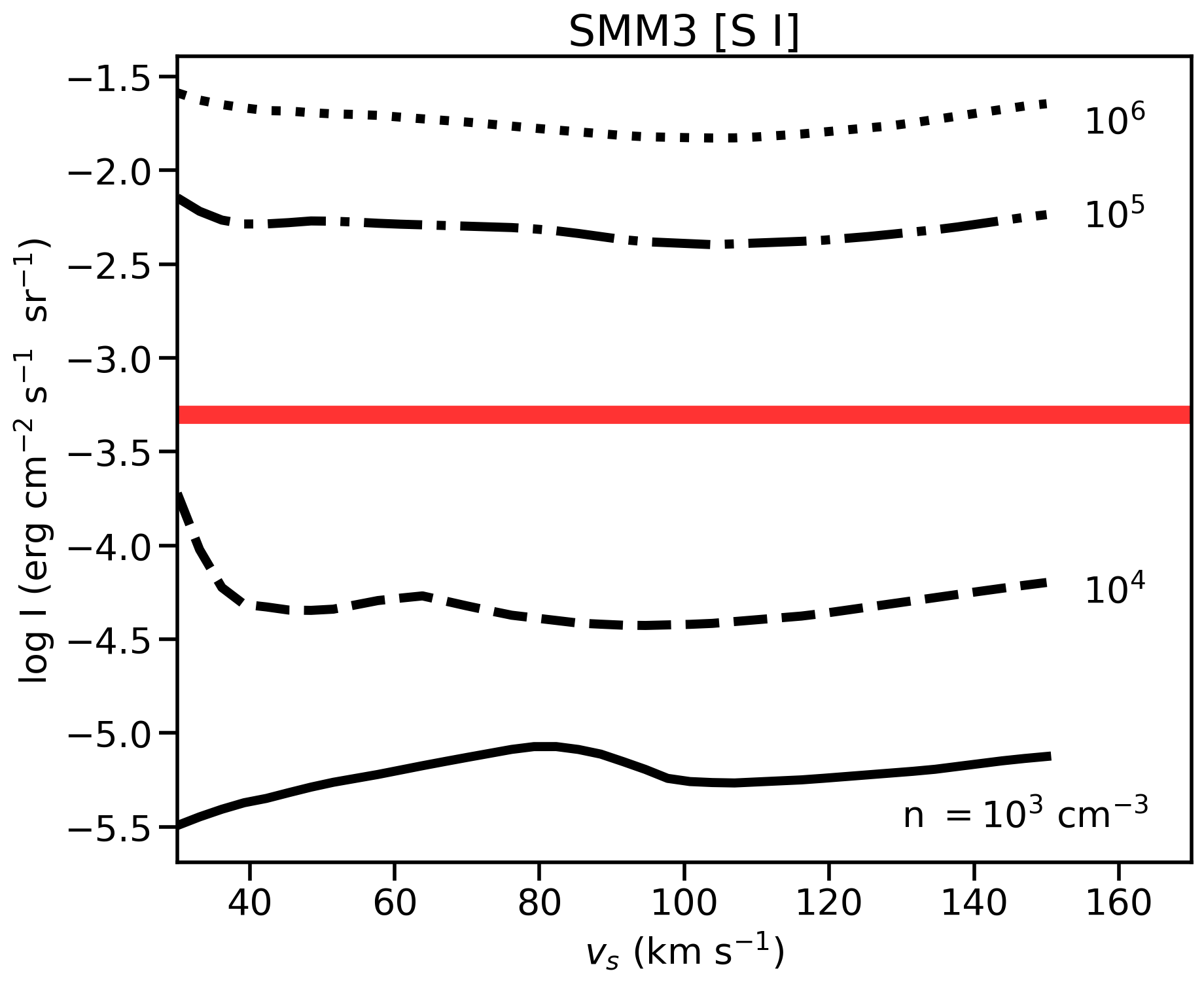}
\caption{Comparison of the extinction-corrected [Ne II] 12 $\mu$m and
  [S I] 25 $\mu$m fluxes for SMM3 in the red outflow lobe with shock
  models of \citet{Hollenbach89} as function of shock velocity and
  pre-shock density. The red horizontal bars indicate the observed
  values for the red lobe of SMM3.}
\label{fig:SMM3vsmodels}
\end{centering}
\end{figure}

To obtain more quantitative information on the physical parameters and
jet properties of SMM3, we follow the analysis methods of \citet{Caratti24}
for HH 211 in slightly modified form. Rather than using [Fe II] and [S
I] line ratios, we take the [Ne II] and [S I] line fluxes to constrain
shock characteristics. Figure~\ref{fig:SMM3vsmodels} compares the
observed extinction-corrected [Ne II] 12.8 $\mu$m and [S I] 25 $\mu$m
fluxes from spectra extracted at the red knot position with the shock
models of \citet{Hollenbach89}. Assuming that all of the line emission
comes from such dissociative shocks, the [Ne II] line constrains the
shock velocity to $40-45$ km s$^{-1}$, whereas the [S I] flux
suggests the pre-shock total hydrogen density $n_0$(H + 2H$_2$) to be
$\sim 10^5$ cm$^{-3}$ (red lobe). The mass loss rate indicated by the
atomic lines is then

\begin{equation}
  {\dot M}_{\rm atomic}= \mu m_{\rm H} n_0 \pi r_j^2 v_{\rm tot},
\end{equation}

\noindent
where $\mu$ is the mean atomic weight (1.35), $m_{\rm H}$ is the
proton mass, $r_j$ is the jet radius, measured to be $\sim$60 au from
the jet diameter at the shortest wavelengths where the MIRI angular
resolution is the highest and sufficient to resolve the width. The total
velocity $v_{\rm tot}$ of $\sim$50 km s$^{-1}$ is obtained from the
measured jet velocity of $\sim$30 km s$^{-1}$ (from ionic lines and
from ALMA CO data in the jet), corrected for an inclination of
$\sim 50^o$ estimated from CO millimeter maps \citep{Yildiz15}. The
inclination could be higher by 10-15$^o$ as estimated from the ALMA
continuum disk image, but is unlikely to be lower. This formula gives
a typical mass loss rate of ${\dot M}_{\rm atomic}=4.5 \times 10^{-8}$
M$_\odot$ yr$^{-1}$ for the red lobe.

An alternative route to getting the mass outflow rates is to use the
flux of the [O I] 63 $\mu$m line, which is the main coolant of the jet
\citep{Hollenbach85}. Using the measured [O I] 63 $\mu$m flux from
{\it Herschel}-PACS of $1\times 10^{-2}$ L$_\odot$ summed over the
$\sim$9$''$ spaxels where the line is detected gives a mass loss rate
$\dot M$ of $\sim (0.6-1)\times 10^{-6}$ M$_\odot$ yr$^{-1}$
\citep{Mottram17,Karska18}, significantly higher than the value inferred
above.

\citet{Watson16} found that the [Fe II] 26 $\mu$m ground
state fine structure line can be used as a proxy for [O I] through the
following relation:

\begin{equation}
 {\dot M}_{\rm atomic}= 1.4 \times 10^{-3} L({\rm Fe}\ 26 \mu {\rm m})/L_\odot .
\end{equation}

Using the extinction corrected 26 $\mu$m [Fe II] luminosities of
$3.5\times 10^{-6}$ and $7.5\times 10^{-6}$ L$_{\odot}$ for the red-
and blue-shifted positions where H$_2$ was extracted gives mass loss
rates of
${\dot M}_{\rm atomic}=0.5 \times 10^{-8}$ and $1.1 \times 10^{-8}$
M$_\odot$ yr$^{-1}$, respectively, more in line with those found from
[Ne II] + [S I]. However, these values seem low compared with the {\it Spitzer}
studies by \citet{Watson16} using this line, also given the
significant luminosity of SMM3, but this could be due to the much
smaller area over which the [Fe II] line flux is measured compared
with {\it Spitzer} data. Summing over the entire $\sim 7'' \times 7''$
MRS channel 4 FoV (comparable to the {\it Spitzer}-IRS aperture) gives
an extinction-corrected [Fe II] 26 line luminosity of
$6.1 \times 10^{-5}$ L$_\odot$, which translates to
${\dot M}_{\rm atomic}=8.6 \times 10^{-8}$ M$_\odot$ yr$^{-1}$. This
value is indeed significantly higher than that derived for a single
position, but is still an order of magnitude lower than that derived
from [O I]. A similar difference was found for HH 211.

\begin{table*}[t]
\begin{centering}
\caption{Overview of mid-infrared lines commonly detected in protostellar systems.}
\label{tab:lines}
\begin{tabular}{lccccc}
  \hline
  \hline
  Species & Transition & $\lambda$ & $E_{\rm u}/k$ & Type$^a$ & Origin \\
          &  $u-l$       & ($\mu$m)  & (K)  &   \\
  \hline
 [Fe II] & a$^4{\rm F}_{9/2}$-a$^6{\rm D}_{9/2}$ &  5.3402 & 2694 &R & Jet \\
         & a$^4{\rm F}_{7/2}$-a$^4{\rm F}_{9/2}$ & 17.9232 & 3496 && Jet \\
         & a$^4{\rm F}_{5/2}$-a$^4{\rm F}_{7/2}$ & 24.5019 & 4083 && Jet \\
         & a$^6{\rm D}_{7/2}$-a$^6{\rm D}_{9/2}$ & 25.9884 & 554 && Jet, Entrained gas \\
 {[Fe I]}  & $^5{\rm D}_{3}$-$^5{\rm D}_{4}$ & 24.0423  & 600 &R& Jet \\
 {[Ni II]} & $^2{\rm D}_{3/2}$-$^2{\rm D}_{5/2}$ &  6.6360 & 2168 &R& Jet \\
 {[Ne II]} & $^2{\rm P}_{1/2}$-$^2{\rm P}_{3/2}$ & 12.8135 & 1123 &V& Jet, 
          Photoionized gas \\
 {[Ne III]} & $^3{\rm P}_{1}$-$^3{\rm P}_{2}$ & 15.5551 & 925 &V& Jet, 
          Photoionized gas \\
 {[Ar II]} & $^2{\rm P}_{1/2}$-$^2{\rm P}_{3/2}$ & 6.9853 & 2060 &V& Jet, 
          Photoionized gas \\
 {[S I]} & $^3{\rm P}_{1}$-$^3{\rm P}_{2}$ & 25.2490 & 570 & SR & Jet, 
  Wind? \\
  \hline
  \noalign{\smallskip}
  H$_2$$^b$ & S(1) & 17.0348 & 170 & V & Wind, Entrained gas \\
            & S(5) &  6.9095 & 2504 & V & Wind, Jet \\
  CO & $v=1-0$ $\geq$P(24) & 4.9--5.2 & 4600--8000 & V  & Jet, Shock$^d$ \\
  H$_2$O   & $v_2$=1-0 rovib & 5.5--7.5 & 2300--$>$4000 & V
           & Shock, Hot core$^d$\\
          & $v$=0-0 warm & 11--18$^c$ & 2600--$>$8000 & V
           & Shock, Hot core$^d$ \\
          & $v$=0-0 cold & 20--27$^c$ & 1100--4000 & V & Hot core$^d$ \\
  CO$_2$ & $v_2$=1-0 & 14.4--15.6 & 960--1500 & V & Hot core$^d$, Outflow \\
  HCN & $v_2$=1-0 & 13.5--14.5 & 1000--1500 &V  & Hot core$^d$ \\
  C$_2$H$_2$ & $v_5$=1-0 & 13.0--14.5 & 1000--2000 & V & Hot core$^d$ \\
  OH & $v$=0-0 & 9--27 & 10000--40000 & V & Shock,
                                            UV-irradiated gas$^d$ \\ 
\hline
\end{tabular}
\end{centering}

$^a$ R=Refractory, SR=Semi-Refractory, V=Volatile. \\
$^b$ Only some example H$_2$ lines in $v$=0 are listed \\
$^c$ Lines from warm H$_2$O also at longer wavelengths \\
$^d$ Some contribution from young disk possible in Class I sources \\

\end{table*}

We note that the [O I] determination should be considered as an upper
limit for various reasons. The \citet{Hollenbach85} model assumes a
single knot, whereas there may be many more in the beam, as is known
for the case of SMM3 \citep{Tychoniec21} (Le Gouellec et al., in
prep.).  For the case of SMM3, the [O I] 63 $\mu$m flux is also likely
affected by extended emission from the nearby SMM6 source in the large
{\it Herschel}-PACS aperture: the peak [O I] emission indeed seems
somewhat off centered from the SMM3 source position \citep{Karska13}.
Finally, optical depth and foreground absorption of the [O I] 63
$\mu$m line can also affect the estimate. In summary, this example
illustrates the uncertainties and pitfalls associated with the
different methods and the need to recalibrate some of the relations
for the higher spatial resolution data provided by JWST.

\section{Which line traces what?}
\label{sec:app_diagnostics}

Table~\ref{tab:lines} summarizes the commonly observed mid-infrared
lines in protostellar systems and their proposed origin based on the
JWST data.

\section{IRAS 18089 continuum, silicate, and H$_2$O ice subtraction} 
\label{sec:app_IRAS18089}

In order to analyze the 7--9 $\mu$m range for absorption of various
ices, first the continuum and silicate features need to be subtracted from
the IRAS 18089 spectrum presented in
Fig.~\ref{fig:IRAS18089spectrum}. The method consists of applying the
following equation to simultaneously fit both components

\begin{equation}
 {F = F_0 {\rm exp} (-\tau_{\rm dust}) + F_{\rm cold},} 
\end{equation}

where $F_0$ is assumed to be a blackbody emission with a temperature
of 700 K, originating from the warm envelope region. $\tau_{\rm dust}$
is the optical depth from the dust along the line of sight, which is
given by $\tau_{\rm dust} = \rho_{\rm dust} \kappa_{\rm dust}$, where
$\rho_{\rm dust}$ is the dust column density in cm$^{-2}$ g, and
$\kappa_{\rm dust}$ is the dust mass absorption coefficient in cm$^2$
g$^{-1}$. Finally, an additional source of emission coming from a cold
envelope at $T=35~$K is used to fit (by eye) the spectrum at
wavelengths above 20 $\mu$m where cold dust emission starts playing a
major role.

The dust properties used in this paper follow previous studies by
\citet{Sargent09,Boogert08,McClure23} and \citet{Rocha24}. They
consist of adopting olivine (MgFeSiO$_4$) and pyroxene
(Mg$_{0.7}$Fe$_{0.3}$SiO$_3$) stoichiometries as representative of the
dust. The optical constants of these materials were taken from
\citet{Dorschner95}. In addition, a grain-shape correction was
performed by adopting a distribution of hollow spheres
\citep[DHS][]{Min05}, and a size distribution ranging from
$a_{\rm min} = 0.1$ $\mu$m to $a_{\rm max} = 1.0$ $\mu$m. The
calculations of the dust opacities with those properties are executed
with the OpTool software \citep{Dominik21}.

\begin{figure}
\begin{centering}
\includegraphics[width=9cm]{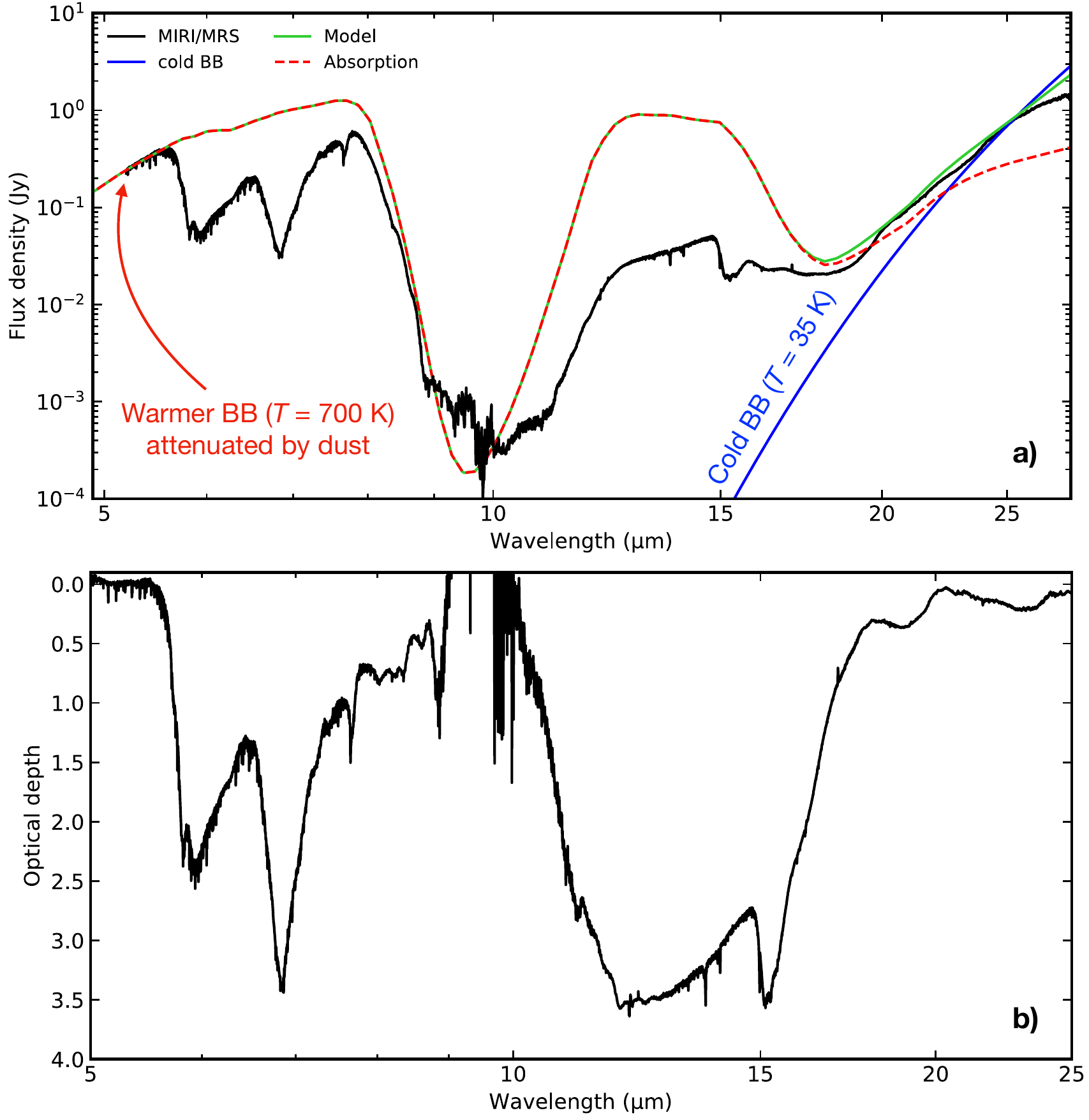}
\caption{Spectrum of IRAS 18089 before and after global continuum and silicate
  subtraction. Panel a (top) shows the simplified model for IRAS 18089 given
  by a warm (red line) and cold (blue line) blackbody emission. Panel
  b (bottom) displays the optical depth scale of IRAS 18089 with both continuum
  and silicate subtracted.}
\label{fig:IRAS18089BBfit2}
\end{centering}
\end{figure}

Figure~\ref{fig:IRAS18089BBfit2} (top) shows the continuum and silicate
absorption profiles for IRAS 18089. The contributions of the two
silicate (pyroxene + olivine) bands at 9.7 and 18 $\mu$m are seen. The
9.7 $\mu$m absorption is slightly overestimated in the model to
account for the excess between 8 and 9 $\mu$m. Part of this missing
absorption at 9--10 $\mu$m may be due to some superposed silicate
emission from an inner warm envelope or disk. Since our primary
interest is in the 7--9 $\mu$m range, the precise details do not
matter. The 18 $\mu$m band matches well the observations but the long
wavelength wing is overpredicted by the model. This is compensated by
the cold blackbody emission above 20 $\mu$m.

Once the continuum and silicate are estimated, the spectrum is
converted to an optical depth scale using the equation
$ \tau = -$ln($F_{\rm obs}/F_{\rm cont}$), where $F_{\rm obs}$ is the
observed flux and $F_{\rm cont}$ is the continuum flux which in this
approach includes the silicate absorption profile. The result
is presented in Figure~\ref{fig:IRAS18089BBfit2} (bottom).

Based on this continuum model with the silicate features, we obtain
$\tau_{9.7}$ = 9.2. Assuming an extinction law similar to dense
star-forming regions in Perseus ($A_V = 28.6 \times \tau_{9.7}$,
\citealt{Boogert13}), the derived visual extinction is $A_V \sim$263
mag (see Appendix~\ref{sec:app_extinction} for more details). While
such a huge extinction correction would be required for any emission
features arising from the inner region \citep[see][for a detailed
discussion]{vanGelder24overview}, this is not the case for absorption
bands from the ices/dust. For example, independent analyses of
$L-$band spectra from protostars in the Serpens molecular cloud were
performed by \citet{Pontoppidan04} with extinction correction and
\citet{Perotti20} without correction. Both methods result in
equivalent optical depths after continuum removal.

\begin{figure}
\begin{centering}
\includegraphics[width=9cm]{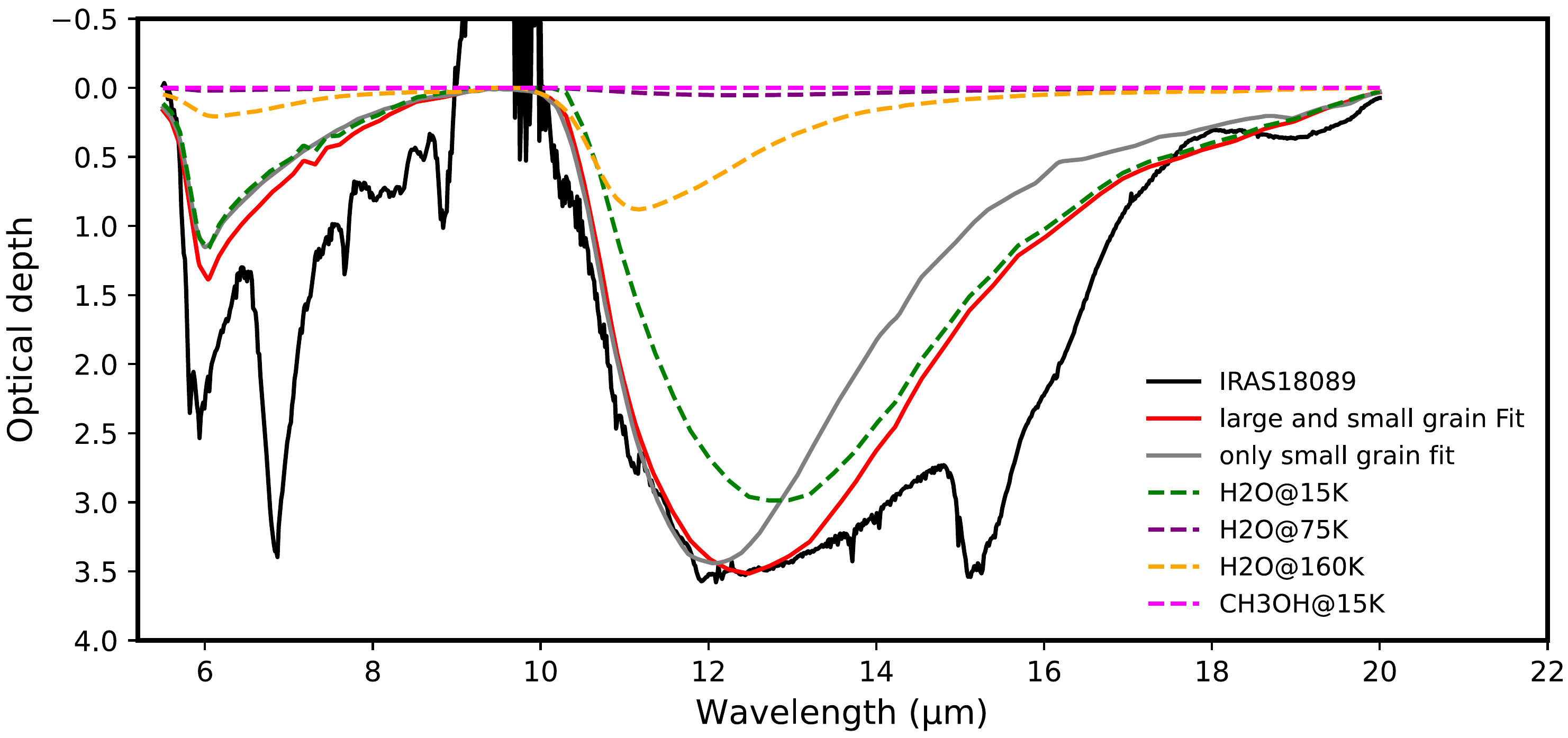}
\caption{Simplified H$_2$O ice model for IRAS 18089 after subtraction
  of the global continuum and silicate band. The small-grain only fit
  is indicated by the gray line, and the combined (large and small)
  grain population is shown by the red line. The dashed lines show the
  H$_2$O ice components for the combined population model.}
\label{fig:IRAS18089H2Ofit}
\end{centering}
\end{figure}

The next step is to fit and remove the H$_2$O ice features.  In the
MIRI range, H$_2$O ice has two prominent and broad features, the
bending mode around 6 and the libration mode at 12 $\mu$m. Similarly
to the dust properties, the analysis of the H$_2$O ice features has
been performed using a distribution of grain-shape corrected spectra
adopting the DHS approach and different temperatures to account for
the temperature gradient of the source. The result is presented in
Figure~\ref{fig:IRAS18089H2Ofit}. The fit using particles ranging from
0.1 to 1.0 $\mu$m in size is indicated by the gray line. Despite the
good fit to the short wavelength range of the 12 $\mu$m band, it
leaves a strong absorption excess unexplained at longer
wavelengths. For this reason, we considered an additional size
distribution of larger grains, ranging from 1.0 to 2.0 $\mu$m. In
particular, we assumed that the hot ice (160 K) component is dominated
by smaller grains, whereas the cold ice (15 K) is represented by
larger grains. In this combined case, the fit is equally good at short
wavelengths and reduces the excess at longer wavelengths.

The fit to the various ice features in the 7--9 $\mu$m range is
presented in Figure~\ref{fig:IRAS18089COMfit}. The fit to the 6.85
$\mu$m feature, consisting of both NH$_4^+$ and CH$_3$OH ice, is
contained in Figure~\ref{fig:IRAS18089NH4}. Following
\citet{Slavicinska25sulfur}, ammonium hydrosulfide salt NH$_4$SH
provides the best fit to the feature; the adopted band strength is
3.6$\times$ 10$^{-17}$~cm~molecule$^{-1}$.  This allows us to quantify
the N-budget in ices. In this calculation, the overall N abundance is
assumed to be $6.3\times 10^{-5}$ \citep{Przybilla08} and the total
hydrogen column along the line of sight $N_{\rm H}=3.6\times 10^{23}$
cm$^{-2}$ based on the above extinction $A_V$=263 mag and
$N_{\rm H}/A_V = 1.37\times 10^{21}$ cm$^{-2}$ mag$^{-1}$
\citep{Evans09}.

\begin{figure}
\begin{centering}
\includegraphics[width=8cm]{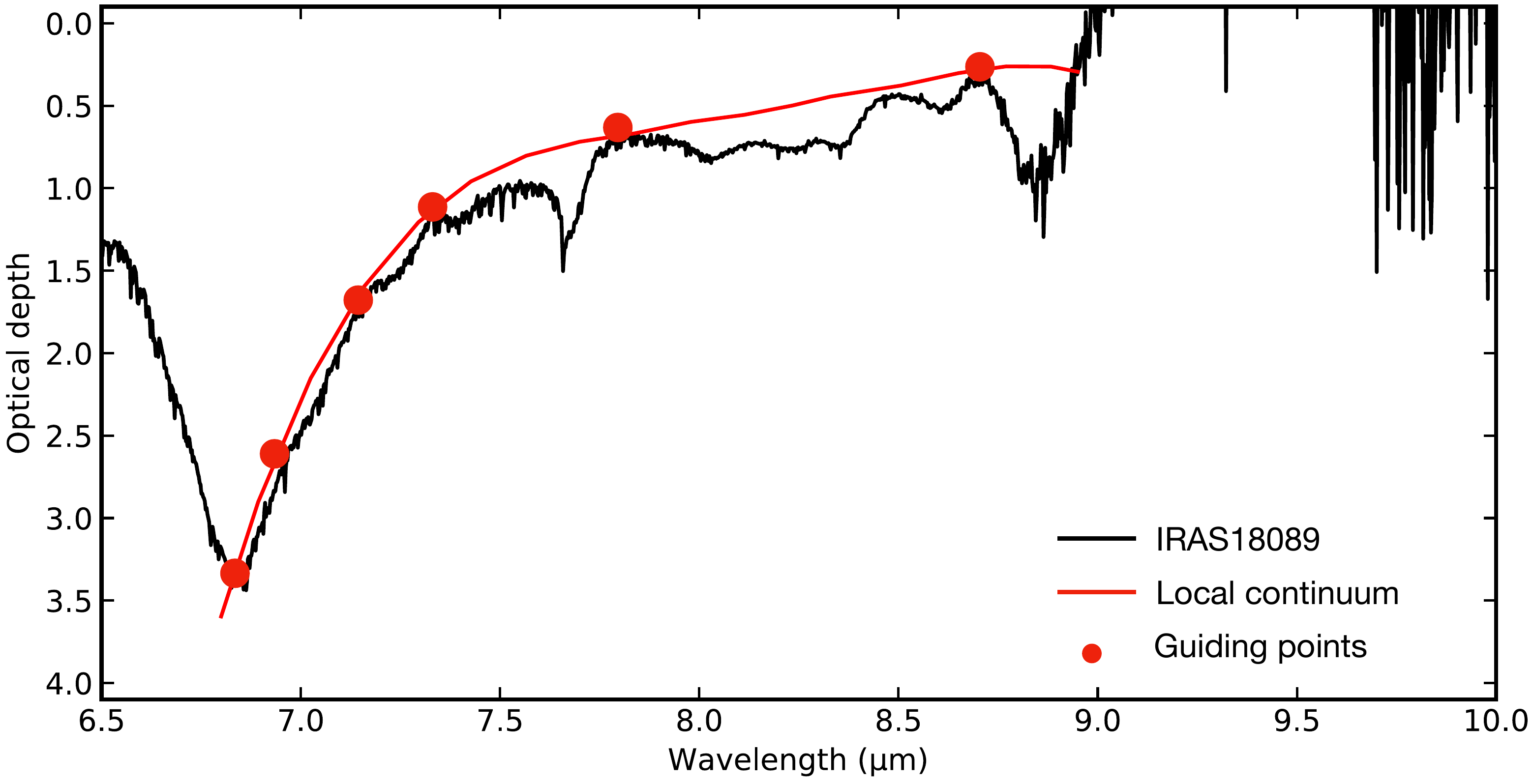}
\caption{Local continuum fit and guiding points for the 7--9 $\mu$m
  range of the IRAS 18089 spectrum, after subtraction of the global
  continuum, silicate feature and water libration band.}
\label{fig:IRAS18089localfit}
\end{centering}
\end{figure}

Table~\ref{tab:ices} summarizes the column densities of the molecules
used to produce the fit in Figures~\ref{fig:IRAS18089COMfit} and
\ref{fig:IRAS18089NH4}. The column density used to fit the newly
detected 8.38 $\mu$m feature will be presented in a future paper
dedicated to the analysis of this band (Rocha et al., in prep.).

\begin{figure}
\begin{centering}
\includegraphics[width=9cm]{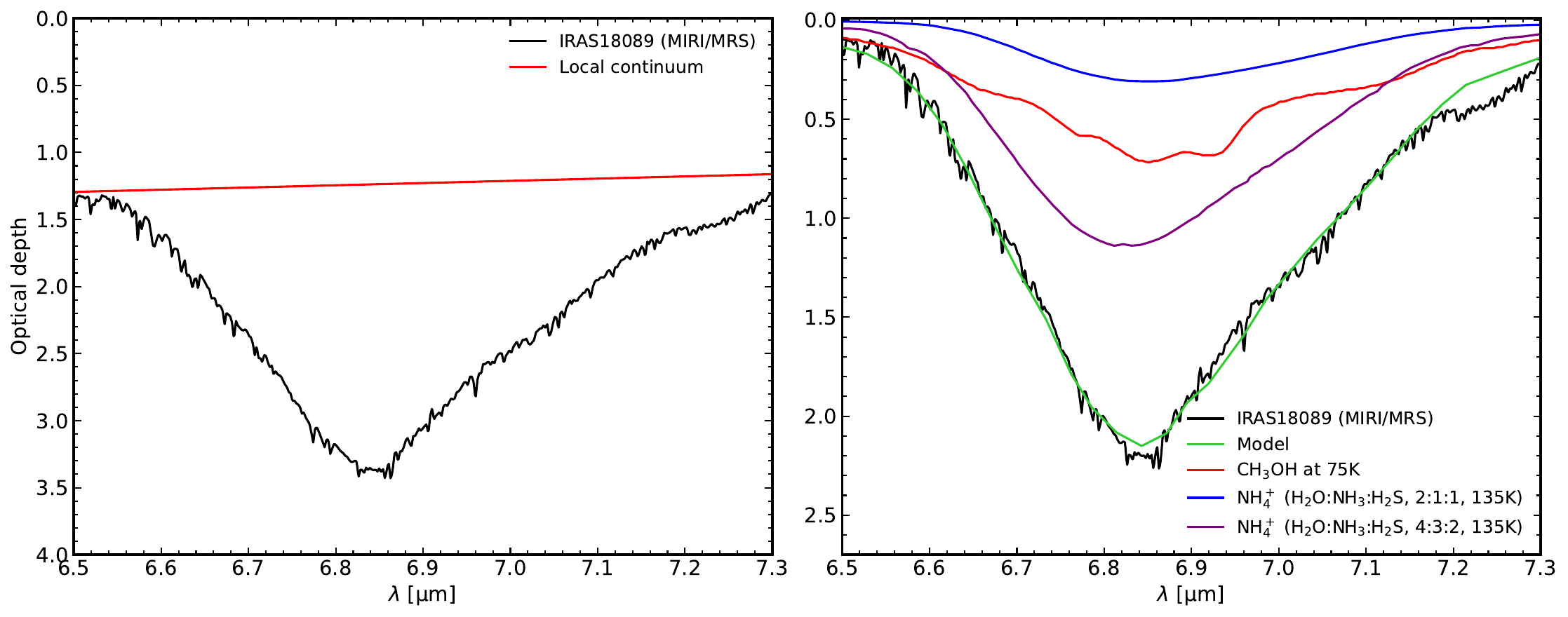}
\caption{Ice features at 6.85~$\mu$m toward IRAS18089. The left panel shows
the local continuum subtraction, and the right panel displays the best fit and
the individual NH$_4^+$ components and warm CH$_3$OH ice.}
\label{fig:IRAS18089NH4}
\end{centering}
\end{figure}

\begin{table}[t]
  \caption{Ice column densities corresponding to the best fit of the IRAS 18089
  spectrum.}
\begin{tabular}{llc}
\hline
\hline
Species & Name & $N$ \\
       & &     ($10^{17}$ cm$^{-2}$) \\
\hline
H$_2$O & water          & 282.3$\pm$70.5 \\
CH$_4$ & methane        & 5.3$\pm$1.3 \\
SO$_2$ & sulfur dioxide & 0.3$\pm$0.1 \\
OCN$^-$ & cyanate ion   & 5.3$\pm$1.5 \\
NH$_4^+$ & ammonium ion & 33.0$\pm$4.0 \\
HCOO$^-$ & formate ion  & 1.1$\pm$0.3 \\
H$_2$CO & formaldehyde  & 28.2$\pm$7.1 \\
CH$_3$OH & methanol     & 50.4$\pm$12.7 \\
HCOOH & formic acid     & 2.9$\pm$0.7 \\
CH$_3$CHO & acetaldehyde &1.9$\pm$0.5 \\
CH$_3$CH$_2$OH & ethanol & 4.2$\pm$1.1 \\
CH$_3$OCHO & methyl formate & 0.9$\pm$0.2 \\
CH$_3$COOH & acetic acid & $<$2 \\
CH$_3$OCH$_3$ & dimethyl ether & 2.8$\pm$0.7 \\
CH$_3$COCH$_3$ & acetone & 0.7$\pm$0.2 \\
\hline
\end{tabular}
\label{tab:ices}
\end{table}

\section{Extinction determinations}
\label{sec:app_extinction}

Most lines and diagnostics require a correction for extinction, which
is substantial for these deeply embedded sources. As discussed in
detail by \citet{vanGelder24,vanGelder24overview} the correction
consists of two parts: (i) an absolute extinction ($\tau_{\rm dust}$)
based on an extinction law; and (ii) differential extinction caused by
various ice absorption bands and silicates ($\tau_{\rm ice,
silicates}$) at specific wavelengths.

Several formulations of the extinction law exist in the literature
\citep[e.g.,][]{Weingartner01,Gordon23}; they are usually optimized
for diffuse cloud dust from UV to infrared and include already some
form of silicate absorption. The JOYS+ papers to date use either the
\citet{McClure09} law or the \citet{Pontoppidan24ext} law (KP5), which
have been developed for the mid-infrared extinction due to dust in
dense clouds. The KP5 relation contains also some generic ice features
and has a deeper silicate absorption than the \citet{McClure09} law.

In the JOYS+ papers, the local ice extinction is usually estimated
from the observed spectrum itself and can be substantial: for example,
the gaseous CO$_2$ $P-$branch lines can be fully extincted by the
CO$_2$ ice band, but this does not hold for the $R-$branch lines
\citep{Francis24}. This example also illustrates that the appropriate
extinction to use depends on the origin of the emission. For example,
hot core emission from molecules such as CO$_2$ and H I recombination
lines originating close to the protostellar embryo experience more
extinction than lines coming from an outflow cavity.

One way to determine the absolute extinction in units of $A_K$ or
$A_V$ is to use one of the above global extinction laws, following the
procedure as described in, for example, Appendix C of
\citet{vanGelder24}.  There are at least three other methods that can
and have been adopted.  The first method is to measure just the
optical depth of the silicate feature at 9.7 $\mu$m, $\tau_{\rm 9.7}$,
which represents the total extinction along the line of sight to the
dust sublimation radius, very close to the protostellar embryo. Its
relation with $A_K$ has been well calibrated by comparing
$\tau_{\rm 9.7}$ from {\it Spitzer} spectra against extinction
measurements of reddened background stars behind dense
clouds. \citet{Boogert13} (their equations (4) and (5)) find that the
relation changes between diffuse cloud ($A_V\lesssim 10$~mag) and
dense cloud dust in the Lupus star-forming region. These relations
have more recently been confirmed for other clouds by
\citet{Madden22}. Assuming $A_V/A_K$=7.7 as appropriate for $R_V=5.5$
\citep{Cardelli89}, the \citet{Boogert13} diffuse cloud relation leads
to $A_V=18.5 \times \tau_{\rm 9.7}$.  For dense cloud dust, it becomes
$A_V=28.6 \times \tau_{\rm 9.7}$.  The latter relation is appropriate
for deeply embedded low-mass Class 0 and high-mass sources, and has
indeed been used for IRAS 18089 in Appendix~\ref{sec:app_IRAS18089}. Class
I sources have lower extinctions but generally still well above 10
mag. However, use of the dense cloud relation has been found to lead
to unreasonably high accretion rates, and hence the diffuse cloud relation
is adopted for Class I sources in Sect.~\ref{sec:accretion}.

A second, related method is to use the observed relations of ice
optical depths with $A_V$ \citep{Whittet01,Boogert15}. The best-known
relation is that of the 3 $\mu$m water ice feature, $\tau_{3.0}$, versus
$A_K$ calibrated by \citet{Boogert13,Madden22} for dense clouds. Since
the 3 $\mu$m ice band is often saturated in the most deeply embedded
JOYS+ sources, an alternative is to use the 15 $\mu$m CO$_2$ ice band
$\tau_{15}$ \citep{Bergin05,Whittet09} which can now be readily
observed with JWST. Further calibration of the $\tau_{15}$ versus $A_K$
relation for dense clouds is warranted.

A third measure of the extinction can be obtained from the excitation
of H$_2$: rotational diagrams using lines in the MIRI wavelength range
show that the S(3) line flux at 9.66 $\mu$m clearly falls below that
of the neighboring H$_2$ lines due to enhanced silicate extinction, as
has also been found in ISO and {\it Spitzer} data. Various papers show
how both the extinction and the H$_2$ OPR can be
fitted simultaneously to H$_2$ rotation diagrams to obtain a value of
$\tau_{\rm S(3)}$ which can then be converted to $A_V$ using an
extinction law \citep[e.g.,][]{Neufeld06,Giannini11,Francis25HD,Okoda25}. This
extinction value should be appropriate for any emission lines
originating from the outflow and its cavity. If NIRSpec data are
available, one can add higher excitation H$_2$ lines at shorter
wavelengths to this analysis.

A more effective way to use NIRSpec data may, however, be to obtain an
extinction estimate from atomic or molecular lines lines that
originate from the same upper level; the MIRI range does not contain
line pairs that can readily be used for this purpose. The intrinsic
ratio of these line pairs depends only on their Einstein $A$ values
and frequencies, so the difference between observed and theoretical
flux ratios directly gives the local reddening at the corresponding
wavelengths. For H$_2$, the 1--0 S(1) line at 2.12 $\mu$m and 1--0
O(5) line at 3.23 $\mu$m are a well-known line pair that can be used
for this purpose: entire outflow extinction maps can even be obtained
using narrow-band filters containing these lines with NIRCam
\citep{Ray23,Caratti24}. Similarly [Fe II], H I or other atomic lines
originating from the same upper level in the 0.4--2 $\mu$m range can
be useful extinction tracers in regions that are not too extincted
\citep[e.g.,][]{Giannini15}. We note that for JOYS+ sources, high
spectral resolution NIRSpec data are limited to $>$3
$\mu$m. \citet{Assani25} have recently introduced a procedure for
comparing extinction estimates from [Fe II] NIRSpec and MIRI data that
warrants further exploration.

As noted in Appendix~\ref{sec:app_IRAS18089}, no extinction correction
is applied to absorption features of ices or silicates since they
directly measure the column along the line of sight. Molecular
absorption lines originating in hot core gas lying interior to the
outer icy envelope would require an extinction correction, but in
practice only the differential correction due to ices (e.g., the
CO$_2$ gas + ice case) has been applied in the JOYS papers
\citep[e.g.,][]{vanGelder24overview}. Also, a covering fraction of
unity by the absorbing material is assumed (see also
Sect.~\ref{sec:gasanalysis}). Since the main interest is in relative
column density ratios, rather than absolute ones, this should not
change significantly any conclusions unless the lines are heavily
saturated.

In this paper all of these methods to determine extinction have been
used, often to be consistent with earlier JOYS papers. A systematic
comparison between them is left for a future study. It should be noted
that differences in absolute values of, most notably, accretion rates
can be substantial depending on the adopted method. 

\end{appendix}

\end{document}